\documentclass[aps,epsfig,prd,superscriptaddress]{revtex4-2}
\usepackage{bbm}
\usepackage{comment}
\usepackage {chngcntr}
\usepackage{graphicx}
\usepackage{amsmath,mathtools}
\usepackage{latexsym}
\usepackage{amsfonts}
\usepackage{amssymb}
\usepackage{array}
\usepackage{placeins}
\usepackage[dvipsnames]{xcolor}
\usepackage{dsfont}
\usepackage{verbatim}
\usepackage{ulem}
\usepackage{graphicx} 
\usepackage[english]{babel}
\usepackage{forest} 
\usepackage{braket}

\usepackage[colorlinks=true, allcolors=blue]{hyperref}

\newcommand{\ii}{\mathrm{i}}

\begin{document}

\title{Quantum Phase Estimation Beyond the Gaussian Limit}
\author{Kimin Park}
\email{park@optics.upol.cz}
\affiliation{
Department of Optics, Palacký University, 17. Listopadu 1192/12, 77146 Olomouc, Czech Republic}
\author{Tanjung Krisnanda}
\affiliation{Centre for Quantum Technologies, National University of Singapore, Singapore 117543, Singapore}
\author{Yvonne Gao}
\affiliation{Centre for Quantum Technologies, National University of Singapore, Singapore 117543, Singapore}
\affiliation{Department of Physics, National University of Singapore, Singapore 117542, Singapore}
\author{Radim Filip}
\email{filip@optics.upol.cz}
\affiliation{
Department of Optics, Palacký University, 17. Listopadu 1192/12, 77146 Olomouc, Czech Republic}
\date{\today}

\begin{abstract}
{Quantum metrology aims to enhance measurement precision beyond the standard quantum limit (SQL), the benchmark set by classical resources, enabling advances in sensing, imaging, and fundamental physics. 
A critical milestone beyond the SQL is surpassing the Gaussian bound—the fundamental precision limit achievable with any Gaussian state, such as optimally squeezed states. 
Certain non-Gaussian states, specifically asymmetric superpositions of coherent states (SCS) and superpositions of a vacuum and a Fock state (ON states), can outperform this Gaussian bound within an intermediate energy range.}
In particular,  asymmetric SCS emerge as a highly practical resource for near-term quantum sensing architectures operating beyond the Gaussian limit {due to their efficient preparation {and processing} via a constant-complexity} {protocol}.
Our comprehensive analysis under realistic loss, noise, and detection schemes quantifies the critical trade-off between achievable precision and the operational range of the non-Gaussian advantage.
This work sheds light on the fundamental impact of non-Gaussianity and asymmetry on metrological tasks, and offers insights on how to leverage such resources in realistic near-term quantum enhanced sensors beyond the Gaussian limit.

\end{abstract}

\maketitle

\section{Introduction}

Quantum metrology harnesses   nonclassical states of light and matter to achieve accuracies beyond the standard quantum limit (SQL), {the benchmark set by classical resources}, where the maximum extractable information scales linearly with quantum  average particle number \cite{Giovannetti2004SQL,Paris2009Estimation,GiovannettiNatPho2011,ZhangPRL2015,VahlbruchPhysRevLett2016Squeezed.117.110801,DemkowiczNatureComm2012Heisenberg,Degen2017RevModPhys.89.035002Sensing,Nolan2017PhysRevLett.119.193601,PirandolaNatPho2018PhotonicSensing,KwonPhysRevLett2019.122.040503,LenNatComm2022ImperfectMeasurement,Maleki2023SpeedLimitMetrology, PezzeRevModPhys.90.035005}. 
Overcoming the SQL is a well-established goal and crucial first step toward the Heisenberg limit, a fundamental quadratic scaling of sensitivity.
The next critical milestone is surpassing the precision of the more stringent {Gaussian bound—the benchmark for optimal Gaussian strategies} 
(typically involving displaced squeezed states \cite{PinelPRA2013Gaussian,abbottLIGO2016observation,Lawrie2019SqueezedSensing}), under relevant constraints like fixed average {number of quanta}.  
For practical and efficient quantum sensors  reliably exceeding {the Gaussian bound}, deterministic sources of non-Gaussian states  robust to noise and loss  are essential.
Non-Gaussian resources, such as superpositions of vacuum and a single Fock state (ON states) \cite{WangPhysRevLett2017ConvertingON, Wang2019,McCormick2019, WOLF2021100271,huang2025fastsidebandcontrolweakly},  multi-mode entangled states (e.g., NOON states \cite{Dowling01032008,NielsenPhysRevLett2023.130.123603NOON,huang2025fastsidebandcontrolweakly} and entangled coherent states \cite{JooPhysRevLett2011ECS,JooPRA2012ECS,Cernotik2024PhysRevResearch.6.033074SWAP,ChenPhysRevA2024AECS}), and superpositions of coherent states (SCS) \cite{MunroPRA2002SCS,JooPhysRevLett2011ECS, JooPRA2012ECS,KnottPhysRevA.89.053812ECS,TatsutaPhysRevA2019.100.032318GeneralizedCat,LeePhysRevA2020.101.012332ECS,CottePRR2022.4.043170,Pan2024,tatsuta2024generationmetrologicallyusefulcat} 
{are explored for their potential to achieve high phase sensitivity}~\cite{OszmaniecPhysRevX2016.6.041044Random,tan2019nonclassical,Xu2022PhysRevLett.128.150501NonGaussianSuperconducting,Rahman2024Fock,Fadel2024}.

Recent experiments in various quantum systems, such as superconducting circuits \cite{Touzard2019,SchoelkopfNature2020,Kwon2021Superconduting,MaSciBull2021SupCond,Ripoll2022SuperconCircuit}, trapped ions \cite{Leibfried2003,KienzlerPRL2016,Bruzewicz2019APRtrappedion,FluhmannHomeNature2019,HackerCatNatPho2019,BrownNRM2021TrappedIon,SchuppPRX2021TrappedIon}, quantum electromechanics \cite{TouzardPhysRevLett2019RabiElectromechanics}, and spin-mechanics \cite{Dareau2018PhysRevLettSpinmechanics}, have enabled {the deterministic and fast creation of large, adaptable SCS}. 
These experiments leverage readily available ultra-strong {qubit-oscillator interactions (Rabi-type gates) to create SCS with a fixed, small number of operations}.
This {positions SCS as a practical resource for} advanced quantum metrology and highlights their potential for precision exceeding  Gaussian bounds, {comparable to other non-Gaussian states like} displaced Fock states \cite{deng2023heisenberglimited} and traditional Fock state superpositions {(including ON states)} \cite{Lee2019,McCormick2019,Wang2019}.

SCS are a superposition of two coherent states where the superposition weight controls the relative amplitude.
A special case is the symmetrical SCS with equal weights, which have been studied for their potential in metrology \cite{MunroPRA2002SCS,JooPhysRevLett2011ECS, JooPRA2012ECS,KnottPhysRevA.89.053812ECS,LeePhysRevA2020.101.012332ECS,CottePRR2022.4.043170}. 
More broadly, asymmetrical SCS, where the weights are unequal, offer further flexibility and control over the state's properties \cite{Pan2024}, and  are  key to unlocking a quantum advantage over the Gaussian bound using SCS. 
Both asymmetric SCS and ON states offer powerful, distinct paths to surpass the Gaussian bound. 
This work theoretically shows and compares their performance, showing that optimized asymmetric SCS are highly competitive, particularly when considering the trade-off between metrological gain and preparation efficiency in realistic single-mode scenarios \cite{BraunRMP2018}.  
Under ideal conditions, optimized asymmetrical SCS and ON states enhance estimation precision and approach the Heisenberg limit's quadratic scaling.
Crucially, while {even} symmetric SCS can surpass practical Gaussian benchmarks {by suboptimal measurements} under certain practical conditions, {we show that} asymmetry is essential to overcome the fundamental Gaussian bound.

In contrast to idealized theoretical scenarios, practical quantum sensors must operate in noisy environments \cite{Zhou2023PRXQuantumOptimalNoisyMetrology}.
This work addresses this challenge by investigating the performance of SCS {and ON states} under realistic conditions, demonstrating their robustness and potential for practical quantum metrology beyond the Gaussian bound. 
In realistic scenarios with ambient loss and decoherence, the performance of SCS is highly dependent on the parameters  and particularly their relative weighting, and the quadratic scaling is quickly reduced to a linear scaling despite ideal state preparation~\cite{EscherNatPhys2011ScalingTransition}. 
{In practice, this non-Gaussian advantage may be confined to a very narrow operational range of phase angles, or may depend critically on the chosen detection scheme.}
Therefore, careful optimization of parameters in terms of classical Fisher information (CFI) reflecting practical detection schemes is crucial for practical quantum metrology.
{Under realistic loss and noise, we show that the CFI for optimized asymmetric SCS predicts a lower mean estimation error surpassing that achievable by any Gaussian state, and is competitive with other leading non-Gaussian probes, such as ON states. 
This advantage holds for an intermediate range of average quanta while maintaining a consistent estimation operational range.}

\begin{figure}
    \centering
    \includegraphics[width=1.0\linewidth]{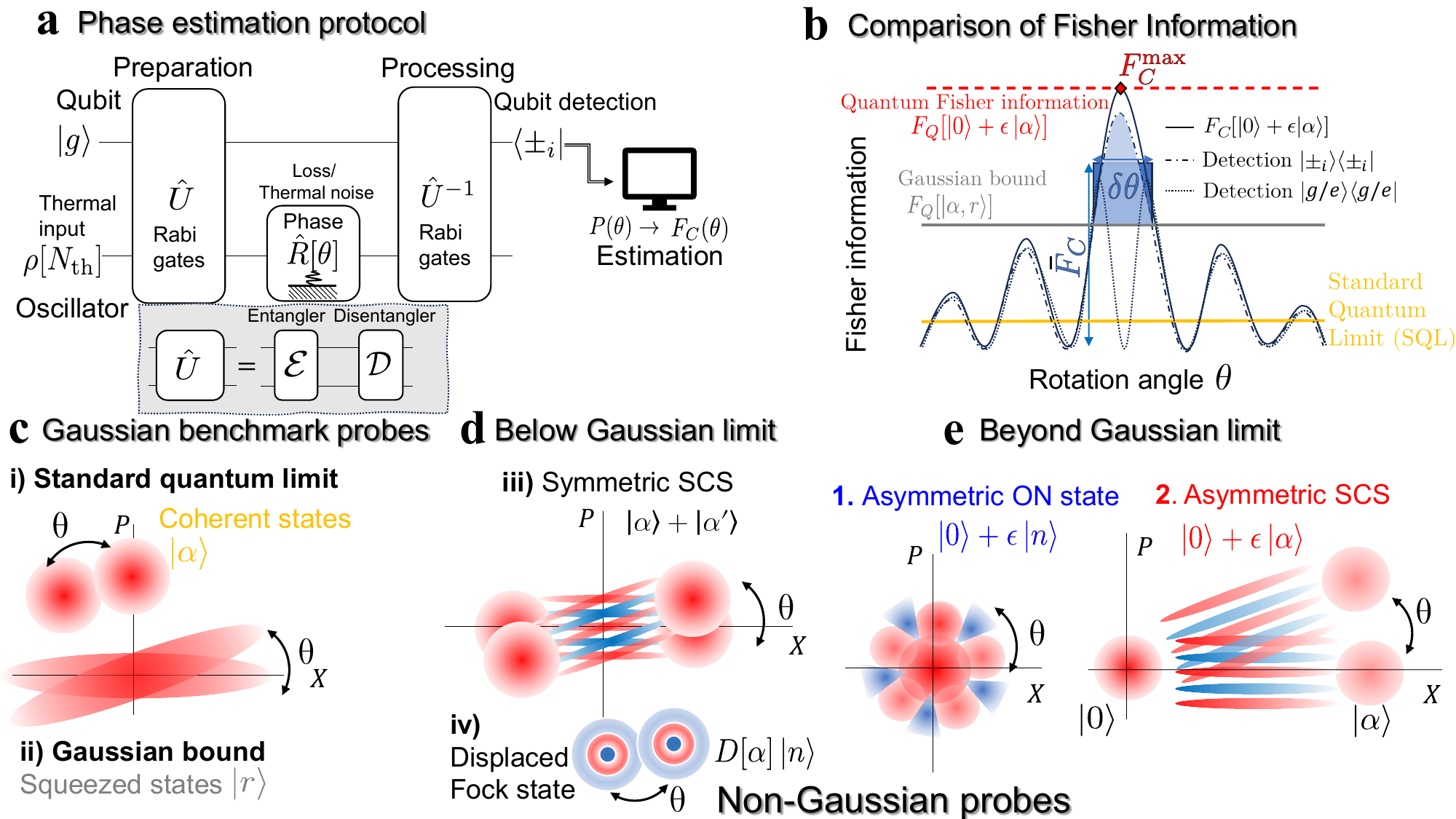}
\caption{Phase sensing with asymmetric SCS outperforming the Gaussian bound. 
\textbf{a}) {Deterministic} phase estimation protocol. 
The oscillator and an ancillary qubit, initially in the ground state (in general, a thermal input), undergoes two controlled-displacement operations (Rabi gates) $\mathcal{E}$ (entangler) and $\mathcal{D}$ (disentangler), deterministically preparing an asymmetric SCS {with the qubit left nearly in a ground state}.
An unknown phase rotation by $\theta$, the target of the sensing, affects the oscillator, {where loss or thermal noise may occur simultaneously}. 
Subsequently, the inverse operations and a qubit projective measurement in the Pauli matrix $\hat{\sigma}_y$ basis follows. 
The estimator then extracts $\theta$ from the measurement data $P(\theta)$ {whose precision dictated by CFI $F_C(\theta)$}. 
Other non-Gaussian states often involves more complex setups. 
\textbf{b}) Fisher information comparison. 
The CFI for an optimized asymmetric SCS {using the robust $\ket{\pm_i}\bra{\pm_i}$ detection basis (dot-dashed)} surpasses both the standard quantum limit (SQL, yellow) and the more stringent Gaussian bound (gray) over a specific non-Gaussian enhancement range $\delta\theta$. 
{Under realistic noise, performance with the standard $\ket{g/e}\bra{g/e}$ basis (dotted) degrades significantly.}
The peak CFI, $F_C^{\text{max}}$, coincides with the QFI $F_Q$ in the ideal case. 
Imperfect control and realistic environmental effects {(loss, or thermal noise)} reduce the CFI (blue dot-dashed and dashed curves), while optimal measurement scheme utilizing qubit detection can help to preserve the central peak \cite{Zhou2023PRXQuantumOptimalNoisyMetrology}. 
The mean CFI, $\bar{F}_C$, over the range $\delta\theta$ provides a measure of overall estimation performance. 
\textbf{c}) Wigner function of Gaussian probes for benchmarks: i) Coherent states representing the standard quantum limit, ii) Squeezed states representing the Gaussian bound. 
\textbf{d}) Non-Gaussian probes for comparison:  iii) Symmetric SCS $\ket{\alpha}+\ket{\alpha'}$, iv) Displaced Fock states. 
\textbf{e}) Probes surpassing the Gaussian bound: 1. {asymmetric} ON states, 2. asymmetric SCS. 
The asymmetry is key to their high performance for a fixed {average number of quanta}.
}
    \label{fig:setup}
\end{figure}

We focus on SCS that  can be prepared and measured each with two Rabi gates available in many qubit-oscillator systems {as in Fig. \ref{fig:setup}\textbf{a}.} 
Unlike an interferometric setup \cite{Park2023quantumrabi}, the mutual coherence with the qubit after the preparation is not exploited.
By accounting for various forms of decoherence  and proposing measurement strategies to mitigate  degradation from increasing loss and noise, the robustness of precision achieved by optimized SCS probes demonstrates their superiority beyond Gaussian benchmarks. 
We compare our schemes under such realistic considerations with cutting-edge non-Gaussian probes such as superposition of {ON states} \cite{WangPhysRevLett2017ConvertingON, Wang2019,McCormick2019, WOLF2021100271,huang2025fastsidebandcontrolweakly} and displaced Fock states \cite{deng2023heisenberglimited} to clarify their non-Gaussian advantages. 





\section{Results}

Here, we demonstrate the central result of this work: asymmetric SCS probes can achieve a phase estimation precision that surpasses not only the SQL but also the more stringent Gaussian bound. 
We show that their performance is highly competitive with other leading non-Gaussian resources, such as ON states, and offers distinct advantages in practical {phase estimation} scenarios. 
Specifically, the advantage of asymmetric SCS is prominent when considering their efficient preparation with a fixed number of gates.

We first analyze the fundamental precision limits using the QFI. 
We then evaluate the practical performance under realistic detection schemes using the CFI. 
This dual analysis allows for a direct comparison of the strengths and weaknesses of different quantum probes under identical conditions, {bridging the gap between the ultimate theoretical sensitivity  and the practically achievable precision}. 
Because the effects of bosonic loss, thermal noise, and the estimated phase rotation are all phase-insensitive, they commute; their order of application is therefore inconsequential to the final state.

\subsection{Quantum Fisher information and Cramér-Rao bound}

{The Cramér-Rao bound (CRB) establishes the fundamental asymptotic limits of precision} \cite{braunsteinPRL1994distance,Hervas2025PhysRevLett.134.010804}. 
The variance of an unbiased estimator $\hat\theta$  about the phase $\theta$ is bounded by the reciprocal of the Fisher information, which quantifies the sensitivity of a probability distribution to changes in the parameter, expressed as:
\begin{equation}
\text{Var}(\hat{\theta}) \geq \frac{1}{F_C},
\end{equation}
where $F_C$ is the CFI, a measure of practical estimation precision,  defined later in \ref{sec:CFI}. 

In the quantum framework, the QFI, $F_Q$, {generalizes} this idea to quantify the maximum achievable precision over all possible quantum measurements~\cite{braunsteinPRL1994distance}. 
The quantum generalization of the CRB is given by:
\begin{equation}
\text{Var}(\hat{\theta}) \geq \frac{1}{F_Q[\rho_\theta]}\equiv (d\theta)^2,
\end{equation}
where $F_Q[\rho_\theta]$ quantifies the distinguishability between quantum states $\rho_\theta$ and $\rho_{\theta + \mathbbm{d}\theta}$ with an infinitesimal difference $\mathbbm{d}\theta$. 
{For a pure state $|\psi\rangle$ undergoing a unitary phase rotation $R[\theta]=\exp[-\ii\theta \hat{G}]$ generated by a Hermitian operator $\hat{G}$, the QFI is given simply by 
\begin{align}F_Q[|\psi\rangle]=4(\Delta \hat{G})^2,\label{eq:FQG}\end{align} the variance of the generator \cite{Paris2009Estimation}. 
In our work, $\hat{G}=\hat{n}$, and the phase $\theta$ is a rotation angle.}
%
For a mixed state with a rotated density matrix $\rho_\theta=\exp[\ii \theta \hat{n}]\rho\exp[-\ii \theta \hat{n}]$ where $\hat{n}$ is the number operator,  QFI is given as 
\begin{align}F_Q[\theta]=\lim_{\mathbbm{d}\theta\theta\to 0}\frac{4(1-\mathbb{F}[\rho_\theta,\rho_{\theta+\mathbbm{d}\theta}])}{\mathbbm{d}\theta^2}, \label{eq:FQfid}\end{align} 
where $\mathbb{F}$ is the Uhlmann's fidelity \cite{braunsteinPRL1994distance,Jozsa1994}.  
{While other equivalent expressions exist (e.g., using the spectral decomposition of $\rho$ \cite{Paris2009Estimation}), the fidelity-based definition offers a physical intuition.}
Crucially, the CFI is bounded by the QFI for any given probe state, as the QFI represents the maximum information obtainable over all possible quantum measurements (POVMs), while the CFI quantifies the information extracted by one \textit{specific} choice of measurement. This establishes the fundamental hierarchy $F_C \le F_Q$. The equality, $F_C = F_Q$, is achieved only if the measurement is optimal for the given estimation task. This relationship is central to our work, as we aim to find practical measurement schemes whose CFI approaches the fundamental QFI limit, thereby maximizing the extractable precision.

The QFI comparison across states depends heavily on the applied constraints, as shown in Appendix \ref{append:CFIrange}. 
Traditional choice is the average number of quanta contained in the {prepared} state, $N_\mathrm{av}$, directly 
impacting both scalability, susceptibility to noise, and the disturbances it produces. 
As a high quality of the prepared probe state is generally required, we denote the $N_\mathrm{av}$ of the initial pure probe state before $\hat{R}[\theta]$, loss- and noise-independent. 
For quantum non-Gaussian states of our interest, a higher $N_\mathrm{av}$ generally leads to increased sensitivity but also greater experimental complexity and faster decoherence due to stronger environmental interactions. 
This regime of a large $N_\mathrm{av}$ is vital for pushing the boundaries of quantum-enhanced metrology, exploring the trade-off between enhanced signal strength and increased noise. 
Conversely, a small $N_\mathrm{av}$ minimizes expected heating and disturbances in the phase process where $\hat{R}[\theta]$ is estimated, offering robustness at the cost of lower precision, while many copies of them can provide comparable precision under the average quanta number constraint. 
{We model environmental loss and noise as occurring primarily during the phase sensing interval $\hat{R}[\theta]$, after state preparation. 
This represents a common scenario where a well-prepared quantum probe interacts with an external, uncontrolled system to acquire a phase shift. 
Mathematically, this model is robust because the phase rotation operator commutes with the bosonic loss and thermal noise channels. 
Decoherence is treated as a single effective channel acting on the prepared probe, providing a clear framework to compare the intrinsic resilience of different quantum states. 
The validity of this effective model for complex preparation sequences, such as for SCS, is further supported by the numerical simulations in Appendix~\ref{app:pulse_simulations}.}
Loss  and noise presented only during phase rotation merely rescales the average number of quanta in the state $N_\mathrm{av}$ linearly, affecting all states equally as 
\begin{align}N_\mathrm{av}[\eta]=\eta N_\mathrm{av}+(1-\eta)N_\mathrm{th}, \label{eq:Navloss}\end{align} 
with  the transmission coefficient $\eta$, that justifies our idealizations of the pure initial states and simplify following description {in the limit of $\eta\approx 1$ and $N_\mathrm{th}\ll 1$.}

\subparagraph{Benchmarks: SQL and Gaussian bound for QFI}
{Two key benchmarks define the landscape of quantum precision enhancement.}
A coherent state $\ket{\alpha}$, the closest quantum analog to a classical state, exhibits Poissonian statistics and has an average quanta number $N_\mathrm{av} = |\alpha|^2$.
For coherent states, the QFI is $F_Q[\ket{\alpha}]=4N_\mathrm{av}$, setting the benchmark for phase sensing at the SQL \cite{HelstromBook1976}. The linear scaling of QFI of this state signifies the classical nature of it, {and surpassing the SQL signifies a non-classical advantage}.
However, these ideal QFI values are naturally degraded by realistic, even small, imperfections such as loss and thermal noise.
In our model, we consider bosonic loss as a linear attenuation of the signal mode, and thermal noise as the addition of a thermal state to the signal mode. These effects are modeled mathematically to understand their impact on the QFI and CFI of various probe states.
The state under a loss remains as a coherent state with a diminished coherent amplitude. 
{We note that $N_\mathrm{av}$ is loss- or noise-independent. 
The average number of quanta after loss is related in a state-independent manner, as in (\ref{eq:Navloss}).}
The QFI is then reduced as $F_Q[\ket{\alpha}]=4\eta N_\mathrm{av}$. 
With the addition of thermal noise, it is given as \cite{AspachsPhysRevA2009.79.033834thermalGaussian,PezzeRevModPhys.90.035005}
\begin{align}
F_Q[\ket{\alpha}]=4  \lambda_C  N_{\text{av}},
\label{eq:FQcohthermal}
\end{align}
and therefore,  loss and thermal noise do not change the linear scaling vs $N_\mathrm{av}$, only the monotonously decreasing multiplicative suppression prefactor $\lambda_C\equiv\eta(2(1-\eta)\eta N_\mathrm{th}+1)^{-1}$ against $\eta$ and  environmental thermal occupation number $N_\mathrm{th}$.

The second, more stringent sensing benchmark is the \textit{Gaussian Bound}. To surpass the SQL, one can employ 
the entire class of Gaussian states, typically optimized {over pure} displaced squeezed states $\ket{\alpha,\zeta}$ (with displacement amplitude $\alpha$ and squeezing parameter $\zeta$).
These states reduce quantum noise in one quadrature at the expense of increased noise in the conjugate quadrature.
It has average quanta $N_\mathrm{av}=\sinh^2|\zeta|+|\alpha|^2$, and quanta number variance $V_\mathrm{sq}=\frac{4 \alpha  \alpha ^* \cosh (2 \zeta )-2 \left(\alpha ^2+\left(\alpha ^*\right)^2\right) \sinh (2 \zeta
   )+\cosh (4 \zeta )-1}{4}$
, surpassing coherent state variance at non-zero squeezing. 
Under the constraint of fixed $N_\mathrm{av}$, the optimal QFI is achieved with a squeezed vacuum state with $\alpha=0$ for $\eta=1$ {regardless of $N_\mathrm{th}$}, yielding the \textit{ideal Gaussian bound} \cite{MonrasPRA2006optimal,PinelPRA2013Gaussian,PezzeRevModPhys.90.035005}: 
\begin{align} 
    F_Q^\mathrm{Gauss} = 8 N_\mathrm{av}^2 + 8 N_\mathrm{av}. \label{eq:GaussianLimit}
\end{align}
In this ideal noiseless and lossless case, squeezed states are the optimal Gaussian states, offering a quadratic scaling of QFI with $N_\mathrm{av}$.
This bound is fundamental for \textit{all} Gaussian resources; as presented as a numerical conjecture in Appendix~\ref{append:mixtures}, even classical mixtures of Gaussian states cannot surpass this limit. 
This solidifies the Gaussian bound as a stringent benchmark that can only be overcome with genuine non-Gaussian states. 
However, when ambient loss and noise are present in the sample after even ideal state preparation, the Gaussian bound has to be re-derived for general Gaussian state.
However, even under  weak loss $\eta\approx 1$ and low noise $N_\mathrm{th}\ll 1$ {during the phase rotation}, the quadratic Heisenberg scaling of the Gaussian bound with respect to $N_\mathrm{av}$  degrades to a classical linear scaling \cite{EscherNatPhys2011ScalingTransition}, primarily due to the rapid decrease of QFI in the asymptotic large $N_\mathrm{av}$ limit, as shown below:  
\begin{align}
    F_Q^\mathrm{Gauss}[\eta]\approx   
    4 \lambda_G  N_{\text{av}}, ~ \lambda_G=\begin{cases}
    \frac{2 \eta ^2}{2 (1-\eta )^2 N_{\text{th}}^2+2(1-\eta) N_{\text{th}}+1}, & N_\mathrm{av}\ll 1,\\
    \frac{ \eta  }{(1-\eta ) \left(2 N_{\text{th}}+1\right)}, & N_\mathrm{av}\gg 1,\\
    \end{cases} 
    \label{eq:FQGausseta}
\end{align}
where at $N_\mathrm{th}=0$, it gives the scaling under loss. 
{As in all cases,  $N_\mathrm{av}$ is independent of loss and noise.}
In such noiseless limit, this equation shows a transition in scaling behavior around $N_\mathrm{av} \approx N_\mathrm{av}^\mathrm{trans}$, where $\lambda_G$ takes the middle value between the two asymptotic scalings {in (\ref{eq:FQGausseta})}, given at $N_\mathrm{av}^\mathrm{trans}=\frac{1}{2\eta(1-\eta)}$, occurring  around $N_\mathrm{av}^\mathrm{trans}\approx 50$ at $\eta=0.99$. 
The divergence at $\eta=1$ in a large average quanta number limit $ N_\mathrm{av}\gg 1$ is due to the change of  quadratic scaling {in Eq. (\ref{eq:GaussianLimit})} into linear scaling {in Eq. (\ref{eq:FQGausseta})}.  
This $N_{\text{av}}^{\text{trans}}$ intuitively signifies the threshold where, for Gaussian states, the metrological advantage of squeezing is increasingly undermined by loss. 
Beyond this point, the optimal Gaussian strategy tends to shift from squeezed vacuum towards coherent states as the cost of maintaining squeezing under decoherence begins to outweigh its benefits. 
This shift in optimal strategy underpins the change in the dominant behavior of the Gaussian bound (\ref{eq:FQGausseta}) as the average photon number changes from large to small values. 
Detailed analysis of Gaussian states under realistic conditions, including bosonic loss and thermal noise (Appendix \ref{sec:GaussianQFI}), reveals an expected significant degradation in their quantum advantage. These imperfections cause the ideal quadratic scaling of Gaussian states to diminish, leading to a more limited practical enhancement. This QFI described in (\ref{eq:FQGausseta}) provides a more realistic bound under small ambient loss and noise for both small and large $N_\mathrm{av}$.

In the limit of small loss and noise, although the asymptotic scaling vs $N_\mathrm{av}$ drops to linear as for the SQL, the multiplicative suppression factor $\lambda_G$
for the Gaussian states is increasing {for intermediate values of} $N_\mathrm{av}$, and is still substantially larger than the factor $\lambda_C$ of the SQL. 
Therefore, Gaussian states can substantially overcome coherent states {irrespective of $N_\mathrm{av}$, whenever  environmental effects are below a critical threshold}: $\eta\ge \eta_G[N_\mathrm{th}]$ \cite{SciPostPhys2025optimalGaussian}. 
For purely lossy environment $N_\mathrm{th}=0$, $\eta_G$ is given as $\frac{\sqrt{2 N_{\text{av}}+1}-1}{2 N_{\text{av}}}$, which has decreasing value from $0.5$ at $N_\mathrm{av}=0$. 
In this lossy environment, $\lambda_C=\eta$, and $\lambda_G=2\eta^2$, and in short, Gaussian prefactor is larger by $2\eta$ times than SQL prefactor. In general, {the Gaussian linear scaling factor is larger than classical one} $\lambda_G\ge\lambda_C$ for all $N_\mathrm{av}$ for $\eta\ge \eta_G[N_\mathrm{th}]\approx 0.5$.  
The maximum value of $\lambda_G$ is achieved in the large $N_\mathrm{av}$ limit asymptotically.
In the results below, we primarily use the idealistic Gaussian bound (\ref{eq:GaussianLimit}) {but we can also compare to realistic Gaussian bound in (\ref{eq:FQGausseta}) that gives applicable results to non-Gaussian states.} 
Overcoming the Gaussian bound requires non-Gaussian resources and demonstrates a more profound quantum enhancement. 
We now analyze the QFI for specific states in relation to these bounds.

\subparagraph{QFI of displaced Fock states}

While squeezed states represent the optimal Gaussian resource ideally, quantum non-Gaussian states {can be explored for} further enhancement possibilities. 
One such example is the displaced Fock state of the form $\ket{\alpha,n}=D[\alpha]\ket{n}$ that has an average number of quanta given as $N_\mathrm{av}=n+|\alpha|^2$, {assuring the inequality $N_\mathrm{av}\ge n$}. 
This state has been investigated recently experimentally \cite{deng2023heisenberglimited}, largely due to the increasing experimental availability of Fock states with high photon numbers $n$. 
{To be shown below, they ultimately fall short of the Gaussian bound.}


For all values of $N_\mathrm{av}$, the QFI of this state takes piecewise values given as:
\begin{align}
    F_Q[\ket{\alpha,n}]=4 \left(\left\lfloor\frac{1}{4} (3-2 N_\mathrm{av})\right\rfloor+N_\mathrm{av}\right) \left(2 \left\lfloor\frac{1}{4} (2N_\mathrm{av}+1)\right\rfloor+1\right),
    \label{eq:dispFockopt}
\end{align}
where the term in the first parenthesis is positive for all $N_\mathrm{av}$. This piecewise expression exhibits a clear \textit{quadratic} scaling with $N_\mathrm{av}$, approximately proportional to $\frac{(1+2N_\mathrm{av})^2}{2}$.
It approximately follows the approximate curve  very closely for all values of $N_\mathrm{av}$, differing maximally by only $2$ at $N_\mathrm{av}=2n-\frac{1}{2}$ for all integers $n$, other than in the limit of $N_\mathrm{av}\ll 1$, where it is approximated by $F_Q\approx 4N_\mathrm{av}$. 
{Eq. (\ref{eq:dispFockopt}) shows}  a quadratic scaling of QFI vs $N_\mathrm{av}$, implying {an ideal} quantum enhanced scaling. 
However, the assumption of increasing Fock number indefinitely is not practical due to the experimental limitation of requiring a large experimental resource of number of gates $\mathcal{O}(n)$, and usually there is a maximum Fock number $n_\mathrm{max}$. 
Furthermore, its QFI is only about $25\%$ of the Gaussian bound  (\ref{eq:GaussianLimit}) for all $N_\mathrm{av}$, even without imperfection, as shown in Appendix \ref{Appendix:QFI}.
{Under loss, they are below the lossy Gaussian bound (\ref{eq:FQGausseta}) at all $\eta$.}

\subparagraph{QFI of asymmetric ON states}
\label{Append:ON}

{An important class of non-Gaussian states are the ON states, which are superpositions of the vacuum and a single Fock state: $\ket{\psi}_\mathrm{ON} \propto \ket{0} + \epsilon \ket{n}$, where $\epsilon\in [-\infty,\infty]$. Their potential for high-precision phase estimation is governed by the QFI, benefiting from the intuitively large quanta number variance due to the extreme number distribution.}
Under the primary constraint of a fixed average number of quanta $N_\mathrm{av}$, their QFI is best expressed as:
\begin{align}
F_Q[\ket{\psi_\mathrm{ON}}] = 4N_{\mathrm{av}}(n_{\text{max}} - N_{\mathrm{av}}), \quad \text{where } n_{\text{max}} \geq N_{\mathrm{av}} \label{eq:QFION_revised}
\end{align}
Here, $n_{\text{max}}$ is the maximum Fock number component, which acts as a second, practical resource parameter. It is related to the state's average number of quanta $N_\mathrm{av}$ and asymmetry $\epsilon$ via $N_\mathrm{av} = n_\mathrm{max}\frac{\epsilon^2}{1+\epsilon^2}$, from which the condition $n_{\text{max}} \geq N_{\mathrm{av}}$ follows for all real $\epsilon$.

This formula {enables} theoretically unbounded precision, {rooted in that the QFI for phase rotation is proportional to the photon number variance from (\ref{eq:FQG}) as $F_Q= 4(\Delta \hat{n})^2$}. 
For a \textit{fixed} energy cost $N_\mathrm{av}$, the QFI grows linearly with $n_{\text{max}}$, considered as an experimentally feasible or reliable parameter to prepare and utilize within the protocol, serving as a practical limit due to factors like finite gate fidelity, decoherence affecting higher Fock states, or control limitations. 
By making the state highly asymmetric (decreasing $\epsilon \to 0$), one can, in principle, increase $n_{\text{max}}$ indefinitely while holding $N_\mathrm{av}$ constant, {thereby increasing the number variance~\cite{Fadel2024}}, implying arbitrarily high precision. 
In the ideal case, the ON state surpasses the Gaussian bound (\ref{eq:GaussianLimit}) when $n_{\text{max}} \geq 3N_{\mathrm{av}} + 2$.
Therefore, symmetric ON state which satisfies $n_\mathrm{max}=2N_\mathrm{av}$ does not violate the Gaussian bound.

However, achieving this {asymmetry and utilizing it for sensing} is practically challenging. 
Preparing states with high $n_{\text{max}}$ is  resource-intensive, typically requiring a number of {sequential qubit-oscillator interactions} that scales polynomially with $n_{\text{max}}$, $\mathcal{O}(n_{\text{max}})$ \cite{McCormick2019, huang2025fastsidebandcontrolweakly}. 
This leads to significant accumulated errors under realistic decoherence, imposing a practical limit on the achievable precision.
This contrasts with the $\mathcal{O}(1)$ gate complexity for preparing SCS via the two-gate protocol discussed later, a key practical distinction.




Asymmetric ON state under loss  modeled as a bosonic loss channel (with transmission efficiency $\eta$) {only during phase rotation} is described in detail in Appendix \ref{Appendix:QFI}, analyzing their tolerance to ambient weak loss and  the realistic scaling of QFI with $N_\mathrm{av}$. 
For a weak loss $1-\eta\ll 1$,  $ F_Q[\rho_\theta]$ does not increase indefinitely {anymore} by increasing $n$ but also suffers an attenuation factor $\eta^{2n}$, and therefore has a maximum value 
\begin{align}
     F_Q^\mathrm{opt}[\rho_\theta]\simeq\frac{4n_\mathrm{opt} (n_\mathrm{opt}-N_\mathrm{av}) N_\mathrm{av}\eta ^{ n_\mathrm{opt}}}{N_\mathrm{av} \eta ^{ n_\mathrm{opt}}+n_\mathrm{opt}-N_\mathrm{av}},
     \label{eq:9}
\end{align}
at optimal $n=n_\mathrm{opt}$ {that needs to be found numerically in general. In the asymptotic limits of a large average number of quanta and weak loss, we can predict the optimal values:}
\begin{align}
n_\mathrm{opt}\stackrel{N_\mathrm{av}\gg1}{\approx} \lfloor N_\mathrm{av}+\frac{1}{2}-\frac{1}{\log \eta}\rfloor\approx  N_\mathrm{av}-\frac{1}{\log \eta}\stackrel{1-\eta\ll 1}{\approx}N_\mathrm{av}+\frac{1}{1-\eta}, 
\label{eq:n_opt}
\end{align}
where the floor function $\lfloor\cdot\rfloor$   ensures that the optimal Fock number is a physically realistic integer value. Therefore, at this optimal Fock number, the ideal {indefinite} increase of the QFI by $n$ is  inaccessible already under a small loss {during the phase rotation}.  
The approximate optimal QFI is obtained by substituting (\ref{eq:n_opt}) into (\ref{eq:9}) as 
\begin{align}
 F_Q^\mathrm{opt}[\rho_\theta]\approx \frac{4 N_{\text{av}} \eta ^{N_{\text{av}}} \left(N_{\text{av}} \log \eta -1\right)}{\log \eta  \left(e-N_{\text{av}}  \eta ^{N_{\text{av}}}  \log \eta \right) },\end{align}
 asymptotically reduced to zero for a large $N_\mathrm{av}$  due to the attenuation factor $\eta^{N_\mathrm{av}}$. 
{At small $N_\mathrm{av}$ this expression still gives a good approximation, and can be further simplified to linear scaling} as 
\begin{align} 
F_Q^\mathrm{opt}[\rho_\theta]\stackrel{N_\mathrm{av}\ll 1}{\approx}-\frac{4}{e \log \eta}N_\mathrm{av}\stackrel{1-\eta\ll 1}{\approx} \frac{4}{e (1-\eta)}N_\mathrm{av}. 
\label{eq:FQoptNavsmall}
\end{align}
This result {enables overcoming} Gaussian bound (\ref{eq:GaussianLimit}) by optimized ON states for moderate values of $N_\mathrm{av}$, which is not possible by the displaced Fock states.
{The ON state QFI under loss provides non-Gaussian enhancement within a certain range $0\le N_\mathrm{av}\le N_\mathrm{av}^\mathrm{MAX}(\eta)$, where $N_\mathrm{av}^\mathrm{MAX}(\eta)\approx -\frac{0.172}{\log \eta }-0.937$ represents the \textit{threshold average quanta number} for ON states under loss $\eta\ge\eta_\mathrm{min}\approx 0.832$. Below this $\eta_\mathrm{min}$, Gaussian states offer better performance than ON states for all $N_\mathrm{av}$.}
Against the purely lossy Gaussian bound in the limit of $N_\mathrm{av}\ll 1$ given as $8\eta^2N_\mathrm{av}$ in (\ref{eq:FQGausseta}),  the threshold loss is given at $\eta_\mathrm{min}=1/\sqrt{e}\approx 0.607$.
The QFI under thermal noise is considered only numerically. 
{Again, the symmetric ON state does not violate even the lossy Gaussian QFI in (\ref{eq:FQGausseta}).}



\subparagraph{QFI of asymmetric SCS}

Building upon these concepts, we now focus on SCS. Unlike the Fock or ON states with definite Fock numbers, SCS has a well-defined phase, deterministically generated from the oscillator ground state using the Rabi gates. 
While ON states also {provides} non-Gaussian enhancement, their preparation often involves different resource scaling (e.g., $\mathcal{O}(n)$ gates \cite{McCormick2019}) compared to the $\mathcal{O}(1)$ gate protocol for SCS which is a focus of this work.
One of the simplest non-Gaussian extensions of the pure Gaussian states is to superpose them in the form of $c_1\ket{\alpha,r}+c_2\ket{\alpha',r'}$. This general form that includes nonzero squeezing $r,r'$, however, is non-trivial to prepare  with small number of Rabi gates on the oscillator ground state, which is our main resource.  
Even an analytical and numerical investigation of their QFI is not straightforward. 
From our brief analysis, squeezing with non-zero $r$  generally seems to improve the QFI at the cost of many Rabi gates \cite{HastrupPRL2020squeezing}, therefore we leave this broader aspect as a future exploration beyond the scope of this work.
Therefore, we narrow the focus to the simplest non-squeezed case $c_1\ket{\alpha}+c_2\ket{\alpha'}$, a superposition of coherent states. 
First consider a symmetric SCS $|\psi\rangle_\mathrm{SCS} = \frac{|\alpha\rangle + |-\alpha\rangle}{\sqrt{2\left(1+e^{-2|\alpha|^2}\right)}}$, with phase shifts generated by the photon number operator \(\hat{n}\). 
For this state, one obtains the QFI \cite{MunroPRA2002SCS}:
\[
F_Q[|\psi\rangle_\mathrm{SCS}] = 4\left[|\alpha|^2\,\tanh |\alpha|^2 + |\alpha|^4\,\mathrm{sech}^2 |\alpha|^2\right].
\]
The average photon number is given as: 
$N_\mathrm{av} = |\alpha|^2\,\tanh |\alpha|^2,$
in the limit of large \( |\alpha|^2 \) one recovers:
\[
F_Q[|\psi\rangle_\mathrm{SCS}]  \stackrel{|\alpha|^2\gg 1}{\to} 4N_\mathrm{av},
\]
saturating to classical SQL, 
{and therefore not surpassing the ideal Gaussian QFI bound for all $|\alpha|^2$. 
Nevertheless, as we will discuss for the CFI, even symmetric states can outperform a practical Gaussian benchmark when the measurement for the latter is suboptimal, which might be relevant for practical applications (see Appendix \ref{append:CFIrange}).}

Now shifting the main focus to the  binary asymmetric SCS, we can explore the impact of asymmetry between $\alpha,\alpha'$ and $c_1, c_2$ on their metrological performance.  
While a displaced symmetric cat state can surpass the SQL, it has not been shown to overcome the Gaussian bound~\cite{Pan2024}.
{Among them, the superposition involving the vacuum state ($\alpha_0=0$) is found to be the most efficient for phase estimation under the primary constraint of fixed average quanta $N_\mathrm{av}$ as detailed in Appendix~\ref{Appendix:QFI}. This highlights the critical role of asymmetry for SCS to achieve genuine non-Gaussian quantum advantage. Intuitively, the $\ket{\alpha_0}$ component adds significantly to the state's {average number of quanta} ($N_\mathrm{av}$) without contributing proportionally to the quantum interference features (related to the phase-space separation between the components) that enhance the QFI.} {Optimizing sensitivity per unit of {average number of quanta} ($F_Q/N_\mathrm{av}$) therefore favors maximizing the contrast (phase-space separation related to $\alpha$) with the component adding the least energy cost, which is the vacuum state.}
Consider the superposition of the vacuum state and a coherent state, defined as
\begin{align}
\ket{\psi}_\mathrm{SCS} = \frac{1}{\sqrt{2 e^{-\frac{\alpha^2}{2}} \epsilon + \epsilon^2 + 1}} \left( \ket{0} + \epsilon \ket{\alpha} \right),
\label{eq:2SCS}
\end{align}
where $\alpha$ is the coherent amplitude, assumed to be real without loss of generality due to the equivalence of the results by the rotated states, and $\epsilon$ is a weight parameter of the coherent state component in the superposition. The average number of quanta  for this state is calculated as $N_\mathrm{av} (\alpha,\epsilon)= \frac{\alpha^2 \epsilon^2}{2 e^{-\frac{\alpha^2}{2}} \epsilon + \epsilon^2 + 1}$. The superposition state in Eq.~(\ref{eq:2SCS}) exhibits interference fringes between two coherent peaks in phase space as shown in Fig. \ref{fig:setup}\textbf{e}. This interference suggests potential quantum advantages for sensing applications involving oscillator phase rotations, as the superposition enhances sensitivity beyond the SQL.
{
This state form maps directly onto efficient preparation from the oscillator ground state via the two-gate protocol, a key motivation for this work.}
In the limit of $\alpha \to 0$, we can approximate the coherent state $\ket{\alpha}$ using its Taylor expansion around $\alpha = 0$, leading to the simplified state
$\ket{\psi}_\mathrm{SCS} \approx \frac{1}{\sqrt{1 + \epsilon'^2}} \left( \ket{0} + \epsilon' \ket{1} \right)$,
where $\epsilon' = \alpha \frac{\epsilon}{\epsilon + 1}$. This approximation reveals a connection to the finite rank ON state (a superposition of zero and one quanta states). 
Correspondingly, in this small $\alpha$ limit, the average number of quanta simplifies to $N_\mathrm{av} \approx \frac{\alpha^2 \epsilon^2}{(\epsilon + 1)^2}$.

Binary SCS can be classified into three regimes based on the weight parameter  $\epsilon$ of the coherent state component: (i) vacuum-like asymmetric ($\epsilon<1$), (ii) symmetric ($\epsilon=1$), and (iii) coherent-state-like asymmetric ($\epsilon>1$). The latters (ii,iii), despite a higher mean quanta, are less sensitive than (i) for equal average quanta numbers, and thus is not {the primary} focus here (the symmetric case is  optimal in terms of QFI for $\epsilon=1$ under fixed $\alpha$ constraint  among all SCS, as was shown in Appendix \ref{Appendix:QFI}). 
Negative $\epsilon$ is not of the main focus due to its higher mean quanta but comparable precision to positive $\epsilon$'s.

\begin{figure}
    \centering
    \includegraphics[width=1.0\linewidth]{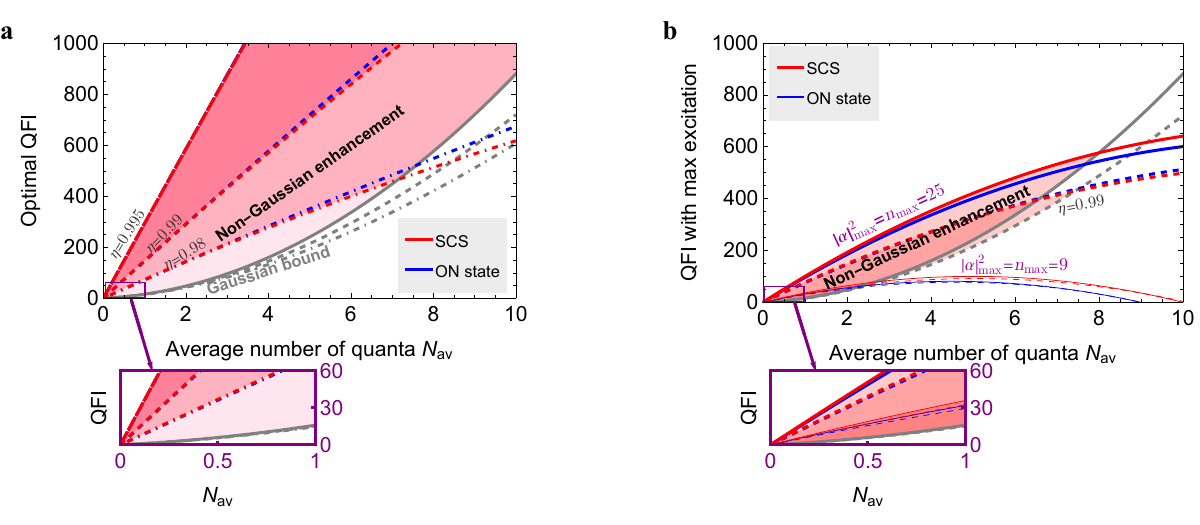}
   \caption{Comparison of QFI for different quantum probes across varying practical loss levels and fundamental excitation constraints $N_\mathrm{av}$. 
   The shaded area highlights the region of non-Gaussian enhancement outperforming the Gaussian bound.  
   All angles are in units of radians ($\mathrm{rad}$), and Fisher information in units of $\mathrm{rad}^{-2}$. \textbf{
   a}) QFI under unconstrained optimization for each $N_\mathrm{av}$. 
   {Different line styles correspond to different transmission efficiencies $\eta$ representing varying levels of loss.}
   Both SCS and ON state exhibits a comparable non-Gaussian enhancement up to $N_\mathrm{av}^\mathrm{NGE}$, effectively unbounded but shrinking as loss increases. 
   \textbf{b}) QFI under  practical constraints on $|\alpha_\mathrm{max}|^2$ and $n_\mathrm{max}$.
   {Here, thin and thick lines correspond to different maximum excitation constraints.}
   Asymmetric SCS maintains consistently surpass  Gaussian bound even with limited resources, and comparable to ON states.  
   (Inset) Low  $N_\mathrm{av}$ regime. 
   QFI  scales linearly for all states as Eqs. (\ref{eq:FQGausseta},\ref{eq:FQoptNavsmall},\ref{eq:FQSCSlinear}).   }
    \label{fig:QFI-comp}
\end{figure}

{For a comprehensive analysis, asymmetric SCS, typically of the form $|\psi\rangle_\mathrm{SCS} \propto (|0\rangle + \epsilon |\alpha_\mathrm{max}\rangle)$, are characterized by three key parameters:
\begin{enumerate}
    \item \textbf{Mean quanta number ($N_\mathrm{av}$):} This is our primary constraint, reflecting energy requirements and susceptibility to disturbances.
    \item \textbf{Maximum coherent state amplitude ($|\alpha_\mathrm{max}|^2$):} This represents a practical upper limit on the achievable coherent state amplitude due to experimental constraints (e.g., Rabi interaction strength, duration before decoherence dominates, available control power, overall calibration precision, or potential oscillator non-linearities at large amplitudes). It dictates the maximum phase-space separation achievable. While fixing \(|\alpha_\mathrm{max}|^2\) provides a useful benchmark, the actual achievable \(|\alpha_\mathrm{max}|^2\) with high fidelity is itself dependent on control resources and platform capabilities, which may introduce implicit dependencies not fully captured by treating it as a simple fixed parameter in all regimes.
    \item \textbf{Asymmetry weight parameter ($\epsilon$):} This parameter in the state superposition directly controls the degree of asymmetry.
\end{enumerate}
These parameters are interconnected. For a fixed $N_\mathrm{av}$, achieving a large $|\alpha_\mathrm{max}|^2$ (which, as we will see, is desirable for high QFI) necessitates a small $\epsilon$, making the state highly asymmetric. 
The absolute-squared coherent amplitude $|\alpha_\mathrm{max}|^2$ is particularly key to preparation costs (similarly as $n_\mathrm{max}$ for ON states) and bounds the probe state's phase space reach.}
The QFI for asymmetric SCS can be expressed on these parameters $(N_{\mathrm{av}}, |\alpha_\mathrm{max}|^2)$ as:
\begin{equation}
    F_Q[\ket{\psi}_{\mathrm{SCS}}] = 4 N_\mathrm{av}(1 + |\alpha_\mathrm{max}|^2 - N_{\mathrm{av}}),
    \label{eq:FQSCSideal}
\end{equation}
{where the weight parameter $\epsilon$ is absorbed in $N_\mathrm{av}$.}
This equation reveals a beneficial increase in QFI by {a term proportional to} $|\alpha_\mathrm{max}|^2$ under the fixed $N_{\mathrm{av}}$ constraint, which is larger than $N_{\mathrm{av}}$ due to physicality constraint, enhancing sensitivity beyond SQL and Gaussian bound. 
Comparing with the QFI of ON states in (\ref{eq:QFION_revised}), {an ideal} unbounded increase has an analogical logic, while $1+|\alpha_\mathrm{max}|^2$ plays similar role for the SCS as $n_\mathrm{max}$ for the ON states.
The non-vanishing additive constant $1$ within the parenthesis of (\ref{eq:FQSCSideal}) {exists because} of phase sensitive coherent state in the SCS, in contrast to the phase-insensitive Fock state in the ON-states.  
This structural baseline enhancement independent of $|\alpha_{\text{max}}|^2$  underpins their enhanced robustness, particularly considering the $\mathcal{O}(1)$ gate complexity for SCS versus $\mathcal{O}(n_\mathrm{max})$ for ON states which impacts susceptibility to cumulative gate errors.
For symmetric SCS at which $N_\mathrm{av}\approx |\alpha_\mathrm{max}|^2/2$, the QFI is approximately $4N_{\mathrm{av}}^2 + 4N_{\mathrm{av}}$, manifesting no non-Gaussian advantage compared to Gaussian bound (\ref{eq:GaussianLimit}).
However,  sufficient asymmetry {allows QFI to exceed}  the Gaussian benchmark when $|\alpha_\mathrm{max}|^2\ge 3N_\mathrm{av}+1$ for SCS as it was $n_\mathrm{max}\ge 3N_\mathrm{av}+2 $ when ON state was considered analogously. 
Equation~(\ref{eq:FQSCSideal}) reveals a key insight: to maximize the QFI for a \textit{fixed} average quanta number $N_\mathrm{av}$, one must maximize the coherent state amplitude $|\alpha_\mathrm{max}|^2$. 
Since $N_\mathrm{av} \approx |\alpha_\mathrm{max}|^2\epsilon^2/(1 + \epsilon^2)$ for large $|\alpha_\mathrm{max}|$, achieving a large $|\alpha_\mathrm{max}|^2$ at fixed $N_\mathrm{av}$ necessitates making the superposition highly asymmetric, i.e., choosing $\epsilon \ll 1$. 
Effectively, $\epsilon$ mediates the trade-off: for a given $N_\mathrm{av}$, a smaller $\epsilon$ allows for a larger $|\alpha_\mathrm{max}|^2$ (enhancing phase-space separation and thus potential QFI), but makes the state more `vacuum-like' with a tiny admixture of a large coherent state.
This strategy concentrates the energy cost $N_\mathrm{av}$ primarily in the small probability amplitude of the large-$|\alpha_\mathrm{max}|$ component. Physically, this asymmetry maximizes the phase-space separation between the vacuum $|0\rangle$ and coherent $| \alpha_\mathrm{max} \rangle$ components relative to the average energy, enhancing the state's ability to distinguish phase shifts generated by $\hat{n}$, thereby maximizing $\mathrm{Var}(\hat{n})$ and $F_Q$.
{Crucially, this entire class of advantageous SCS probes can be prepared deterministically with a constant-complexity, two-gate protocol. 
This $\mathcal{O}(1)$ gate complexity is a key distinction from common preparation schemes for high-$n$ ON states, which typically involve a number of operations with polynomial scaling \cite{McCormick2019,huang2025fastsidebandcontrolweakly}.}

Under a weak loss $1-\eta\ll 1$, the analytical expression for QFI becomes {too} complex.  
However, a highly precise and concise form can be derived in the limit of a large $|\alpha|^2\le |\alpha_\mathrm{max}|^2$, as detailed in Appendix \ref{Appendix:QFI}:
\begin{align}
     F_Q[\ket{\psi}_{\mathrm{SCS}}]\stackrel{|\alpha|^2\gg 1}{\simeq} 4 \eta  N_{\text{av}} \left(\eta  e^{-|\alpha| ^2 (1-\eta)} \left(|\alpha| ^2-N_{\text{av}}\right)+1\right).
\end{align}
For a fixed $\eta$ and $N_\mathrm{av}$, the QFI does \textit{not} increase boundlessly by increasing $|\alpha|^2$, analogously to ON states, and there exists an optimal $|\alpha|^2$ equal to $|\alpha_\mathrm{opt}|^2=\frac{N_{\text{av}}(1-\eta)+1}{1-\eta }$ {that is different from and can be smaller than $|\alpha_\mathrm{max}|^2$,  and}  gives the maximum $F_Q$:
\begin{align}
    F_Q[\ket{\psi}_{\mathrm{SCS}}^\mathrm{max}]\simeq 4 \eta  N_{\mathrm{av}} \left(1+\frac{\eta  e^{-(1-\eta ) N_{\mathrm{av}}-1}}{1-\eta }\right).
    \label{eq:FQSCSmax}
\end{align}
{For weak loss ($1-\eta\ll 1$), the QFI approaches the following piecewise linear scaling with $N_\mathrm{av}$:}
\begin{align}
F_Q[\ket{\psi}_{\mathrm{SCS}}^\mathrm{max}]\approx \begin{cases}4 \eta  N_{\mathrm{av}}, & N_\mathrm{av}\gg 1,\\
4 \eta  N_{\text{av}}\left(1  +\frac{ \eta  }{e(1- \eta) }\right), & N_\mathrm{av}\ll 1,\end{cases}
\label{eq:FQSCSlinear}
\end{align}
implying a change from ideal quadratic Heisenberg scaling to a more classical-like linear scaling with increasing average photon number, reduced to that of coherent states for a large $N_\mathrm{av}$ limit. 
This nearly linear QFI dependence on $N_\mathrm{av}$ under loss advantageously avoids saturating to zero as in the case of ON states, and remains above the SQL even at large $N_\mathrm{av}$.  
While smaller than even the reduced Gaussian bound (\ref{eq:FQGausseta}) incorporating same ambient loss in the asymptotic large average number of quanta $N_\mathrm{av}\gg 1$,  the non-Gaussian advantage persists in an \textit{intermediate} range of $N_\mathrm{av}$ {due to the larger scaling in the limit of $N_\mathrm{av}\ll 1$}.  In this intermediate range of $N_\mathrm{av}$, the transition between scalings in $N_\mathrm{av}$ takes place at $N_\mathrm{av}^\mathrm{tran}=\log2/(1-\eta)$ (which takes value around $N_\mathrm{av}^\mathrm{tran}\approx 69$ at $\eta=0.99$). 
Despite the asymptotic scaling being suppressed from quadratic to linear similarly as for the ON states, the multiplicative prefactor provides the gain over the Gaussian states, from $N_\mathrm{av}=0$ up to the  \textit{threshold average quanta number}, $N_\mathrm{av}^\mathrm{NGE}[\eta]\stackrel{1-\eta\ll 1}{\simeq} \frac{-\eta ^2+3 \eta -2 W\left(\frac{1}{2} e^{\frac{1}{2} (\eta -3) \eta } \eta ^2\right)-2}{2 (1-\eta )}\stackrel{\eta\to 1}{\approx} \frac{0.157}{1-\eta}$ where $W[z]$ is the principal solution $w$ satisfying $z=w  e^w$. 
{This non-Gaussian advantage regime for SCS, achievable with an $\mathcal{O}(1)$ gate protocol, is particularly relevant for near-term quantum sensors.}
{Expectedly, this non-Gaussian enhancement threshold, $N_\mathrm{av}^\mathrm{NGE}$, is non-zero only under weak loss ($\eta\ge\eta_\mathrm{min}\approx 0.802$). This condition defines the operational regime where asymmetric SCS maintain their metrological advantage. Notably, SCS tolerate a higher level of loss than ON states, expanding this practical operating window.}
This implies that for high-fidelity systems where loss can be kept low, the threshold can be substantial; for instance, with 1\% loss ($\eta=0.99$), the non-Gaussian enhancement persists up to $N_\mathrm{av} \approx 16$, a range encompassing relevant operational points for current state-of-the-art experiments preparing non-Gaussian states (e.g., $N_\mathrm{av}$ values of 3 and 5 shown in our figures) \cite{SaleemPRA2024PhysRevA.109.052615Heisenberg}. 
Therefore, while the non-Gaussian advantage is not indefinite and vanishes for asymptotically large $N_\mathrm{av}$, the threshold is sufficiently high for low-loss scenarios to make asymmetric SCS a practically relevant resource for surpassing Gaussian bounds within an currently accessible intermediate energy regime. 
This finding also underscores the critical importance of minimizing environmental loss to maximize the operational range of SCS-based quantum sensors. 
When the loss is strong beyond this threshold $\eta\le \eta_\mathrm{min}$, Gaussian states are always better than the SCSs. 
{This linear QFI scaling in SCS explains} the linear behavior of optimal CFI in Appendix  \ref{append:CFIrange}. 

In Fig. \ref{fig:QFI-comp}\textbf{a}, the optimal QFI of the states of main interest, i.e. SCS, ON state, and Gaussian states is compared {over all state parameters for each $N_\mathrm{av}$ }. Without loss or noise, the optimal QFI of SCS and ON states can attain arbitrarily high values without upper bounds. Under a realistic loss or noise, however, the optimal QFI takes finite values that linearly increase by   a larger $N_\mathrm{av}$. For all loss level, the optimal QFI of SCS is comparable to that of ON states, and have non-Gaussian advantage over a range of  $N_\mathrm{av}$ up to a maximum tolerable loss level $\eta_\mathrm{min}$.
{At a large $N_\mathrm{av}$,  ON states have an advantage over SCS when the loss is weak. At $\eta=0.99$, their ratio of QFI, $ F_Q[\ket{\psi}_{\mathrm{ON}}^\mathrm{max}]/F_Q[\ket{\psi}_{\mathrm{SCS}}^\mathrm{max}]$ can get up to about $2.5$ times at $N_\mathrm{av}\approx 280$,  persisting up to  $N_\mathrm{av}\lesssim 530$.  Beyond this limit, SCS have advantage, asymptotically their ratio of QFI reduced to $0$. However, in such a macroscopic limit, both lose non-Gaussian advantages beyond Gaussian limit.}

In Fig. \ref{fig:QFI-comp}\textbf{b}, the QFI of these states is compared over a range of $N_\mathrm{av}$ when $|\alpha_\mathrm{max}|^2$ and $n_\mathrm{max}$ is fixed. The asymmetric SCS still shows precision surpassing the Gaussian threshold. 
Remarkably, even for small, fixed $N_{\mathrm{av}} \ll 1$, the QFI of the optimal asymmetric SCS in (\ref{eq:FQSCSmax}) can significantly exceed the Gaussian benchmark $F_Q^{\text{Gauss}} = 8N_{\mathrm{av}}^2 + 8N_{\mathrm{av}}$, up to the aforementioned asymmetry limit $N_{\mathrm{av}} \le  (|\alpha_{\text{max}}|^2 - 1)/3$, and up to the loss regime where $\eta\gtrsim 0.8$. 
When comparing the ON state and SCS state with identification $n_{\text{max}} = |\alpha_{\text{max}}|^2$, there is a small but noticeable QFI enhancement of additional $4N_{\mathrm{av}}$ for the SCS over the ON state. If $|\alpha_{\text{max}}|^2$ and $n_{\text{max}}$ are treated as independent parameters, this performance boost occurs when $|\alpha_{\text{max}}|^2 \geq n_{\text{max}} - 1$. 
Under a strong loss, this advantage persists. In short, the optimal QFI of the SCS surpasses the Gaussian benchmark, close to that of ON states, surpassing it in a heavy loss limit.
{While these analytical approximations 
provide valuable scaling insights, the performance evaluations presented throughout this work  utilize numerical calculations based on the full density matrix descriptions, confirming the persistence of these trends and advantages beyond the strict asymptotic or weak-loss regimes. This $\eta_{min} \approx 0.802$ threshold aligns well with numerical results indicating the crossover point where Gaussian states regain optimality.}
{A comprehensive analysis of the effect of thermal noise, detailed in Appendix \ref{appendix:thermal}, further substantiates the robustness of SCS compared to ON and Gaussian states under thermal noise.}

\subsection{Classical Fisher information}
\label{sec:CFI}

The preceding QFI analysis reveals that while Gaussian states like squeezed states offer enhanced precision beyond the SQL, non-Gaussian states such as ON states, and particularly asymmetric SCS, demonstrate the potential for even greater sensitivity, over the Gaussian bound. 
However, the QFI represents an idealized scenario, assuming optimal measurements, and in many cases, prior knowledge. 

In practical settings, the achievable precision is constrained by the specific measurement apparatus and procedure employed. 
{We therefore focus on a more realistic benchmark of the CFI, which quantifies the information extractable by a \textit{particular} measurement setup accounting for its limitation, inherently bounded by QFI. 
The CFI can therefore determine the practical utility of a non-Gaussian state, validating a genuine non-Gaussian enhancement of our protocol beyond any Gaussian strategy.}

The CFI for a probe state $\rho_\mathrm{in}$ depends on the chosen detection setup, and is upper-bounded by the QFI. It is computed for a detector with a finite number of outcomes $N_{\mathrm{det}}$ as 
\begin{align} 
F_C[\rho_\mathrm{in};\theta]= \sum_n^{N_\mathrm{det}} P(n;\theta)^{-1} (\partial_{\theta} P(n;\theta))^2. \label{eq:CFIdef} 
\end{align}
Here,  $P(n; \theta)$ represents the probability of obtaining the $n$-th measurement outcome by a detector with $N_\mathrm{det}$ possible outcomes. This function takes values that depend on $\theta$ \cite{MukhopadhyayPRApp2025Globalsensing}, and the largest peak in height of the CFI $F_C^{\text{max}}\equiv F_C[\rho_\mathrm{in};\theta_\mathrm{max}]$ at a point of $\theta=\theta_\mathrm{max}$  is commonly of the primary interest, as depicted in Fig. \ref{fig:setup}\textbf{b}. However, $\theta$ is only partially controllable  due to the incomplete knowledge of it {due to a finite number of estimation runs}, and we cannot use any estimation protocol to perform perfectly at $\theta_\mathrm{max}$, but only close to it asymptotically. The primary peak width of the CFI indicates the range of parameter values with sufficiently high precision, where a narrower width suggests a smaller range. 
If data or information is insufficient or if there is instability in the estimation setup, the van Trees inequality \cite{Gill1995} aids by including prior information, predicting estimation precision in relation to the integration of CFI over the range of the peak. 

In our work, the discussion centers on single detector with data dimensions $N_{\mathrm{det}}=2$ as illustrated in Fig. \ref{fig:setup}\textbf{a}, considering a general estimation scheme involving qubit detection after  inverse state-preparation  steps for processing. This setup is a single-mode analogue of an interferometric protocol in \cite{Park2023quantumrabi}.
The probability of detection outcome, when the probe preparation and processing is kept abstract for other probes such as squeezed states, displaced Fock states, and ON states, is determined by $P[\theta]=\mathbb{F}[\hat{R}[\theta]\rho \hat{R}[-\theta],\ket{\psi}_\mathrm{measure}\bra{\psi}]$ for input state $\rho$ and $\ket{\psi}_\mathrm{measure}\bra{\psi}$ is the state that can be prepared by the ideal inverse processing setup, where $\hat{R}[\theta]=\exp[\ii \theta \hat{n}]$ represents oscillator rotation \cite{Pan2024}. The  detailed derivation of CFI of Gaussian states is provided in Appendix \ref{sec:GaussianQFI}, although the detection can be suboptimal \cite{Oh2019NPJQIOptimal,McIntyrePhysRevA2024HomodyneOptimalInterferometryECS}, but can be implemented in the considered experimental systems. 
{We define the non-Gaussian enhancement range, denoted $\delta \theta$ (as illustrated in Fig. \ref{fig:setup}\textbf{b}), as the interval of phase angles $\theta$ over which the CFI of the non-Gaussian probe state exceeds the QFI of the optimal Gaussian state (e.g., squeezed vacuum) under the same constraints (such as equal average quanta number $N_\mathrm{av}$). 
The uncertainty in the estimation   denoted as $d\theta$, differs from the non-Gaussian enhancement range $\delta\theta$, the range which the estimation method of interest overcomes all  Gaussian strategies, as shown in Figure \ref{fig:setup}\textbf{b}.
To assess the overall performance across the operational range where non-Gaussian states provide an advantage, we consider the average CFI, denoted $\bar{F}_C$. Unless otherwise specified, this average is calculated over the non-Gaussian enhancement range $\delta\theta$:
\begin{equation}
    \bar{F}_C = \frac{1}{\delta\theta} \int_\Theta F_C(\theta') d\theta',
    \label{eq:avg_cfi_def}
\end{equation}
where $F_C(\theta')$ is the classical Fisher Information at phase angle $\theta'$, and $\Theta$ is the   interval(s) constituting $\delta\theta$. 
This metric quantifies the practical sensitivity advantage offered by the non-Gaussian probe compared to the Gaussian benchmark within its effective range.
}

\subparagraph{Preparation and measurement of asymmetric SCS using Rabi gates}

We employ the preparation and processing sequence  illustrated in Fig. \ref{fig:setup}\textbf{a}. 
Both preparation and measurement protocols utilize only two Rabi gates, comprised of an entangler $\mathcal{E}$ and a disentangler $\mathcal{D}$ {that localizes the state onto the oscillator}, along with qubit rotations and oscillator displacements, or their inverses. 
{This protocol, where the qubit is decoupled during the sensing interval, enhances robustness against qubit decoherence.}
We describe mainly the preparation, while the  processing is nearly the inverse of the preparation.

Initially, the qubit $q$ and oscillator $O$ is prepared in the ground state $\ket{g}_q$  and  the vacuum state $\ket{0}_O$ as $\ket{\psi_0}=\ket{g}_q\ket{0}_O$.
First, for the preparation of general weight of SCS of the form $\ket{0}_O+\epsilon \ket{\alpha}_O$, the qubit is rotated by a Pauli rotation as $\exp[\ii (\phi_\epsilon+\pi/2) \hat{\sigma}_y ]\ket{\psi_0}=\left(\cos[\phi_\epsilon]\ket{e}-\sin[\phi_\epsilon]\ket{g}\right)\ket{0}_O$. 
Entangler $\mathcal{E}$ and disentangler $\mathcal{D}$ are made of  Rabi interactions \cite{Forn-DiazRevModPhys2019.91.025005,KockumNature2019Ultrastrong}, or  ECD gates~\cite{SchoelkopfNature2020,eickbusch2022}, effectively applying controlled displacement on the oscillator. Using a Rabi gate of the form $\exp[\ii \frac{\alpha}{\sqrt{2}}\hat{\sigma}_z \hat{X}]$ as the entangler $\mathcal{E}$, we have an entangled state:
\begin{align}
\exp[\ii \frac{\alpha}{\sqrt{2}}\hat{\sigma}_z \hat{X}](\cos[\phi_\epsilon]\ket{e}_q-\sin[\phi_\epsilon]\ket{g}_q)\ket{0}_O=\cos[\phi_\epsilon]\ket{e}_q\ket{\ii \frac{\alpha}{2}}_O-\sin[\phi_\epsilon]\ket{g}_q\ket{-\ii \frac{\alpha}{2}}_O.
\end{align}
Now the disentangler $\mathcal{D}$ is made of a different type of Rabi gate $\exp[\ii \frac{\beta}{\sqrt{2}}\hat{\sigma}_y \hat{P}]$, or another ECD gate.  To achieve the near-separability between the qubit and the oscillator, we set the strength as $\beta= \frac{\pi}{2\alpha}$. These sequence of gates asymptotically prepares a state of the form $\ket{-}_g \left(\cos[\phi_\epsilon]\ket{\ii \frac{\alpha}{2} }_O+\sin[\phi_\epsilon]\ket{-\ii \frac{\alpha}{2}}_O\right)$ in a limit of a large $\alpha$, with the approximation of {a coherent state } $\ket{\ii  \Gamma+\delta }\approx e^{-\ii \Gamma \delta}\ket{\ii \Gamma}$  {for a real and imaginary part of the coherent amplitude $\Gamma,\delta$} when $\Gamma\gg \delta$. After an additional displacement to place one of the coherent peak to the phase space origin making it a vacuum, the oscillator state is approximately in a state
    $\cos[\phi_\epsilon]\ket{0}_O+\sin[\phi_\epsilon]\ket{-\ii \alpha}_O$, {where we can equate the weights $\epsilon=\tan[\phi_\epsilon]$.}
Due to the rotational symmetry, we can denote coherent amplitude $-\ii \alpha$ as simply a real $\alpha$ without loss of generality, which can be achieved with additional auxiliary phase rotation.    This prepared state undergoes a phase rotation $\hat{R}[\theta]$ {with an unknown target angle $\theta$}
, the unknown target process of the estimation. 
{This straightforward two-gate scheme is a practical advantage over more complex preparation methods for other non-Gaussian states such as high-$n$ ON states.}
While Appendix \ref{app:pulse_simulations} demonstrates high state preparation fidelity with optimized pulses, the direct impact of cumulative gate infidelity on the ultimate metrological CFI beyond state fidelity is an important consideration for future detailed investigation.

The processing gate sequence is {nearly} an inverse of the preparation $\mathcal{E}^\dagger\mathcal{D}^\dagger$, taking place {after the unknown phase rotation}.
{By this sequence, in the absence of the signal phase, the state goes back to the initial state.}
The final detection data is obtained by a qubit detection on an optimized basis at the end of the setup. 
As detailed in Appendix~\ref{sec:qubitmeasurement}, the choice of qubit detection basis significantly impacts the robustness of the estimation against environmental noise. 
While a measurement in the $\hat{\sigma}_z$ basis ($\{\ket{g},\ket{e}\}$) might seem standard, its corresponding CFI peak around $\theta=0$ is highly susceptible to suppression by loss or dephasing. 
This occurs because noise tends to drive the outcome probabilities towards certainty (0 or 1) at $\theta=0$, diminishing the crucial derivative of the probability $\partial P/\partial \theta$. 
To preserve sensitivity under realistic conditions, we choose to perform the projective measurement on the qubit in the Pauli $\hat{\sigma}_y$ basis, whose eigenstates are $\{\ket{\pm_i} = (\ket{e}\pm\ii\ket{g})/\sqrt{2}\}$. 
This choice avoids the saturation issue and maintains a non-zero CFI peak near $\theta=0$ even in the presence of moderate noise, enhancing the protocol's practical robustness (see Appendix~\ref{sec:qubitmeasurement} for a detailed analysis and comparison). Ideally in the limit of $\alpha\gg 1$, this corresponds to the projective measurement {on the oscillator state after the phase rotation onto} the balanced SCS $\ket{0}+\ii \ket{\alpha}$.
{The collected data is then processed using an estimation protocol, such as Bayesian estimation, to estimate the target parameter $\theta$. The ultimate precision of this protocol is fundamentally bounded by the CFI via the Cramér-Rao bound. }

\subparagraph{CFI of SCS}

Our CFI analysis in this study focuses on  superpositions of vacuum and coherent states prepared as described earlier, specifically within the vacuum-like asymmetric SCS regime ($\epsilon\ll 1$ or $\phi_\epsilon\ll 1$). This focus is aimed at optimizing precision under a fixed average quanta constraint.
In the regime where the displacement $\alpha$ is significantly larger than 1, the coherent state component $\ket{\alpha}$ becomes well-separated from the vacuum state $\ket{0}$ in phase space, resulting in a negligible overlap between them. 
In this limit, change in the detection probability is primarily governed by the overlap between the rotated coherent state $\ket{\alpha e^{\ii\theta}}$ and the original coherent state $\ket{\alpha}$, which is captured by the term $\braket{\alpha|\alpha e^{\ii \theta}}=e^{-\alpha^2(1-e^{\ii\theta})}$. 
{For practical analysis under specific conditions, we can derive asymptotic forms for the CFI. For instance, }
the analytical expression of CFI under loss or thermal noise can be obtained in Appendix \ref{sec:qubitmeasurement} and is analyzed below, while its full form is not presented in this paper due to its highly complex form.
When expressed in terms of the parameter pair $(N_\mathrm{av},\alpha^2)$, we have the asymptotic CFI of the form
\begin{align}
    F_C[\ket{\psi}_\mathrm{SCS}] \stackrel{\alpha\gg 1}{\approx} \frac{4 \alpha ^4 \eta  N_{\text{av}} e^{-2 \alpha ^2 \left(1-\sqrt{\eta } \cos \theta \right)} \cos ^2\left(\alpha ^2 \sqrt{\eta } \sin \theta \right)}{ \alpha
   ^2+N_{\text{av}}},
\end{align}
{where the detection scheme implemented by the processing as in Fig. \ref{fig:setup}   projects  on symmetric SCS $\ket{0}+\ket{\alpha}$.}
{This choice is not arbitrary; as analyzed in Appendix \ref{append:CFIrange}, a symmetric projection maximizes the signal contrast for any probe asymmetry, providing a robust, phase-independent measurement strategy, analogously to ON state's case.}
The maximum of this CFI is given as  $F_C^\mathrm{max}[\ket{\psi}_\mathrm{SCS}]\approx\frac{4 \alpha ^4 \eta  N_{\text{av}} e^{-2 \alpha ^2 \left(1-\sqrt{\eta } \right)} }{ \alpha
   ^2+N_{\text{av}}}$, reducing to $\frac{4 \alpha ^4   N_{\text{av}}  }{\alpha
   ^2+N_{\text{av}}}$ at $\eta=1$, coinciding with QFI (\ref{eq:FQSCSideal}) in the limit of $\alpha \gg 1$.
{While this approximation accurately captures the peak height and width, it introduces a small, artifactual asymmetry in $\theta$ not present in the exact numerical result. The true CFI is symmetric around its peak, as expected.}
The first CFI zero of the primary peak occurs at around $\pm\sin^{-1}\left[\frac{\pi}{2\alpha^2\sqrt{\eta}}\right]$, and the half of the peak height occurs at around $\pm\sin^{-1}\left[\frac{\pi}{4\alpha^2\sqrt{\eta}}\right]$, {roughly giving the scale of the estimation range}.

\subparagraph{CFI of ON state}

For variable weight ON states, the measurement is assumed to be projection onto another ON state $\sqrt{w'}\ket{0}+\sqrt{1-w'}\ket{n}$ and its complementary outcome \cite{Pan2024}.
From a simple analysis, we find that the optimal measurement 
{that maximizes the signal contrast, without prior knowledge of the phase, is a projection onto the balanced ON state ($w'=1/2$), regardless of the probe's initial asymmetry. 
This makes it a robust and practical choice for detection, as detailed in Appendix \ref{append:CFIrange}.} 

The CFI for ON states with $n=n_\mathrm{max}$ in ideal scenario  is given exactly by
\begin{align}
    F_C[\ket{\psi}_{ON}] = \frac{4 n_\mathrm{max}^2 N_\mathrm{av} (n_\mathrm{max}-N_\mathrm{av}) \sin ^2(\theta  n_\mathrm{max})}{n_\mathrm{max}^2+4 N_\mathrm{av} (N_\mathrm{av}-n_\mathrm{max}) \cos ^2(\theta  n_\mathrm{max})}
\end{align}
expressed in terms of parameter set $(N_\mathrm{av}, n_\mathrm{max})$. The average number of quanta of the ON state is related to other parameters as $N_\mathrm{av}=n_\mathrm{max}\frac{\epsilon^2}{1+\epsilon^2}$. This CFI exhibits $2n_\mathrm{max}$ peaks of equal heights and widths over the interval $\theta \in [-\pi, \pi]$.

The CFI for the ON state under loss conditions can be derived analytically in the limit of large $n_\mathrm{max}$, similarly as in the derivation of its QFI in Appendix \ref{Appendix:QFI}. The analytical CFI expression under loss in this large $n_\mathrm{max}$ limit is given as:
\begin{align}
   F_C[\ket{\psi}_{ON}] &\stackrel{n_\mathrm{max}\gg 1}{\simeq}  4 N_{\text{av}} \eta ^{n_\mathrm{max}} \left(n_\mathrm{max}-N_{\text{av}}\right) \sin ^2(  n_\mathrm{max}\theta).
   \label{eq:ONCFIloss}
\end{align}
The maximum CFI is then given as  \(
F_C^\mathrm{max}[\ket{\psi}_{ON}] \stackrel{n_\mathrm{max}\gg 1}{\simeq} 4 (n_\mathrm{max}- N_\mathrm{av}) N_\mathrm{av} \eta^{n_\mathrm{max}}. 
\)
In the absence of loss, the maximum CFI aligns with the QFI. 
Optimized  $n_\mathrm{max}$ at a fixed $\eta$ and $N_\mathrm{av}$  is given as $-\frac{4 N_\mathrm{av} \eta ^{\text{Nav}}}{e \log (\eta )}$.
The CFI analysis of ON states highlights their multi-peak structure and the trade-off between peak height and width.

{
This enhancement for both SCS and ON states, however, comes at the cost of a reduced signal contrast for simple projective measurement. 
For highly asymmetric states, this contrast scales as $\sqrt{N_\text{av}}/|a_\text{max}|$ for SCS or $\sqrt{N_\text{av}/n_\text{max}}$ for ON states, meaning that the states required for high CFI yield a weak signal.
This creates a fundamental trade-off: maximizing the potential precision (related to the signal’s slope) requires sacrificing signal contrast (visibility), making the measurement more demanding (see Appendix \ref{append:CFIrange} for a detailed analysis). }

\begin{figure}
    \centering
    \includegraphics[width=1.0\linewidth]{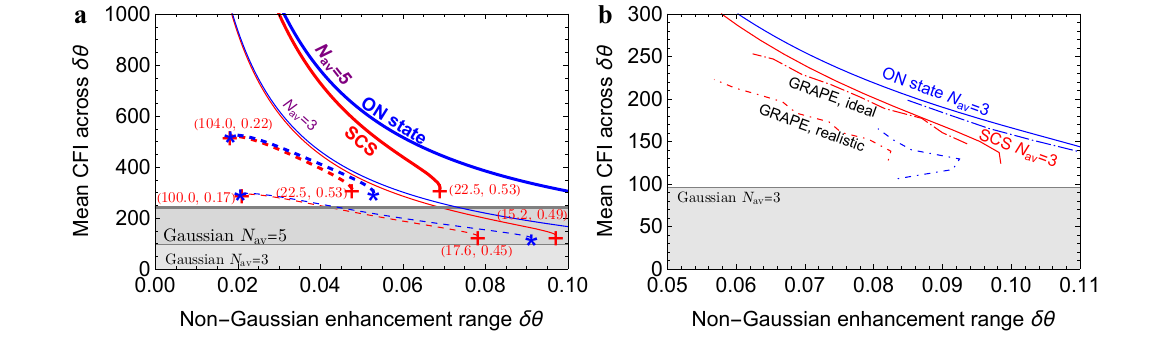}
    \caption{
The Practical Trade-off between Precision and Range for SCS and ON States.
This figure illustrates the performance frontier for phase estimation, the mean CFI against the operational range ($\delta\theta$) over which a probe surpasses the ideal Gaussian bound (gray).
    \textbf{
    a}) 
   {The characteristic arc shape reveals a fundamental trade-off between achievable precision (mean CFI) and operational range.}
    For a fixed average quanta, there is a fundamental trade-off between achievable precision (mean CFI) and operational range. 
    Specific points parameterized by \( (|\alpha|^2, \epsilon) \) for SCS (red)  show how tuning the  asymmetry navigates this frontier.
    {
     Pursuing maximum CFI (top-left) narrows the range, while broadening the range (bottom-right) reduces the peak CFI.
     Loss (dashed lines) predictably lowers this frontier.
    {
    Parameter choices inside this frontier (e.g., those using a very small $|\alpha|^2$) yield a simultaneous reduction in both mean CFI and range \(\delta\theta\) and are therefore not shown.}
    \textbf{b}) Achievable Performance in a Superconducting Circuit. This trade-off is validated using realistic GRAPE-optimized pulses. 
    The performance of a simulated noisy circuit  closely tracks the ideal protocol. 
    This demonstrates that the advantageous performance of asymmetric SCS and ON state is not merely theoretical but is robustly achievable with current experimental control techniques. 
    }
    }
    \label{fig:CFItradeoff}
\end{figure}

\subparagraph{Comparison of CFI of various states}

Figure \ref{fig:CFItradeoff} illustrates our exploration of the enhancement of CFI beyond the Gaussian bound. Specifically, we analyzed the non-Gaussian enhancement range $\delta \theta$ where the CFI of the SCS and ON states exceeds {the QFI} of the squeezed state for the same average quanta number $N_\mathrm{av}$. In lossless conditions, both the SCS and ON states surpass the Gaussian bound. 
The ON state, in particular, exhibits a broader estimation range, or  a higher average CFI for a fixed enhancement range. 
The SCS’s non-Gaussian enhancement tends to increase with higher average number of quanta $N_\mathrm{av}$, even when unaffected by loss, close to that of {an ideally prepared ON state}. 
{The solid curves in Figure \ref{fig:CFItradeoff}\textbf{a} represent the optimal {precision-range} frontier for a given $N_\text{av}$, with the characteristic arc shape revealing a fundamental trade-off.} 
{It implicitly guides experimental choices: for a given $N_{\text{av}}$ and achievable loss, one can select state parameters (e.g., $(|\alpha|^2, \epsilon)$ for SCS) to maximize the mean CFI over a desired $\delta\theta$, balancing peak sensitivity with operational range.}
This frontier is bounded by two competing physical limits. 
On one hand, pursuing maximal precision (high mean CFI) requires a large phase-space separation $|\alpha|$. 
While this boosts the peak CFI {(local sensitivity, or slope)}, it also creates extremely fragile interference features {with low contrast (or visibility)}, thus limiting the frontier at the top-left. 
On the other hand, pursuing the broadest operational range (large $\delta\theta$) requires a smaller $|\alpha|$ to widen these features. 
However, as $|\alpha|$ becomes too small, the peak CFI inevitably drops below the Gaussian bound; the non-Gaussian advantage vanishes, causing $\delta\theta$ to collapse to zero and bounding the frontier on the right. 
These two effects together define the closed arc of optimal performance, showing that neither precision nor range can be maximized indefinitely.
{
The selection of optimal parameters must therefore consider not only the peak CFI but also the required estimation range for the specific application whether it prioritizes maximal sensitivity at a known phase or robust performance over an uncertain phase range, alongside the practical challenges of preparing states with very large \(|\alpha|^2\).}

{Figure~\ref{fig:CFItradeoff}\textbf{b} extends this analysis by specifically considering the trade-off achievable in a realistic superconducting circuit architecture.
Here, both the SCS and ON states are assumed to be prepared using GRAPE (Gradient Ascent Pulse Engineering) optimized control pulses, with further details on the GRAPE methodology and simulation parameters provided in Appendix~\ref{app:pulse_simulations}.
Crucially, this modeling incorporates realistic experimental imperfections such as losses and noise inherent to our experimental setups.
This provides a grounded assessment of the practical performance limits and the achievable balance between mean CFI over the non-Gaussian enhancement range and the width of this range $\delta\theta$ for these advanced quantum probes in a near-term experimental setting.}

\begin{figure}
    \centering
    \includegraphics[width=1.0\linewidth]{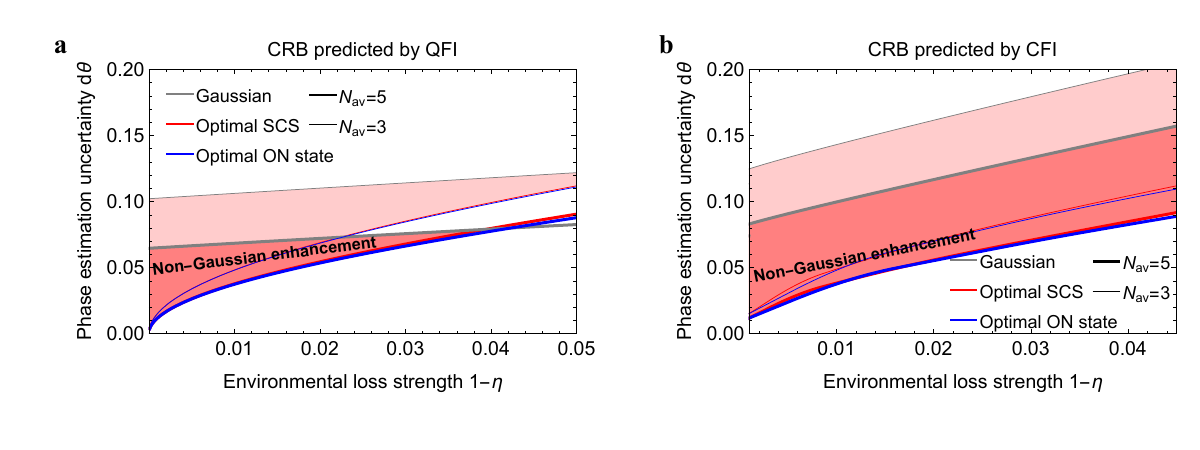}
    \caption{
{Phase estimation uncertainty ($d\theta = 1/\sqrt{F_{Q,C}}$) as a function of environmental loss strength $1-\eta$, compared against the optimal Gaussian bound (gray) for average quanta numbers $N_\mathrm{av}=3$ (thin lines) and $N_\mathrm{av}=5$ (thick lines). 
The shaded regions highlight the parameter space where SCS provide a non-Gaussian advantage. 
\textbf{(a)} Uncertainty predicted by the ultimate QFI. 
This illustrates the trade-off between precision and robustness: higher $N_\mathrm{av}$ states are more powerful but are also more fragile, with their advantage diminishing more rapidly as loss increases.
\textbf{(b)} Uncertainty predicted by the CFI for our specific, experimentally feasible protocol. 
A non-zero uncertainty at zero loss, representing the inherent performance cost of using a practical, fixed measurement scheme instead of a theoretically perfect one.
Our numerical calculation in this limit did not fully capture the infinite CFI approaching QFI.  
Despite this, the protocol preserves a substantial \textit{relative} advantage over the Gaussian CFI in the crucial low-loss regime, demonstrating its robustness and practical value.
The comparison between (a) and (b) quantifies the performance gap between the fundamental quantum limit and what is achievable in a feasible experiment.
}
    }
    \label{fig:dphiSCSvsSqz}
\end{figure}

{Figure \ref{fig:dphiSCSvsSqz} provides a direct comparison of the phase estimation uncertainty ($d\theta = 1/\sqrt{F_{Q,C}}$) for optimal SCS, ON states, and Gaussian states, illustrating their performance under environmental loss. The analysis separates the fundamental limits predicted by the QFI from the practical performance achievable with our protocol, as quantified by the CFI.}

{Fig. \ref{fig:dphiSCSvsSqz}\textbf{a}, based on the QFI, reveals the ultimate precision limits. As expected, higher average quanta $N_\mathrm{av}$ leads to lower fundamental uncertainty. It also reveals a generic trade-off between precision and robustness: states with higher $N_\mathrm{av}$ are more powerful in the low-loss regime but are also more fragile. Their quantum advantage diminishes more rapidly with increasing loss, causing their performance to fall below the Gaussian bound at a smaller loss threshold $1-\eta$. This highlights that the operational range for non-Gaussian advantage shrinks for higher-energy probe states.}

{Fig.  \ref{fig:dphiSCSvsSqz}\textbf{b} presents the practical uncertainty achievable with our specific measurement protocol, calculated from the CFI. The results differ from the QFI predictions in two crucial ways. First, there is a non-zero uncertainty offset even in the ideal lossless case ($1-\eta=0$). This offset is a direct consequence of the practical measurement scheme, which, while efficient, is not perfectly optimal and cannot extract the full quantum Fisher information from the state. This limit, however, may not represent the true optimal precision due to the finite dimension numerical calculation. Second, while higher $N_\mathrm{av}$ still yields lower uncertainty, the practical advantage over the Gaussian bound can be understood in relative terms. For instance, for $N_\mathrm{av}=3$, the SCS protocol offers a threefold reduction in uncertainty over the Gaussian bound at a 1\% loss level. This relative advantage grows significantly as the environment improves, reaching nearly an order-of-magnitude (eightfold) reduction at 0.1\% loss. This demonstrates that even with the inherent limitations of a practical protocol, asymmetric SCS  and ON state offer a robust and substantial quantum non-Gaussian advantage in realistic, low-loss scenarios. Gaussian CFI, possibly from its sub-optimality, loses the advantageous robustness of its QFI, maintaining the non-Gaussian enhancement by SCS and ON states.  The comparison between Fig. \ref{fig:dphiSCSvsSqz}\textbf{a} and \ref{fig:dphiSCSvsSqz}\textbf{b} thus quantifies the performance gap between the fundamental quantum limit and a feasible experimental implementation.}

\section{Discussions and conclusion}

This study highlights the  non-Gaussian enhancement in phase estimation  achieved by asymmetric superposition of coherent states (SCS) prepared via {an efficient two-gate protocol} \cite{ParkNJP2022Slowing,Park2023quantumrabi,KockumNature2019Ultrastrong,SchoelkopfNature2020,eickbusch2022}.  We have established that 
optimized SCS not only surpass the Gaussian bound enhancing sensitivity beyond what is possible with Gaussian states, but  also demonstrate remarkable robustness 
against realistic experimental imperfections such as  loss and thermal noise, establishing them as a pragmatically strong candidate balancing precision gains with implementational feasibility.
This advantage holds within a crucial intermediate energy regime, up to a significant threshold in the average number of quanta.
Furthermore, while asymmetry is essential to surpass the fundamental Gaussian QFI bound, our analysis shows that even symmetric non-Gaussian states can gain a practical advantage. 
This occurs when they outperform a Gaussian benchmark that uses a suboptimal measurement, a scenario demonstrated with the optimized pulse methods also in \ref{app:pulse_simulations}.
This practical advantage stems not just from the state's intrinsic properties, but decisively from its efficient preparation and processing, compared to resource-intensive alternatives.

\begin{table}[h!]\centering\caption{{Comparison of key quantum probe states for phase estimation. The table highlights the trade-offs between ultimate precision under ideal and lossy conditions, and the practical resource cost of state preparation, which is central to this work's conclusion.}}\label{tab:comparison}\begin{tabular}{l|c|c|c}\hline\textbf{Probe State} & \textbf{Ideal QFI Scaling} & \textbf{QFI Scaling (Lossy)} & \textbf{Prep. Gate Complexity} \\
\hline Squeezed Vacuum& $\propto N_{\text{av}}^2$ (Quadratic) & $\propto N_{\text{av}}$ & $\propto \operatorname{arcsinh} \left(1.31 N_\text{av}^{0.443}\right)$ \cite{HastrupPRL2020squeezing} \\ON States            & $\propto (n_{\text{max}} -N_{\text{av}})N_{\text{av}}$ (Unbounded)& $\propto N_{\text{av}}$ (suppressed to $0$ at high $N_\mathrm{av}$)& Polynomial \cite{McCormick2019,huang2025fastsidebandcontrolweakly}, $\mathcal{O}(n_{\text{max}})$ \\Asymmetric SCS       & $\propto (|\alpha_{\text{max}}|^2 -N_{\text{av}}+1)N_{\text{av}}$ (Unbounded)& $\propto N_{\text{av}}$ (suppressed to SQL at high $N_\mathrm{av}$)& \textbf{Constant, $\mathcal{O}(1)$} \\ \hline\end{tabular}\end{table}

While both asymmetric SCS and ON states {can} offer comparable precision beyond the Gaussian bound in the ideal, lossless limit, their  {practical} implementation strategies diverge significantly. 
The SCS {is often implemented with experimentally native operations in dispersive cQED regimes for long lived oscillator. 
Preparing high-$n$ ON states in this regime, {for instance via fast sideband drives \cite{ZhangPhysRevA2024ONstate}, is a complex control task.} 
While the required number of   $\mathcal{O}(n_{\text{max}})$ operations \cite{ZhangPhysRevA2024ONstate, McCormick2019,huang2025fastsidebandcontrolweakly} can make scaling up challenging, the use of short gates can  make the total preparation time comparable to that of SCS.}
{The   resource cost of $\mathcal{O}(1)$ gate protocol for  a large coherent amplitude SCS (e.g., via higher pulse power or longer gate duration) is a critical practical consideration. }
{
}
{As explored in Appendices \ref{sec:appendix_bias} and \ref{app:pulse_simulations}, numerical simulations confirm that the protocol is robust against small systematic gate errors and that high-fidelity SCS can be generated with realistic control pulses, strengthening the protocol's practical viability.}

   {The practical realization of the high-precision asymmetric SCS discussed herein is supported strongly by numerical simulations employing optimized control pulses for the qubit-oscillator interaction. 
   These simulations, detailed in Appendix~\ref{app:pulse_simulations}, confirm that target asymmetric SCS with significant displacement amplitudes ($\alpha \sim 6.0$) are achievable using realistic pulse shapes and durations, consistent with parameters accessible in platforms like superconducting circuits \cite{Touzard2019,SchoelkopfNature2020,Kwon2021Superconduting,MaSciBull2021SupCond,Ripoll2022SuperconCircuit}.} 
  {This contrasts with the potentially more demanding resource scaling for high-$n$ ON states.}
  The efficiency of preparing SCS with few gates is therefore a significant advantage for \textit{near-term experimental realization} compared to states needing complex control and resources scaling with the desired quantum feature size ($n_\mathrm{max}$ or $|\alpha_\mathrm{max}|^2$), making them highly attractive for quantum sensors operating under current technological constraints.  

Both SCS and ON states exhibit linear QFI and CFI scaling with average quanta number $N_\mathrm{av}$ {under loss}, {which implies that while they do not outperform Gaussian states in the asymptotic limit of large $N_\mathrm{av}$, they provide a significant enhancement in an intermediate regime, thus limiting the range where the Gaussian bound can be overcome} \cite{EscherNatPhys2011ScalingTransition}.
However, their enhancement factor over SQL in the low energy regime are given respectively for the SCS as  $1+\frac{ \eta  }{e(1- \eta) }$ in  (\ref{eq:FQSCSlinear}) and  for ON state as $\frac{1}{e (1-\eta)}$ in (\ref{eq:FQoptNavsmall}), both significantly larger than $2\eta$ of squeezed state. 
Under realistic loss and thermal noise, asymmetric SCS, especially with a smaller weight, show comparable  resilience to ON states. 
 Ultimately, this practical divergence stems from the distinct physical nature of the underlying non-Gaussian resource: the phase-space coherence of SCS is  robustly and efficiently generated by classical drives \sout{than the number-space superposition of ON states}, making it a  accessible resource for near-term technologies operating under realistic noise and control limitations.
Current sensing technologies, often constrained by the Gaussian bound, and inherently susceptible to loss and noise, stand to gain significant performance improvements by adopting SCS probes, resulting in more sensitive and accurate measurements.

Future research should explore multi-coherent state superpositions for improved sensitivity or dynamic range \cite{LeeJOSAB2015Multihead, ShuklaPRA2023superpositioncompass}, {investigate alternative encoding and adaptive measurement strategies to improve the trade-off relation of dynamic range and maximum CFI and mitigate the diminishing fringe contrast.}
Developing more sophisticated error mitigation techniques tailored to SCS \cite{ParkNJP2022Slowing} and experimentally realizing these protocols in platforms like superconducting circuits or trapped-ion systems \cite{Pan2024} {will be crucial}. 
In gravitational wave detectors like LIGO~\cite{abbottLIGO2016observation,Lawrie2019SqueezedSensing}, currently employing squeezed states, the use of optimized SCS probes {promises more sensitive measurements}, {which, based on our results (Fig. \ref{fig:dphiSCSvsSqz}\textbf{b}), could correspond to a threefold uncertainty reduction over the Gaussian bound at a $1\%$ loss level. This would enable}  the detection of weaker gravitational signals or improving the signal-to-noise ratio for existing signals. Similarly, in biological sensing, the enhanced phase sensitivity of SCS could enable more precise measurements of minute changes in refractive index. 
Beyond these examples, the robust sensitivity of SCS probes could be beneficial in atomic clocks, optical gyroscopes, and advanced imaging techniques.
Addressing scalability and further optimizing error correction or mitigation techniques tailored to SCS \cite{ZhouNatComm2018Heisenberg,ShettellNJP2021ECLimitmetrology,ParkNJP2022Slowing} will be crucial steps towards realizing the full potential of these non-Gaussian states for next-generation quantum sensors. 

\section*{Acknowledgment}
We thank Marcin Jarzyna for a helpful discussion regarding Fisher information and the dynamic range. 
{We thank Matteo Fadel for comments.}
K.P. and R.F. acknowledge the project No. 21-13265X of the Czech Science Foundation.
YYG acknowledges funding from the Ministry of Education, under grant A-8001507-00-00.

\section*{Data availability}
The numerical data presented in this study is available from the authors upon request.


\section*{Competing interests}
The authors declare that there are no competing interests.

\section*{Author contributions}
K.P.  performed the analytical and numerical calculations with the inputs and feedbacks from R.F.  
T. K. and Y. Y. G. performed the numerical simulation using GRAPE.
All authors discussed and interpreted the results. K.P. and R.F. wrote the manuscript with contributions from all the co-authors. R.F. supervised the project. 


\section*{Ethics}
The authors declare there are no issues in the current work that needs ethical consideration.

\section*{Code availability}
All codes will be available on request.

\appendix
\counterwithin{figure}{section}

\section{Gaussian bound and Gaussian Mixtures}
\label{append:mixtures}

A pertinent question in the context of surpassing the Gaussian bound is whether a classical mixture of Gaussian states can itself approach or exceed this bound. While the Gaussian bound (\ref{eq:GaussianLimit}) is given by the optimal \textit{pure} Gaussian state (a squeezed vacuum state), it is not a priori clear whether a convex combination of two different pure Gaussian states, $\rho_{\text{mix}} = p \rho_1 + (1-p) \rho_2$ for $0\le p < 1$, where $\rho_1$ and $\rho_2$ are general displaced squeezed states, could offer any metrological advantage.


To {further} investigate this, we performed a numerical analysis by generating a large set (over 40,000) of random mixed states of this form. For each state, we calculated its QFI for phase estimation, $F_Q$, and its purity, $\mathcal{P} = \text{tr}(\rho_{\text{mix}}^2)$. The QFI of each mixture was then compared against the ideal Gaussian bound (\ref{eq:GaussianLimit}), $F_Q^{\text{Gauss}}$, corresponding to the same average photon number.

\begin{figure}[hp!]
    \centering
    \includegraphics[width=0.5\textwidth]{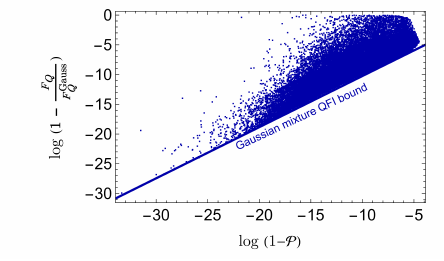} 
    \caption{QFI of a mixture of two Gaussian states relative to the Gaussian bound, as a function of the state's purity. The plot shows $\log(1 - F_Q / F_Q^{\text{Gauss}})$ versus $\log(1 - \mathcal{P})$. Each blue dot represents a randomly generated mixed state $\rho_{\text{mix}} = p \rho_1 + (1-p) \rho_2$ for $0\le p\le 1$, where $\rho_1$ and $\rho_2$ are general displaced squeezed states. The solid blue line represents the derived approximate upper bound on the achievable QFI for a given impurity, indicating that no such mixture can reach the Gaussian bound unless it is perfectly pure.}
    \label{fig:gaussian_mixture}
\end{figure}

The results, plotted in Figure \ref{fig:gaussian_mixture}, demonstrate a clear and fundamental limitation. All numerically generated points lie above a hard boundary, indicating that for any given purity $\mathcal{P} < 1$, there is a maximum achievable QFI that is strictly less than the Gaussian bound. This relationship can be expressed concisely by an approximate linear bound on the log-log scale:
\begin{equation}
\log\left(1 - \frac{F_Q}{F_Q^{\text{Gauss}}}\right) \gtrsim 0.86 \log(1 - \mathcal{P}) - 1.65.
\label{eq:mixture_bound}
\end{equation}

This inequality reveals two key insights. First, the term on the right-hand side is always negative for $\mathcal{P} < 1$, confirming that $F_Q < F_Q^{\text{Gauss}}$. Therefore, a classical mixture of Gaussian states cannot surpass the Gaussian bound. Second, in order to approach the Gaussian bound ($F_Q \to F_Q^{\text{Gauss}}$), the left-hand side of the inequality must approach $-\infty$. This necessitates that $\log(1 - \mathcal{P}) \to -\infty$, which implies that the purity $\mathcal{P}$ must approach 1.

In conclusion, this analysis demonstrates that classical mixtures of Gaussian states do not surpass the Gaussian metrological bound. Any impurity in the state introduces a fundamental deficit in the achievable QFI relative to the bound. This result reinforces the conclusion that genuine non-Gaussian quantum features, such as the interference present in superposition states (e.g., SCS), are necessary to achieve a quantum advantage beyond the limits set by all Gaussian resources, including their classical mixtures.

\FloatBarrier
\section{Scaling of Quantum Fisher Information with Repeated Measurements}
\label{Append:QFIrepetition}

In the context of quantum metrology, the QFI provides a fundamental limit on the precision with which a parameter can be estimated. For a given quantum state, the QFI sets a lower bound on the variance of any unbiased estimator, as dictated by the quantum CRB. When considering scenarios involving repeated measurements with many number of probes, the scaling of the QFI with respect to the number of measurements and the average number of quanta becomes crucial.


For a single measurement on a classical state such as coherent state, the QFI scales linearly with the average number of quanta $N_\mathrm{av}=\langle \hat{n} \rangle$ in the probe state. This linear scaling is characteristic of classical strategies. However, when quantum resources such as entanglement or non-Gaussian states are employed, it is possible to achieve a quadratic scaling of the QFI with respect to $N_\mathrm{av}$. This quadratic scaling is referred to as the Heisenberg limit  expressed as
$F_Q \propto N_\mathrm{av}^2$ and represents a fundamental quantum advantage  for metrology.
This quadratic scaling implies that the precision of the estimation, which is inversely proportional to the square root of the QFI, improves as $1/N_\mathrm{av}$, compared to the $1/\sqrt{N_\mathrm{av}}$ scaling of the SQL.


When considering repeated measurements, the total QFI accumulates. For $N$ independent and identically distributed  measurements, the total QFI is simply the sum of the individual QFIs. If each measurement involves a probe state with average number of quanta $N_\mathrm{av}$, and the QFI for a single measurement scales linearly with $N_\mathrm{av}$ (as in the SQL regime), then the total QFI after $N$ measurements scales as:
\[
F_Q^{\text{total}} = N \times F_Q^{\text{single}} \propto N N_\mathrm{av}
\]
However, this is equivalent to a single measurement with a probe state that has an average number of quanta of $N N_\mathrm{av}$. 
Therefore, in linear scaling regime, simple repetition with $N$ times is effectively equal to using a single probe with $N$ times average number of quanta. 

On the other hand, if a single measurement achieves quadratic scaling (Heisenberg limit), the total QFI after $N$ measurements becomes:
\[
F_Q^{\text{total}} \propto N N_\mathrm{av}^2
\]
Importantly, if we consider the effective average number of quanta used across all measurements as $N_\mathrm{av}^{\text{eff}} = N N_\mathrm{av}$, the total QFI still scales quadratically with this effective value in the Heisenberg limit:
$F_Q^{\text{total}} \propto \frac{1}{N} (N_\mathrm{av}^{\text{eff}})^2$.
This demonstrates that in the Heisenberg limit, repeated measurements effectively degrades the quadratic scaling, and it is beneficial to using a smaller number of probes with a large $N_\mathrm{av}$ than using many number of probes with a small  $N_\mathrm{av}$. 


For the cases discussed in this work, it has been shown in the main text that under 
the presence of loss and noise the QFI of all states degrades to linear scaling. 
In the context of repeated measurements, the use of asymmetric SCS with many copies of SCS probes with a small $N_\mathrm{av}$ is better than using smaller copies of SCS probes with a large $N_\mathrm{av}$. 

\FloatBarrier

\section{The QFI and CFI of general Gaussian states}
\label{sec:GaussianQFI}

{The most general pure Gaussian state is the displaced squeezed state $\ket{\alpha, \zeta}=D[\alpha]S[\zeta]\ket{0}$, where displacement operator $D[\alpha]=e^{\alpha \hat{a}^\dagger-\alpha \hat{a}}$ and squeezing operator $S[\zeta]=e^{-\frac{\zeta}{2}\hat{a}^{\dagger2}+\frac{\zeta^*}{2}\hat{a}^{2}}$. 
The primary precision benchmark is obtained among arbitrary pure Gaussian states, parameterized by displacement amplitude $\alpha$ and squeezing parameter $\zeta$. 
We first assume full displacement resource utilization without any limitation for simplicity. 
Without loss of generality, we treat $\zeta=|\zeta|$ as real, given phase estimation's rotational symmetry. 
The state's average quanta is given as: 
\begin{align}
N_\mathrm{av}=\sinh^2[\zeta]+|\alpha|^2.
\label{eq:NavSqz}
\end{align}
The QFI of the general pure Gaussian states is given by
\begin{align}
    F_Q[\ket{\alpha, \zeta}] = -1 + \cosh(4\zeta) +4|\alpha|^2 e^{2\zeta}
    \label{eq:C2}
\end{align}
To find the Gaussian bound, we maximize the QFI in Eq.~\eqref{eq:C2} for a fixed average number of quanta $N_\text{av}$. 
The two parameters, squeezing $\zeta$ and displacement $|\alpha|^2$, contribute to $N_\text{av}$ as shown in Eq.~\eqref{eq:NavSqz}. The optimal strategy for phase sensing corresponds to maximizing the squeezing contribution, as it provides a quadratic enhancement in QFI, while displacement offers a less effective linear enhancement. The maximum QFI is therefore achieved by allocating all resources to squeezing, which corresponds to setting the displacement to zero ($\alpha=0$).
With $\alpha=0$, the optimal probe state becomes a pure squeezed vacuum state. The constraint from Eq.~\eqref{eq:NavSqz} simplifies to $N_\text{av} = \sinh^2(\zeta)$, and the QFI from Eq.~\eqref{eq:C2} becomes:
\begin{equation}   F_Q[|0, \zeta\rangle] = \cosh(4\zeta) - 1.\end{equation}
To express this QFI in terms of the average quanta $N_\text{av}$, we use the hyperbolic identity $\cosh(4x) = 2\cosh^2(2x) - 1$. First, we find $\cosh(2\zeta)$ using the identity $\cosh(2x) = 1 + 2\sinh^2(x)$:
\begin{equation}   \cosh(2\zeta) = 1 + 2\sinh^2(\zeta) = 1 + 2N_\text{av}.\end{equation}
Now we can write $\cosh(4\zeta)$ in terms of $N_\text{av}$:
\begin{align}   \cosh(4\zeta) &= 2\cosh^2(2\zeta) - 1 \nonumber = 2(1 + 2N_\text{av})^2 - 1 
= 1 + 8N_\text{av} + 8N_\text{av}^2.\end{align}
Substituting this result back into the expression for the QFI gives the final form of the Gaussian bound:}
\begin{align} 
    F_Q^\mathrm{Gauss} = 8 N_\mathrm{av}^2 + 8 N_\mathrm{av},
\end{align}
which serves as an upper bound on the achievable QFI for a fixed $N_\mathrm{av}$ across all Gaussian parameters. 
Its quadratic scaling implies {an ideal} quantum enhanced estimation, so-called Heisenberg scaling.

\subsection*{Gaussian Bound under Bosonic Loss}
In the presence of ambient bosonic losses {during the phase rotation}, the QFI for a {Gaussian} state decreases from its ideal value derived earlier in (\ref{eq:GaussianLimit}) that can be calculated in Gaussian formalism~\cite{WeedbrookRevModPhys.84.621,Banchi2015PhysRevLett.115.260501GaussianFidelity}. At a critical loss $\eta_\mathrm{crit}=\frac{\sqrt{2 N_{\text{av}}+1}-1}{2 N_{\text{av}}}$, the QFI of squeezed state goes below the SQL, causing a {sudden} transition between the optimal classes of Gaussian states from a squeezed state to a coherent state under the $N_\mathrm{av}$ constraint, leaving other states sub-optimal for all  $N_\mathrm{av}$ and $\eta$. 

{

The QFI for a state $\rho$ undergoing a phase shift $\theta$ generated by the number operator $\hat{n}$ is given by the second derivative of fidelity $F_Q = -2 \frac{d^2 \mathbb{F}(\theta)}{d\theta^2}\big|_{\theta=0}$  equivalently to (\ref{eq:FQfid}), where $\mathbb{F}(\theta)$ is the fidelity between the initial state and the state after rotation. For Gaussian states, this can be calculated using their covariance matrices (CMs).

\paragraph{Step 1: State Representation and Transformation.}

The most general pure Gaussian state is the displaced squeezed state $|\alpha, \zeta\rangle = D[\alpha]S[\zeta]|0\rangle$, where $D[\alpha]$ is the displacement operator and $S[\zeta]$ is the squeezing operator. This state is parameterized by the complex displacement amplitude $\alpha$ and the complex squeezing parameter $\zeta$.

In phase space, this state is fully characterized by its first and second statistical moments. The first moment is the displacement vector $\mathbf{d}$, which represents the center of the state's Wigner function. For a general displacement $\alpha$, this vector is $\mathbf{d} = \langle(\hat{q}, \hat{p})\rangle^T = (\sqrt{2}\text{Re}(\alpha), \sqrt{2}\text{Im}(\alpha))^T$. Due to the rotational symmetry of the phase estimation problem, we can, without loss of generality, choose $\alpha$ to be real, simplifying the displacement vector to $(\sqrt{2}\alpha, 0)^T$.


We compare the performance of all the displaced squeezed states.
Such {formulas, which are} omitted due to the lengthy form of the QFI of the general displaced squeezed states, concludes that the optimal state is either the squeezed vacuum state or a coherent state depending on the loss level, whose QFI under loss is $F_{Q, \text{coh}} = 4\eta N_{\text{av}}$. 
{We remind that $N_\mathrm{av}$ is loss independent.}
The squeezed vacuum state $\ket{0,r}$ has average photon number $N_{\text{av}} = \sinh^2(r)$. 
Its CM is:
\begin{equation}
    V_{\text{sqz}} = 
    \begin{pmatrix}
        e^{-2r} & 0 \\
        0 & e^{2r}
    \end{pmatrix}.
\end{equation}
{During the phase rotation,} this state passes through a bosonic loss channel with transmission efficiency $\eta$, which transforms its CM to:
\begin{equation}
    V_{\text{loss}} = \eta V_{\text{sqz}} + (1-\eta)\mathbf{I} = 
    \begin{pmatrix}
        \eta e^{-2r} + 1-\eta & 0 \\
        0 & \eta e^{2r} + 1-\eta
    \end{pmatrix},
\end{equation}
where $\mathbf{I}$ is the identity matrix representing the vacuum's CM. 
A phase rotation by $\theta$ transforms this CM to $V_{\text{loss},\theta} = R_\theta V_{\text{loss}} R_\theta^T$, where $R_\theta$ is the standard 2D rotation matrix.

\paragraph{Step 2: Fidelity and QFI Calculation.}
The fidelity between two zero-displacement single-mode Gaussian states with CMs $V_1$ and $V_2$ is given by $\mathcal{F} = 2 / (\sqrt{\Delta + \Lambda} - \sqrt{\Delta - \Lambda})$, where $\Delta = \det(V_1+V_2)$ and $\Lambda = \det(V_1-i\Omega)\det(V_2-i\Omega)$ with $\Omega$ being the symplectic form. For our case, $V_1=V_{\text{loss}}$ and $V_2=V_{\text{loss},\theta}$. The calculation of $\mathcal{F}(\theta)$ and its second derivative is algebraically intensive. The result of the differentiation yields the QFI in terms of $r$ and $\eta$:
\begin{equation}
    F_{Q, \text{sqz}}(\eta, r) = \frac{2(1-\eta)^2 \sinh^2(2r)}{(1-\eta+\eta^2) + (1-\eta)\eta \cosh(2r)}.
\end{equation}

\paragraph{Step 3: Re-parameterization in terms of $N_{\mathrm{av}}$.}
Using the hyperbolic identities $\cosh(2r) = 1+2N_{\text{av}}$ and $\sinh^2(2r) = 4N_{\text{av}}(1+N_{\text{av}})$, we substitute $r$ for $N_{\text{av}}$ and simplify the expression to obtain:
\begin{equation}
    F_{Q, \text{sqz}}(\eta, N_{\text{av}}) = \frac{8\eta^2 N_{\text{av}}(1+N_{\text{av}})}{1 + 2\eta(1-\eta)N_{\text{av}}}.
\end{equation}

\paragraph{Step 4: Optimization over Gaussian States.}
The Gaussian bound is the maximum QFI achievable by any Gaussian state. 
{The QFI is a strictly convex function over $0\le\alpha\le\sqrt{N_\mathrm{av}}$ for all $ N_\mathrm{av}$ and $\eta$, and thus the maximum is obtained only at the boundaries of $\alpha$. }
The squeezed state is optimal only when $F_{Q, \text{sqz}} \ge F_{Q, \text{coh}}$. The critical point $\eta_{\text{crit}}$ is found by setting them equal. 
Solving this quadratic equation for $\eta$ and taking the physical (positive) root gives:
\begin{equation}
    \eta_{\text{crit}} = \frac{\sqrt{1+2N_{\text{av}}} - 1}{2N_{\text{av}}}.
\end{equation}
For $\eta < \eta_{\text{crit}}$, the coherent state provides a higher QFI. Therefore, the optimal Gaussian QFI under bosonic loss is a piecewise function:
}
\begin{align}
 F_Q^\mathrm{Gauss}[\eta]=
\begin{cases}
    \frac{8 \eta ^2 N_{\text{av}} \left(N_{\text{av}}+1\right)}{2 (1-\eta) \eta  N_{\text{av}}+1}, & \eta\ge \eta_\mathrm{crit},\\
     4\eta N_\mathrm{av}, & \eta< \eta_\mathrm{crit}.
     \end{cases}
    \label{eq:QFIsqzloss}
\end{align}
where $\eta$ is the transmission efficiency parameter (with $0 \leq \eta \leq 1$) representing the fraction of quantas transmitted without loss. 
This also provides an alternative benchmark when loss is ambient in the setup and unavoidable. The reduction in QFI from (\ref{eq:GaussianLimit}), i.e. $F_Q^\mathrm{Gauss}[\eta]\approx F_Q^\mathrm{Gauss}[0]-16(1-\eta)  N_{\text{av}} \left(N_{\text{av}}+1\right)^2$
in the weak loss limit $1-\eta\ll 1$ reflects how losses {cubically} degrade the {quadratically rising QFI} achievable with squeezed states particularly with a large $N_\mathrm{av}$, emphasizing the importance of high transmission efficiency in surpassing the SQL. In {very} large quanta limit $N_\mathrm{av}\gg 1$ the Gaussian bound is reduced to $  F_Q^\mathrm{Gauss}[\eta]\approx \frac{4 \eta  N_{\text{av}}}{(1-\eta)}$ for all $\eta<1$ {even for $\eta\approx 1$}, implying classical scaling {linear in $N_\mathrm{av}$} even under a weak loss.
In a small quanta limit $N_\mathrm{av}\ll 1$, the Gaussian bound becomes $  F_Q^\mathrm{Gauss}[\eta]\approx 8\eta^2 N_\mathrm{av}$, going below the ideal SQL at $\eta=1/\sqrt{2}$.
At all levels of loss, squeezing monotonously increases the QFI, and at all levels of squeezing, loss degrades the QFI monotonously.

\subsection*{Gaussian Bound under thermal noise}
In the presence of ambient thermal noise, the QFI for a general Gaussian state decreases from its ideal value due to interactions with the thermal environment. Accounting for thermal noise, its mathematical expression becomes complicated, and presented in Appendix \ref{appendix:thermal}.
Similarly as under loss, the optimal state under average number of quanta constraint is transitioned from a squeezed state to a coherent state at $\eta_\mathrm{crit}[N_{\text{th}}]=\frac{\sqrt{N_{\text{av}} \left(4 N_{\text{th}} \left(N_{\text{th}}+1\right)+2\right)+N_{\text{th}}
   \left(3 N_{\text{th}}+2\right)+1}-2 N_{\text{av}}+N_{\text{th}}-1}{2
   \left(N_{\text{av}}-N_{\text{th}} \left(N_{\text{th}}+2\right)\right)}$.
For different $\eta$, squeezed states $\alpha=0$ are {and coherent states are}  still the optimal states leaving other Gaussian states sub-optimal, and its QFI is given as:
\begin{align}
    F_Q^\mathrm{Gauss}[\eta;N_{\text{th}}]=\begin{cases}\frac{8 \eta ^2 N_{\text{av}} \left(N_{\text{av}}+1\right)}{2 (1-\eta ) \eta  N_{\text{av}} \left(2
   N_{\text{th}}+1\right)+2 (1-\eta ) N_{\text{th}} \left((1-\eta ) N_{\text{th}}+1\right)+1}, & \eta\ge \eta_\mathrm{crit}[N_{\text{th}}] ~\text{for squeezed states}\\
   \frac{4 \eta  N_{\text{av}}}{2 (1-\eta ) N_{\text{th}}+1}, &\eta< \eta_\mathrm{crit}[N_{\text{th}}]~\text{for coherent states}.
   \end{cases}
\end{align}
In {both limits of $N_{\text{av}}$, when $\eta\ge \eta_\mathrm{crit}[N_{\text{th}}]$ it scales as 
\begin{align}
    F_Q^\mathrm{Gauss}[\eta;N_{\text{th}}]\approx \begin{cases}\frac{4 \eta  N_{\text{av}}}{(1-\eta ) \left(2 N_{\text{th}}+1\right)}, & N_{\text{av}}\gg 1\\
    \frac{8 \eta ^2 N_{\text{av}}}{2 (1-\eta )^2 N_{\text{th}}^2+2 (1-\eta ) N_{\text{th}}+1}, & N_{\text{av}}\ll 1.
    \end{cases}
\end{align}
higher than} the QFI of the coherent states, exhibiting
 a classical linear scaling vs $N_\mathrm{av}$, monotonously decreasing with $N_\mathrm{th}$, even in a very weak loss $1-\eta\ll 1$, implying a transition to a classical linear scaling from quadratic Heisenberg scaling.
This expression reflects the degradation of the QFI due to thermal noise, which introduces additional fluctuations and decoherence, thereby diminishing the precision of phase estimation achievable with a squeezed state. The dependence on $N_\mathrm{th}$ and $\eta$ captures how thermal photons and environmental interactions reduce the effectiveness of squeezing. 
In the limit where thermal noise is negligible ($N_\mathrm{th} \to 0$), this is reduced to the QFI under loss (\ref{eq:QFIsqzloss}), and the first order dependence on $N_\mathrm{th}$ is described as:
\begin{align}
    F_Q^\mathrm{Gauss}[\eta;N_{\mathrm{th}}]\approx F_Q^\mathrm{Gauss}[\eta;0]-N_{\text{th}}\frac{16 \eta ^2 N_{\text{av}} \left(N_{\text{av}}+1\right)  (1-\eta )\left(1+2 \eta 
   N_{\text{av}}\right)}{\left(2 (1-\eta )^2 N_{\text{av}}-2 (1-\eta )
   N_{\text{av}}-1\right)^2}.
\end{align}

{

\begin{figure}[hp]
\includegraphics[width=\textwidth]{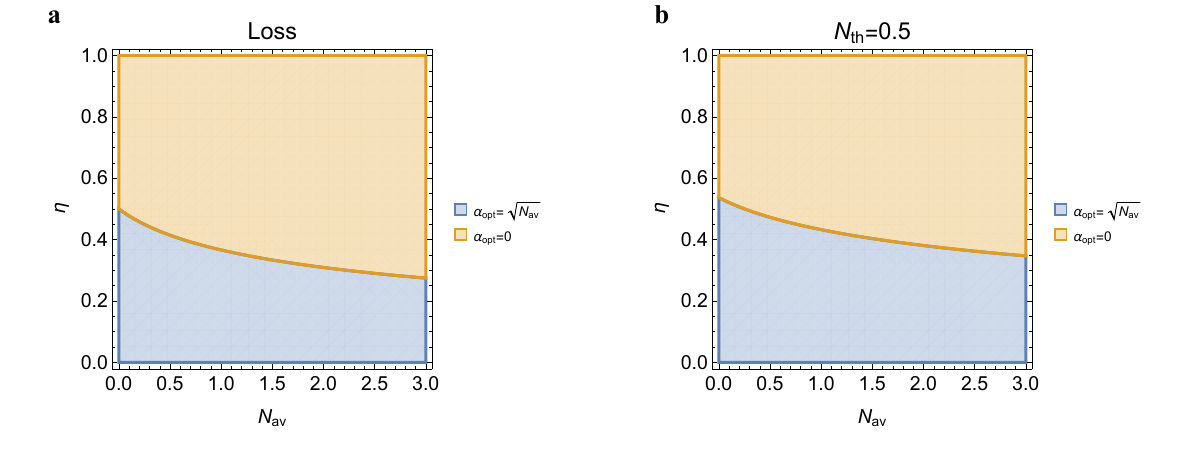}
    
    
    \caption{{\textbf{Phase diagram of optimal Gaussian states under a pure bosonic loss channels.} 
    The regions are shown where each $\alpha_\mathrm{opt}$, such as the squeezed vacuum state ($\alpha_\mathrm{opt}=0$, orange) or the coherent state ($\alpha_\mathrm{opt}=\sqrt{N_\mathrm{av}}$, blue), yields the maximum QFI. 
    In no tested scenario does a displaced squeezed state ($0 < |\alpha| < \sqrt{N_\mathrm{av}}$) become optimal.
    \textbf{(a)} Under a pure bosonic loss channel, the squeezed vacuum is optimal for high transmission efficiency $\eta$. 
    The critical loss $\eta_\mathrm{crit}$ at which the coherent state becomes superior is lower for higher $N_\mathrm{av}$.
    \textbf{(b)} With the addition of symmetric thermal noise ($N_\mathrm{th}=0.5$), the region of optimality for the squeezed vacuum shrinks, as this noise is particularly detrimental to squeezing.
    }    }
    \label{fig:gaussian_phase_diagrams}
\end{figure}

We  now rigorously establish the optimal pure Gaussian states for phase estimation under various realistic noise conditions.
We analyze above four distinct and physically relevant loss or noise models. 
For each model, we numerically maximize QFI over all pure Gaussian states for a given initial state's average number of quanta $N_\mathrm{av}$ and loss or noise strength.
The results are presented as phase diagrams in Fig.~\ref{fig:gaussian_phase_diagrams}, showing which state is optimal in each parameter regime.
Fig.~\ref{fig:gaussian_phase_diagrams}(a) shows the result for pure loss. For high transmission efficiency $\eta$ (low loss), the squeezed vacuum is optimal. As loss increases, its quantum advantage degrades faster than that of a coherent state, and we cross a critical threshold where the coherent state becomes the superior probe. This threshold is lower for states with higher energy ($N_\mathrm{av}$), as they are more susceptible to decoherence.
Fig.~\ref{fig:gaussian_phase_diagrams}(b) shows the effect of a thermal bath with $N_\mathrm{th}=0.5$. The added thermal noise further degrades the squeezed state, causing its region of optimality to shrink. A higher transmission efficiency $\eta$ is now required for the squeezed vacuum to outperform the more robust coherent state. In both symmetric noise models, the optimal state is always one of the two extremal cases.


Our systematic analysis across distinct, physically motivated channels leads to a single conclusion: the optimal pure Gaussian state for phase estimation is always an extremal state—either the \textit{squeezed vacuum state} or the \textit{coherent state}. 
A displaced squeezed state, which represents a compromise between squeezing and displacement, is never found to be the optimal choice. 
This justifies our focus on the Gaussian Bound and the SQL as the two essential benchmarks for evaluating the performance of non-Gaussian quantum states in the main text.
This analysis is conducted using the QFI, while using CFI gives analogous results.

}

\FloatBarrier

\subparagraph{CFI of Gaussian state}

The CFI of the optimal Gaussian  state (squeezed vacuum state), under the average quanta number constraint $N_\mathrm{av}=\sinh[\zeta]^2$, with the projective detection onto itself {(binary detection)}, is expressed in a lengthy form as:
\begin{align}
 &F_C[\ket{\zeta}]=\Bigg(2 \sqrt{2} \eta ^2 \sin ^2(2 \theta ) \left(e^{4 \sinh
   ^{-1}\left(\sqrt{N_{\text{av}}}\right)}-1\right)^4\Bigg)B^{-1}C^{-1},
   \label{eq:CFIgaussianloss}
\end{align}
where {\small
\begin{align*}
    B=&\Bigg(\eta  \cos (2 \theta ) \left(e^{4 \sinh
   ^{-1}\left(\sqrt{N_{\text{av}}}\right)}-1\right){}^2+\left(-4 (1-\eta )^2+6 (1-\eta )-6\right) e^{4
   \sinh ^{-1}\left(\sqrt{N_{\text{av}}}\right)}\\
   &+2 (-\eta -1) (1-\eta ) e^{2 \sinh
   ^{-1}\left(\sqrt{N_{\text{av}}}\right)}+2 (-\eta -1) (1-\eta ) e^{6 \sinh
   ^{-1}\left(\sqrt{N_{\text{av}}}\right)}-\eta  e^{8 \sinh ^{-1}\left(\sqrt{N_{\text{av}}}\right)}-\eta
   \Bigg)^{2}
\end{align*}} 
and 
{\small \begin{align*}C=&2 \sqrt{2}-\Bigg(e^{-4 \sinh ^{-1}\left(\sqrt{N_{\text{av}}}\right)} \Big(-\eta  \cos (2 \theta )
   \left(e^{4 \sinh ^{-1}\left(\sqrt{N_{\text{av}}}\right)}-1\right){}^2+\left(4 (1-\eta )^2-6 (1-\eta
   )+6\right) e^{4 \sinh ^{-1}\left(\sqrt{N_{\text{av}}}\right)}
   \\&-2 (-\eta -1) (1-\eta ) e^{2 \sinh
   ^{-1}\left(\sqrt{N_{\text{av}}}\right)}-2 (-\eta -1) (1-\eta ) e^{6 \sinh
   ^{-1}\left(\sqrt{N_{\text{av}}}\right)}+\eta  e^{8 \sinh ^{-1}\left(\sqrt{N_{\text{av}}}\right)}+\eta
   \Big)\Bigg)^{-1/2}.
   \end{align*}}
The primary peak spans the interval $\theta \in \left[-\frac{\pi}{2}, \frac{\pi}{2}\right]$, showing a broad dynamic range.
The CFI of the squeezed state exhibits a dip at the phase origin and shifts the peak CFI when loss or noise exists, which never reaches the QFI under these environment. However, as a benchmark, we still adopt the QFI instead of the CFI to prove that all Gaussian strategy, even ideal one, is surpassed by the new probes and protocols.

\FloatBarrier

\section{Classical non-Gaussian states}
\label{Appendix:non-Gaussian}
Consider a classical non-Gaussian mixture of the vacuum and a coherent state, represented by the density operator
$\rho_\mathrm{mix} = \frac{1}{2} \left( \ket{0}\bra{0} + \ket{\alpha}\bra{\alpha} \right)$.
This state has an average number of quanta $N_\mathrm{av} = \frac{1}{2} |\alpha|^2$. However, in the limit of large $\alpha$, the quantum Fisher information (QFI) for this mixture is $F_Q[\rho_\mathrm{mix}] =2|\alpha|^2$, which is significantly lower than the QFI achievable with the superposition state. This lower QFI underscores the limited effectiveness of such classical mixtures for enhancing phase estimation sensitivity.
Furthermore, an asymmetric classical non-Gaussian state defined by
$\rho_\mathrm{mix}^{p} = p \ket{0}\bra{0} + (1-p) \ket{\alpha}\bra{\alpha}$,
where $0 \leq p \leq 1$, also yields a QFI of the form $F_Q[\rho_\mathrm{mix}^p] =4(1-p) |\alpha|^2$ in the large $\alpha$ limit. 
The average number of quanta is given as $N_\mathrm{av}=(1-p)|\alpha|^2$. Therefore, the scaling of QFI is equal to classical scaling $F_Q[\rho_\mathrm{mix}^p]=4N_\mathrm{av}$ regardless of $p<1$, showing no non-Gaussian enhancement.
This further emphasizes the superior performance of the superposition state in Eq.~(\ref{eq:2SCS}) over classical mixtures for applications in quantum sensing and highlights the importance of quantum interference effects in achieving quantum advantages.

\FloatBarrier

\section{Analysis of the QFI of various states}
\label{Appendix:QFI}

\begin{figure}
    \centering
    \includegraphics[width=0.5\linewidth]{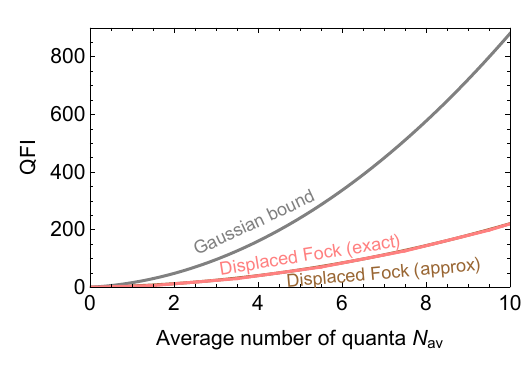}
    \caption{{Comparison of the optimal QFI for displaced Fock states against the Gaussian bound, as a function of the average number of quanta $N_\mathrm{av}$. 
    (Gray) Gaussian bound, the maximum QFI achievable by any Gaussian state. 
   (Red)  the exact QFI for an optimized displaced Fock state; for each $N_\mathrm{av}$, the Fock number $n$ is optimized to maximize the QFI, resulting in a piecewise function due to the integer nature of $n$. 
    (Brown)   the smooth quadratic approximation, which closely tracks the exact value. 
    The figure clearly illustrates that even under optimal conditions, the QFI of a displaced Fock state remains significantly below the Gaussian bound for all values of $N_\mathrm{av}$.}}
    \label{fig:dispFock}
\end{figure}

The QFI of the displaced Fock states $D[\alpha]\ket{n}$, without loss or noise is expressed in the representations of $(n,\alpha)$ and $(N_\mathrm{av},n)$ as:
\begin{align}
    F_Q[\ket{\alpha,n}]=4 (1+2n)|\alpha|^2=4(1+2n)(N_\mathrm{av}-n), ~~~~n\le N_\mathrm{av}.
    \label{eq:QFIdispFock}
\end{align}
{The metrological potential of this state depends critically on the applied physical constraints. 
First, if the Fock number $n$ is held constant due to experimental limitations, Eq.~(\ref{eq:QFIdispFock}) shows that the QFI increases linearly with $N_\mathrm{av}$. This represents a classical-like scaling similar to the SQL in (\ref{eq:FQcohthermal}).}
{However, a more fundamental analysis is performed under the constraint of a fixed total average quanta $N_\mathrm{av}$, which treats $n$ as a tunable parameter to be optimized ($0 \le n \le N_\mathrm{av}$). To find the maximum QFI, we optimize the expression in Eq.~(\ref{eq:QFIdispFock}) with respect to $n$. The optimal Fock number is found to be $n_\mathrm{opt}=\left\lfloor\frac{2N_\mathrm{av}+1}{4}\right\rfloor$, which increases with the average number of quanta $N_\mathrm{av}$, asymptotically approaching $N_\mathrm{av}/2$.}
Due to the discreteness of the Fock number $n_\mathrm{opt}$, this optimal number is achieved at only discrete values of $N_\mathrm{av}$. 
{The QFI of this state optimized under the constraint of the average number of quanta (both exact and approximate (\ref{eq:dispFockopt})), can be compared with the ideal Gaussian bound (\ref{eq:GaussianLimit}), as in Fig.  \ref{fig:dispFock}. 
}

{As illustrated in Figure~\ref{fig:dispFock}, when we compare these results to the Gaussian bound, it is evident that the optimal displaced Fock state does not provide any quantum advantage over optimal Gaussian states. In fact, its QFI is only about 25\% of the Gaussian bound, confirming that this class of states is not a resource for surpassing the Gaussian bound in phase estimation.}

\begin{figure}
    \centering
    \includegraphics[width=1.0\linewidth]{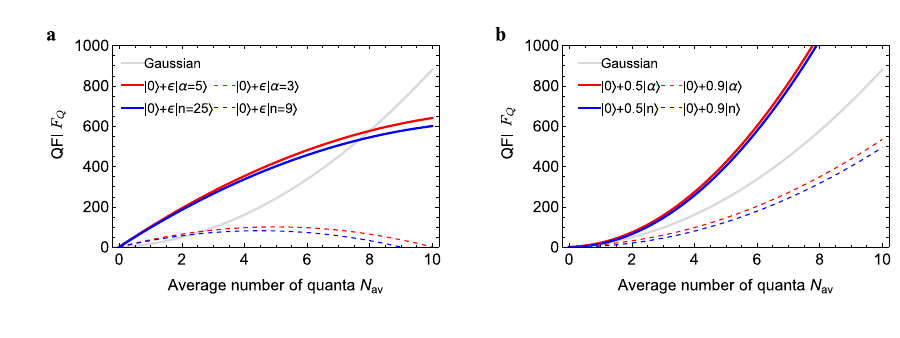}
    \caption{\textbf{a}) Comparison of QFI of the evaluated states for a fixed $n_\mathrm{max}$ and $|\alpha_\mathrm{max}|^2$.  \textbf{b}) Comparison of QFI of the evaluated states for a fixed $\epsilon$. This elucidates how asymmetry contributes to an enhanced QFI, and non-Gaussian enhancement of SCS and ON states. }
    \label{fig:QFI-epsilon}
\end{figure}

Analytical expression of QFI for SCS is given in the parameter pair of $(\alpha,\epsilon)$ without loss or noise as
\begin{align}
F_Q[\ket{\psi}_\mathrm{SCS}]=\frac{4  \alpha ^2 \epsilon ^2 \left( \left(\alpha ^2+\epsilon ^2+1\right)+2 e^{-\frac{\alpha ^2}{2}}\left(\alpha ^2+1\right) \epsilon \right)}{\left(\left(\epsilon^2+1\right)+2e^{-\frac{\alpha ^2}{2}}  \epsilon \right)^2}.
\end{align}
Figure \ref{fig:QFI-epsilon}\textbf{a} shows the non-Gaussian enhancement under the consideration of experimental limitations, while \textbf{b} elucidates the scaling for each class of states for a fixed $\epsilon$.

\begin{figure}
    \centering
    \includegraphics[width=0.7\textwidth]{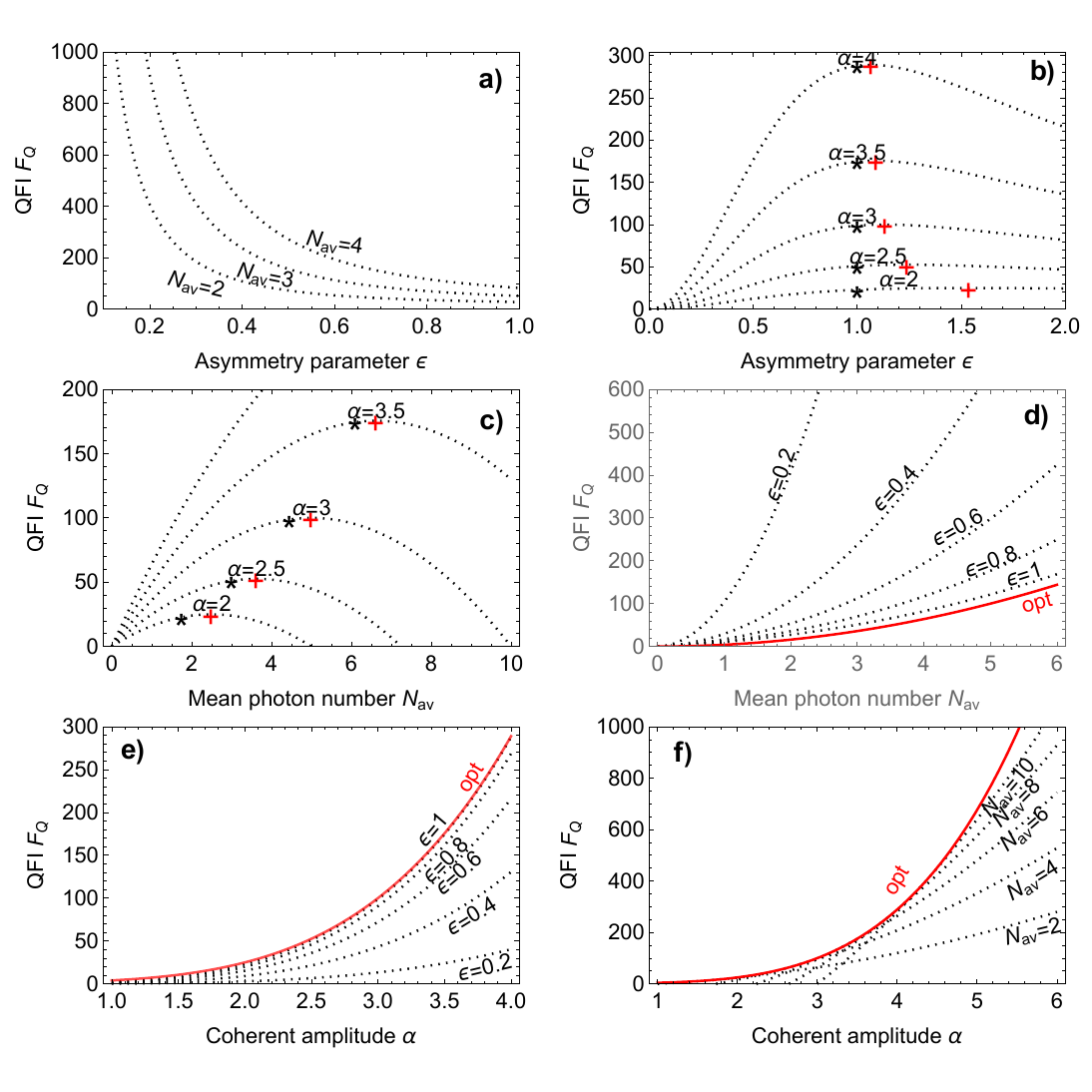}
    \caption{QFI vs asymmetry parameter $\epsilon$ for SCS states under different constraints. 
\textbf{a}) QFI vs. asymmetry parameter $\epsilon$.  
Varying fixed average quanta number $N_\mathrm{av}$ in the parameter pair $(N_\mathrm{av},\epsilon)$, the asymmetric SCS with $\epsilon\ll 1$ exhibits a higher QFI than the symmetric SCS with $\epsilon=1$.
\textbf{b}) QFI vs. asymmetry parameter $\epsilon$.  
Varying coherent state amplitude $\alpha$ in the parameter pair $(\alpha,\epsilon)$, the symmetric SCS gives better QFI than the asymmetric SCS.
The symmetric SCS (with $\epsilon=1$) is marked with ``$*$'', and the optimal SCS  is marked with ``$+$''.
\textbf{c}) With constraint on coherent state amplitude $\alpha$ in the parameter pair $(N_\mathrm{av},\alpha)$, a reparametrization of \textbf{b}.
Here, the large average quanta number limit will have a low QFI due to the classicality, while the low average quanta number limit will have a low QFI due to the small quanta number.
\textbf{d}) With fixed asymmetric $\epsilon$, we get the average quanta number scaling of QFIs.
The collection of optimal points of ``$+$'' in c) for various $\alpha$ (red), that gives a worse scaling than the balanced SCS ($\epsilon=1$), which is a collection of points ``$*$'' in \textbf{c}.
\textbf{e, f}) $\alpha$-scaling of QFI.
The optimal points in \textbf{b,c} has the largest $\alpha$-scaling.}
    \label{fig:QFIvsNavEpsilon}
\end{figure}

Figure \ref{fig:QFIvsNavEpsilon} shows how the optimization of SCS probes depends on the choice of constraint in all possible parameter pairs, $(N_\mathrm{av},\epsilon)$, $(\alpha,\epsilon)$, and $(N_\mathrm{av}, \alpha)$. 
When the average quanta number $N_\mathrm{av}$ is fixed as a constraint in the first pair, asymmetric SCS states with small $\epsilon$ exhibit higher QFI than the balanced cat state, while the analytical solution of optimal QFI depending on $N_\mathrm{av}$ in the limit $\epsilon\to0$ cannot be obtained due to the divergence. 
Under a fixed displacement amplitude $\alpha$ constraint as in \textbf{b,c} in the second and third parameter pair, a slightly more classical cat state with $\epsilon>1$ than the balanced SCS maximizes the QFI, when compared to classical coherent state and a highly asymmetric cat with a very small average quanta number. 
The optimal QFI for a fixed $\alpha$ is given as $F_Q^\mathrm{(opt)}=1+2\alpha^2+\alpha^4$ at the optimal average quanta number $N_\mathrm{av}^\mathrm{(opt)}=\frac{1+\alpha^2}{2}$. 
This can be re-expressed as $F_Q^\mathrm{(opt)}={4N_\mathrm{av}^\mathrm{(opt)}}^2$, and is approached by the symmetric SCS $\epsilon=1$.
We note that this indeed gives the optimal $\alpha$-scaling of QFI, while does not give the optimal $N_\mathrm{av}$-scaling of QFI, and there are SCS with $\epsilon<1$ with a larger $N_\mathrm{av}$-scaling. 
A symmetric SCS has QFI of $F_Q^\mathrm{(sym)}=\frac{2 e^{-\frac{\alpha ^2}{2}} \left(\alpha ^2+1\right) \alpha ^2+\left(\alpha ^2+2\right) \alpha ^2}{\left(e^{-\frac{\alpha ^2}{2}}+1\right)^2}$.
This highlights that the optimal SCS probe depends on the constraints, such as on $N_\mathrm{av}$ or $\alpha$.

A general class of  binary SCS in the form of $\ket{\psi}_\mathrm{SCS}^\mathrm{(gen)}=N(\ket{\alpha_0}+\epsilon \ket{\alpha})$ for $\alpha_0>0$ with a normalization factor $N=(2 \epsilon  e^{-\frac{1}{2} (\alpha -\alpha_0)^2}+\epsilon ^2+1)^{-1/2}$  has an average number of quanta given as $N_\mathrm{av}^\mathrm{(gen)}=\frac{e^{-\alpha  \alpha _0} \left(e^{\frac{1}{2} \left(\alpha ^2+\alpha _0^2\right)} \left(\alpha ^2 \epsilon ^2+\alpha _0^2\right)+2 e^{\alpha  \alpha _0} \alpha  \alpha _0 \epsilon \right)}{e^{\frac{1}{2}
   \left(\alpha -\alpha _0\right){}^2} \left(\epsilon ^2+1\right)+2 \epsilon }$. Its QFI is given in a complex form as 
\begin{align}
        &F_Q^\mathrm{(gen)} = \Bigg(4 e^{-\frac{\alpha ^2}{2}-\alpha _0 \alpha -\frac{\alpha _0^2}{2}} \Big(4 e^{\frac{\alpha ^2}{2}+\alpha _0 \alpha +\frac{\alpha _0^2}{2}} \alpha  \alpha _0 \epsilon ^2+e^{\alpha ^2+\alpha _0^2} \alpha 
   \alpha _0 \epsilon  \left(-4 \alpha ^2 \epsilon ^2+\alpha  \alpha _0 \left(2 \epsilon ^2+2\right)-4 \alpha _0^2+2 \epsilon ^2+2\right)\nonumber\\
   &+e^{\alpha ^2+\alpha _0^2} \epsilon  \left(\alpha ^2 \left(2 \alpha
   ^2+2\right) \epsilon ^2+2 \alpha _0^4+2 \alpha _0^2\right)+e^{\frac{3 \alpha ^2}{2}-\alpha _0 \alpha +\frac{3 \alpha _0^2}{2}} \left(\alpha _0^2 \left(-2 \alpha ^2 \epsilon ^2+\epsilon ^2+1\right)+\alpha ^2
   \epsilon ^2 \left(\alpha ^2+\epsilon ^2+1\right)+\alpha_0^4 \epsilon ^2\right)\Big)\Bigg)\nonumber\\
   &\Big(\left(e^{\frac{1}{2} \left(\alpha -\alpha _0\right){}^2} \left(\epsilon ^2+1\right)+2 \epsilon \right)^2\Big)^{-1}.
\end{align}
For a given constraint of average number of quanta, the best scaling is found by setting $\alpha_0=0$, which is reduced to the asymmetric SCS considered in the main text.

Let us now consider the decoherence of a SCS  undergoing bosonic loss channel $\Gamma_\eta$.
Any density matrix $\rho$ under $\Gamma_\eta$ evolves as $\Gamma_\eta[\rho]=\sum_{l=0}^\infty \frac{(1-\eta)^l}{l!} \eta^{\hat{n}/2} \hat{a}^l \rho \hat{a}^{\dagger l} \eta^{\hat{n}/2}$  with  the loss parameter $\eta=\mathrm{e}^{-\gamma t}\in (0,1]$  and $\gamma t$ is the dimensionless damping parameter, equivalent to Kraus operator notation in \cite{AlbertPRA2018bosoniccode}.
This is equivalent to the solution of the Lindblad equation $\partial_t\rho=L \rho L^\dagger-\frac{1}{2}\{L^\dagger L \rho\}$ with  a Lindblad operator with $L=\sqrt{\gamma}\hat{a}$. The coherence term of SCS then changes by this loss as $\ket{\alpha}\bra{0}\to e^{-\alpha^2(1-\eta)/2}\ket{\sqrt{\eta}\alpha}\bra{0}$, while classical terms undergoes only the amplitude decay.  The full density matrix of a SCS under loss is therefore described as
\begin{align}
    \rho_\mathrm{SCS}=N\left(\ket{0}\bra{0}+\epsilon^2\ket{\sqrt{\eta}\alpha}\bra{\sqrt{\eta}\alpha}+\epsilon e^{-\alpha^2(1-\eta)/2}\ket{\sqrt{\eta}\alpha}\bra{0}+\epsilon e^{-\alpha^2(1-\eta)/2}\ket{0}\bra{\sqrt{\eta}\alpha}\right),
    \label{eq:rhoscs}
\end{align}
for a normalization factor $N=(2+2\epsilon e^{-\alpha^2/2})^{-1/2}$.

Under the loss the probe state becomes mixed, and the density matrix holds a target parameter $\theta$ as $\rho_\theta$.
In our case the target parameter $\theta$ is the angle of the phase rotation of the oscillator.
We use the expression of a QFI for density matrix as $F_Q[\theta]=\lim_{d\theta\to 0}\frac{4(1-\mathbb{F}[\rho_\theta,\rho_{\theta+d\theta}])}{d\theta^2}$ where $\mathbb{F}[\rho,\sigma]=\left(\text{tr}\sqrt{\sqrt{\rho}\sigma\sqrt{\rho}}\right)^2$ is the fidelity \cite{braunsteinPRL1994distance}. The rotated density matrix is simply given as $\rho_\theta=\exp[\ii \theta \hat{n}]\rho \exp[-\ii \theta \hat{n}]$.
However, the square root of a density matrix makes an analytical formulas challenging to be obtained. 

This difficulty can be overcome using the fact that the small number of terms needed to describe the density matrices in Eq. (\ref{eq:rhoscs}). For this, first we make an assumption $\alpha \gg 1$, which is a parameter region of interest where the numerical simulation is difficult. For a small $d\theta\ll 1$, the effect of rotation on coherent state is described approximately as
\begin{align}
\exp[\ii d\theta \hat{n}]\ket{\alpha}=\ket{\alpha e^{\ii d\theta}}\approx \mathcal{N}\Big((1+\ii d\theta \alpha^2)\ket{\alpha}+\ii d\theta \alpha \ket{\alpha^\perp}\Big),
\end{align}
where $\mathcal{N}=(1+ d\theta^2 \alpha^2+d\theta^2 \alpha^4)^{-1/2}$ is the normalization factor and $\ket{\alpha^\perp}\equiv \hat{a}^\dagger \ket{\alpha}-\alpha \ket{\alpha}$ is an normalized state orthogonal to $\ket{\alpha}$ and $\ket{0}$. This representation is advantageous in that now the fidelity between the density matrices can be fully represented in the three dimensional space $\{\ket{0},\ket{\alpha},\ket{\alpha^\perp}\}$.
The QFI found from this representation is neatly expressed as
\begin{align}
    F_Q\approx\frac{4 \alpha ^2 \eta  \epsilon ^2 \left(\alpha ^2 \eta  e^{\alpha ^2 (\eta-1)}+\epsilon ^2+1\right)}{\left(\epsilon ^2+1\right)^2}. 
\end{align}
This expression can improve further to cover the low $\alpha$ cases by considering the overlap  between the vacuum and coherent state and the orthogonalization, which results in a more complex expression.  Interestingly, $\ket{\alpha^\perp}$ still stays orthogonal to these states.
For a symmetric cat where $\epsilon=1$, it is reduced to $ F_Q\approx   4N_\mathrm{av}^2\eta^2  e^{\alpha ^2 (\eta-1)}+4 N_\mathrm{av} \eta $, which is below the lossy Gaussian bound in Eq. (\ref{eq:FQGausseta}).

The QFI of the ON state is calculated using the fidelity between the state before and after the phase rotation under non-ideal scenarios. However, there is a computational challenge in obtaining  the exact analytical expression due to the large Fock dimension required in the numerical simulation of cases with a large $n$.

Similar challenge in the analytic derivation can be partially overcome on ON states and the approximate expression of QFI of the ON state can be calculated analytically by using a simple model of the density matrices in a large $n$ limit, which parameter regime is important for scaling up.
Under an oscillator loss, the density matrix after the phase rotation is described more briefly as:
\begin{align}
    \rho_\theta=\frac{1}{1+\epsilon^2}\ket{0}\bra{0}+\eta^{n/2} \frac{\epsilon}{1+\epsilon^2}e^{-\ii \theta n}\ket{0}\bra{n}+\eta^{n/2} \frac{\epsilon}{1+\epsilon^2}e^{\ii \theta n}\ket{n}\bra{0}+\eta^{n} \frac{\epsilon^2}{1+\epsilon^2}\ket{n}\bra{n}+\rho^\perp.
\end{align}
Here, $\rho^\perp$ term contains all the diagonal terms in the subspace in $\{\ket{0},\ldots \ket{n-1}\}$, and in a small loss limit of $1-\eta\ll 1$ and large $n$ limit, it can be reliably assumed to be in $\{\ket{1},\ldots \ket{n-1}\}$, who does not have any overlap with the other terms. We approximate this term as $\rho^\perp\approx\left(1-\frac{1}{1+\epsilon^2}-\eta^{n}\frac{\epsilon^2}{1+\epsilon^2}\right)\ket{\mathbbm{1}}\bra{\mathbbm{1}}$ where $\ket{\mathbbm{1}}$ is an abstract state orthogonal to all the other states  for simplicity in the calculation. Now the entire density matrix is described as a qutrit state, and the fidelity $\mathbb{F}[\rho_\theta,\rho_0]$ is neatly summarized as
\begin{align}
    \mathbb{F}[\rho_\theta,\rho_0]=\left(\frac{\epsilon ^2 \left(1-\eta ^{ n}\right)}{\epsilon ^2+1}+\sqrt{\frac{\epsilon ^2 \eta ^{4 n}+2 \epsilon ^2 \eta ^{n} \cos (\theta  n)+1}{\left(\epsilon ^2+1\right)^2}}\right)^2.
\end{align}
The QFI then follows straightforwardly from this fidelity as:
\begin{align}
   F_Q[\rho_\theta]= \frac{4 n^2 \epsilon ^2 \eta ^{n}}{\left(\epsilon ^2+1\right) \left(\epsilon ^2 \eta ^{ n}+1\right)}=\frac{4n (n-N_\mathrm{av}) N_\mathrm{av}\eta ^{n}}{N_\mathrm{av} \eta ^{n}+n-N_\mathrm{av}}.
\end{align}
In the lossless case of $\eta=1$, this expression is reduced to the previous expression $\frac{4n^2 \epsilon^2}{(1+\epsilon^2)^2}$. This analytical expression matches numerical results faithfully.
{For the symmetric ON state, the QFI is reduced as $F_Q[\rho_\theta]=\frac{8N_\mathrm{av}^2 \eta ^{2N_\mathrm{av}}}{ \eta ^{2N_\mathrm{av}}+1}$, which is below the lossy Gaussian bound in (\ref{eq:FQGausseta}).}

\FloatBarrier

\section{Qubit measurement basis and the effect of environmental effects on CFI of SCS}
\label{sec:qubitmeasurement}

Initially, we consider the qubit measurement basis $\{\ket{g}\bra{g}, \ket{e}\bra{e}\}$ for simplicity. Utilizing this basis results in a prominent central peak in the CFI as a function of the phase parameter $\theta$ in an ideal scenario without loss, and noise. This peak signifies high sensitivity of the estimation protocol around $\theta = 0$.

However, in realistic scenarios, environmental perturbations such as bosonic loss or qubit dephasing can significantly impact the measurement outcomes. Specifically, these environmental effects cause the probability $p(\theta)$ at $\theta = 0$ to approach unity, making the derivative $\frac{\partial p(\theta)}{\partial \theta}$ near zero. Since the CFI is proportional to the square of this derivative, the central peak in the CFI rapidly diminishes under even minor environmental disturbances that drops $p(\theta=0)$ from unity. This leads to a marked decrease in estimation sensitivity, effectively nullifying the advantage provided by the initial measurement basis in this limit.

To mitigate this issue, we propose modifying the final qubit detection basis to $\{\ket{+_i}\bra{+_i}, \ket{-_i}\bra{-_i}\}$, where $\ket{\pm_i} = \frac{1}{\sqrt{2}}(\ket{e}\pm \ii\ket{g})$. This change has following effects:
In the absence of environmental noise, the maximum CFI at $\theta = 0$ is slightly reduced from the CFI utilizing basis $\{\ket{g}\bra{g}, \ket{e}\bra{e}\}$. 
{In the limit of $\alpha\gg 1$, it realizes the projective measurement onto the balanced SCS $\ket{0}+\ii\ket{\alpha}$.}
However, when small environmental effects are present, the central peak of the CFI does not vanish entirely. With linear dependence of $p$ on $\theta$ as $p(\theta) \approx \frac{1}{2}+C \theta$, the derivative $C=\frac{\partial p(\theta)}{\partial \theta}|_{\theta=0}=2F_C[\theta=0]^{1/2}$  remains non-zero, preserving a significant portion of the CFI peak. This ensures that the estimation sensitivity is maintained even in the presence of modest environmental disturbances.

The actual CFI  can be computed from the following exact detection probability of $\ket{+_i}_q\bra{+_i}$:
\begin{align}
   & P_{+_i}
[\theta] \stackrel{\alpha\gg 1}{\approx}\frac{e^{-\alpha ^2-\left(\alpha '\right)^2} }{\left(\epsilon ^2+1\right)^2}\nonumber\\
&\Bigg(e^{\alpha ^2} \epsilon ^2+\epsilon ^4 \exp \left(2 \alpha  \alpha ' \cos (\theta )\right)+\ii \epsilon ^3 e^{\frac{1}{2} \alpha  \left(\alpha +2 e^{-\ii \theta } \alpha '\right)}-\ii \epsilon ^3 e^{\frac{1}{2} \alpha  \left(\alpha +2 e^{\ii
   \theta } \alpha '\right)}+\epsilon ^2 e^{\left(\alpha '\right)^2}+e^{\alpha ^2+\left(\alpha '\right)^2}\nonumber\\&+\ii f \epsilon ^2 \exp \left(\frac{1}{2} \left(\alpha ^2+2 \alpha  e^{-\ii
   \theta } \alpha '
   +\left(\alpha '\right)^2\right)\right)+f \epsilon ^3 e^{\frac{1}{2} \alpha ' \left(\alpha '+2 \alpha  e^{-\ii \theta }\right)}+f \epsilon ^3 e^{\frac{1}{2}
   \alpha ' \left(\alpha '+2 \alpha  e^{\ii \theta }\right)}-\ii f \epsilon ^2 e^{\frac{1}{2} \left(\alpha ^2+2 \alpha  e^{\ii \theta } \alpha '+\left(\alpha '\right)^2\right)}+2 f
   \epsilon  e^{\alpha ^2+\frac{\left(\alpha '\right)^2}{2}}\Bigg)
\end{align}
which is utilized in all the results presented below. Here,  $\alpha'\equiv\alpha\sqrt{\eta}$ and $f\equiv e^{-\frac{\alpha ^2 (1-\eta )}{2} }$ are adopted for notational simplicity. 
In the large $\alpha$ regime $\alpha \gg 1$, this CFI is approximately given from a simple algebra from $P_{+_i}$. 

 In summary, switching to the $\{\ket{+_i}\bra{+_i}, \ket{-_i}\bra{-_i}\}$ measurement basis offers a trade-off: while it slightly reduces the maximum CFI in ideal conditions {compared to the fragile $\sigma_z$ basis (whose central peak collapses rapidly under noise, see Appendix \ref{append:CFIrange}, it crucially preserves the central CFI peak structure and ensures non-zero sensitivity near $\theta=0$ even under moderate loss and dephasing. {This robustness justifies its use for practical estimation where the parameter may be small, as it simplifies the requirements for the subsequent estimation algorithm. By stabilizing the peak near $\theta=0$, simpler estimators can be used effectively without needing adaptive strategies \cite{RodriguezGarcia2024adaptivephase,deNeevePRR2025Timeadaptive} to track a potentially shifting peak maximum.} It}  substantially enhances the robustness of the estimation protocol against environmental noise. This modification ensures that the high sensitivity of the CFI is preserved within a broader range of $\theta$, thereby improving the overall reliability of qubit-based phase estimation in realistic settings.

 \begin{figure}
     \centering
     \includegraphics[width=1.0\linewidth]{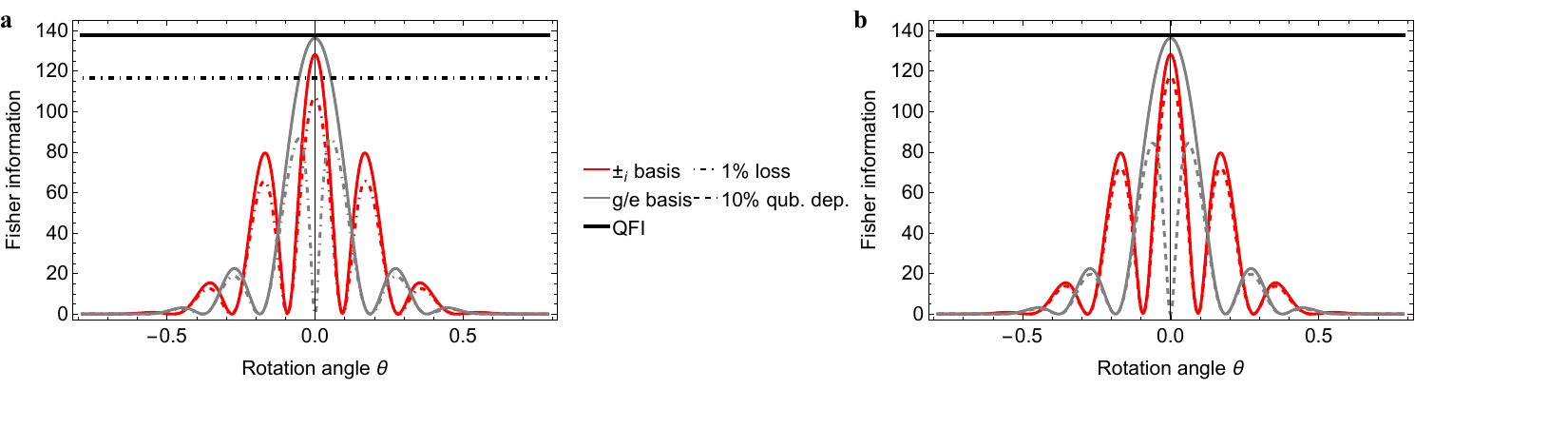}
     \caption{
Fisher information dependence on measurement basis using a protocol approximately preparing a probe $\ket{0}+0.172\ket{\alpha=4i}$. \textbf{a}) Fisher information  under 1\% bosonic loss. \textbf{b}) Fisher information under 10\% qubit dephasing. In both scenarios, the $g/e$ measurement basis (gray curves) significantly diminishes the central peak of the Fisher information, while the $\pm$ measurement basis (red curves) maintains its structure. The $\pm$ basis proves to be more robust under environmental effects. Notably, due to near-disentanglement during phase encoding, qubit dephasing has less severe impact on the CFI. The QFI (black solid line) provides the theoretical maximum across both cases.
}
     \label{fig:basisdep}
 \end{figure}

Figure~\ref{fig:basisdep} illustrates the dependence of the Fisher information on the choice of measurement basis under environmental decoherence effects, such as bosonic loss and qubit dephasing. 
Specifically, the $g/e$ measurement basis (gray curves) leads to a significant suppression of the central Fisher information peak, particularly in the presence of bosonic loss or dephasing noise. In contrast, the $\pm$ measurement basis (red curves) effectively preserves the central peak, ensuring greater robustness against environmental effects.

Interestingly, qubit dephasing has a less pronounced impact on the CFI relative to bosonic loss, attributed to the near-disentanglement during phase encoding. This robustness under dephasing conditions highlights the suitability of the $\pm_i$-measurement basis for maintaining measurement sensitivity in noisy environments. The figure demonstrates  the advantage of tailoring the measurement scheme to counter environmental decoherence effects to protect the maximum CFI peak.

Our theoretical analysis assumes ideal gate operations for both the SCS preparation ($\mathcal{E}, \mathcal{D}$) and the inverse processing steps ($\mathcal{D}^\dagger, \mathcal{E}^\dagger$). In any experimental realization, imperfections stemming from control errors (e.g., pulse timing, amplitude inaccuracies affecting $\alpha$ and $\epsilon$), decoherence during the finite gate execution time, or imprecise calibration will inevitably lead to deviations from the target unitary operations. Such gate infidelities will result in reduced preparation fidelity of the SCS probe and imperfect state recombination during processing, consequently lowering the achievable QFI and CFI compared to the ideal predictions presented herein. While the choice of a deterministic protocol with only two main qubit-oscillator gates is advantageous in minimizing susceptibility to cumulative errors compared to potentially longer sequences required for other states like high-$n$ ON states, a detailed analysis quantifying the sensitivity to specific, platform-dependent gate error models is beyond the scope of this current theoretical work. 
{A crucial next step is to directly quantify the degradation of the CFI as a function of gate infidelity, partially explored in Appendix \ref{sec:appendix_bias} and \ref{app:pulse_simulations}. 
This would provide a more complete picture of the protocol's practical robustness and would be a valuable subject for future studies focused on optimizing end-to-end metrological performance.}



\FloatBarrier

\section{Estimation properties of CFI by SCS}
\label{append:CFIrange}

\subparagraph{Dynamic ranges, and offset of CFI}

For precise parameter estimation, a high CFI is desirable alongside a broad dynamic range, defined as the range of phase values, $\theta$, where the CFI exceeds a specified threshold (typically half the peak CFI or the Gaussian bound). A wide dynamic range is crucial for practical applications where the true parameter value is uncertain. The average CFI over this dynamic range becomes a key metric, balancing sensitivity with usability.

Ideally, the CFI forms a central peak around zero phase, with its width, often quantified by the full width at half maximum (FWHM), defining the dynamic range. However, realistic imperfections like loss or decoherence can introduce offsets in the maximum CFI, shifting or splitting the central peak. These shifts complicate estimation, as the maximum CFI is no longer at zero phase. Notably, SCS states at high average quanta achieve a better balance between peak CFI and dynamic range than squeezed states.

Two methods are proposed to mitigate these offsets:
1. An auxiliary rotation $\theta_0$ to realign the offset peak with the zero-rotation angle.
2. Adjusting the measurement protocol, such as modifying the qubit detection basis, to avoid instabilities and neutralize offsets. This enhances the dynamic range by preventing offset creation.

\subparagraph{Trade-off between the average CFI and dynamic range }

\begin{figure}
    \centering
    \includegraphics[width=1.0\linewidth]{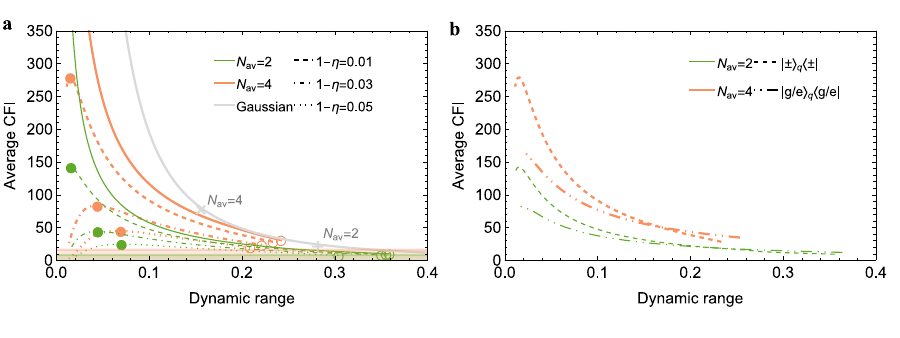}
  \caption{Trade-off between the average CFI and the dynamic range for asymmetric SCS. \textbf{a}) Average CFI versus dynamic range for different average photon numbers ($N_\mathrm{av}=2$ in Green, $N_\mathrm{av}=4$ in orange) and loss levels ($1-\eta$). The dynamic range is quantified by the full width at half maximum (FWHM) of the central CFI peak. Solid lines represent the Gaussian bound for the corresponding $N_\mathrm{av}$ values, while the shaded areas depict the region below the standard quantum limit (coherent states) for each $N_\mathrm{av}$. Filled circles mark the optimal asymmetry parameter $\epsilon$ that maximizes the average CFI for a given $N_\mathrm{av}$ and loss level. Open circles correspond to the symmetric SCS ($\epsilon = 1$). The optimal asymmetric SCS consistently outperform the Gaussian bound and offer a better trade-off between average CFI and dynamic range. As loss increases, the curves shift towards lower average CFI and narrower dynamic range. This trade-off analysis underscores the advantage of asymmetric SCS in balancing precision and estimation range compared to Gaussian and ON states. \textbf{b}) Mitigation of loss-induced shifts. Average CFI versus dynamic range for different measurement bases: $\{\ket{g}_q\bra{g},\ket{e}_q\bra{e}\}$ (dash-dotted lines) and $\{\ket{+}_q\bra{+},\ket{-}_q\bra{-}\}$ (dashed lines), for $N_\mathrm{av} = 2$ (Green) and $N_\mathrm{av}=4$ (orange). The latter basis, equivalent to an offset prevention by an additional qubit rotation, recovers a higher average CFI and a generally wider dynamic range.}
    \label{fig:CFIvsRange}
\end{figure}


Figure \ref{fig:CFIvsRange} presents the inherent trade-off between the average CFI over the primary peak and FWHM  across various SCS parameters, in the presence of bosonic loss. It is observed that probes with higher average quanta numbers $N_\mathrm{av}$ offer advantages in the trade-off. SCS with an highly vacuum-like asymmetry, while showcasing increased CFI, suffer from a diminished dynamic range. The infinite average CFI in this limit is decreasing critically from the effect of loss, leading to the presence of an optimal average CFI value at non-vanishing dynamic range. 
{This optimum arises from balancing the drive towards high asymmetry (increasing interference contrast for high peak CFI) against the increased fragility of highly asymmetric states to loss (which degrades the peak and narrows the useful range).}
The figure highlights specific cases displaying the peak average CFI for each $N_\mathrm{av}$, denoted by marked dots. In comparison to optimal Gaussian techniques using squeezed states for the same $N_\mathrm{av}$ values, an improvement in CFI is evident beyond this benchmark, particularly under moderate loss conditions  within a restricted dynamic range.


\begin{figure}
    \centering
    \includegraphics[width=0.5\linewidth]{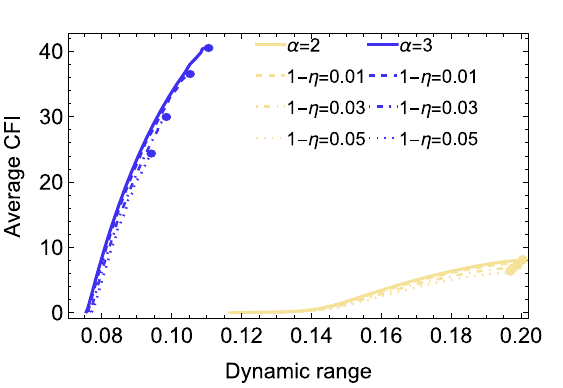}
    \caption{
Average CFI over the primary peak as a function of dynamic range (measured by Full Width at Half Maximum - FWHM) under a fixed coherent state displacement amplitude \( \alpha \) constraints. Results are shown for varying loss levels \( 1 - \eta \) and weight parameters \( \epsilon \). Curves corresponding to \( \alpha_\mathrm{max} = 2 \) (yellow) and \( \alpha_\mathrm{max} = 3 \) (blue) reveal that asymmetric cat states (\( \epsilon = 1 \)) maintain the highest CFI over broad dynamic ranges, demonstrating superior probe performance over ON state across different loss regimes. The plot highlights the optimal balance of CFI and dynamic range for each probe and trade-off of dynamic range and average CFI, where curves for a larger \(\alpha_\mathrm{max} \) exhibit a steeper decay as the loss increases, showcasing a higher sensitivity to environmental effects. In the highly asymmetric regime where the state approaches the vacuum, the central peak and its dynamic range converges to non-zero finite value, much slower than the reduction of the central peak height.   
}
    \label{fig:CFIscalingalpha}
\end{figure}

\subparagraph{Scaling of average CFI under loss in the $\alpha$-constraint}

With a constant $\alpha_\mathrm{max}$ constraint illustrated in Fig. \ref{fig:CFIscalingalpha}, the symmetric cat state ($\epsilon=1$) stands out by optimizing both the average CFI and dynamic range, differing from optimization approaches based on average quanta number $N_\mathrm{av}$. This highlights how the choice of optimal probes is influenced by the constraints applied.

\begin{figure}
    \centering
    \includegraphics[width=0.5\linewidth]{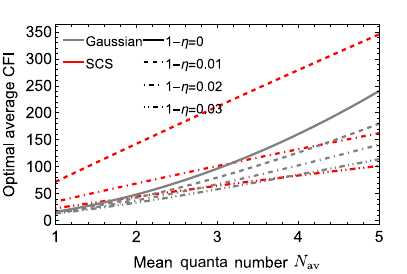}
    \caption{
Optimal average CFI as a function of the mean number of quanta \( N_\mathrm{av}\) for varying levels of loss \( 1 - \eta \). The CFI scales nearly linearly with \(  N_\mathrm{av}\) for optimal SCS probes, even as losses increase from \( 1 - \eta = 0.01 \) to \( 1 - \eta = 0.03 \), shown in different line styles. Higher loss levels display a slightly lower CFI but still maintain a robust performance. The gray represents the Gaussian benchmarks by QFI in (\ref{eq:GaussianLimit}). Despite increasing losses, SCS probes with larger \( N_\mathrm{av}\) continue to exhibit their advantageous phase sensitivity, making them resilient choices under realistic experimental conditions. 
}
    \label{fig:aCFIscale}
\end{figure}

In optimized SCS in $\alpha$ and $\epsilon$ showing the largest average CFI over the primary peak, Fig. \ref{fig:aCFIscale} demonstrates the average CFI increases nearly linearly with the average quanta number $N_\mathrm{av}$, leading to enhanced sensitivity, although the non-Gaussian enhancement manifests up to a threshold $N_\mathrm{av}$. 
The dynamic range measured by FWHM  remains mostly consistent, experiencing only a slight reduction for probes associated with higher CFIs.


The total area under the CFI curve as a function of the target parameter (in this work the angle for the oscillator phase rotation) $\theta$ plays a crucial role in quantifying the efficiency and reliability of parameter estimation methods. To capture the essence of sensitivity and utility of estimation across the entire spectrum of $\theta$ values, this area metric offers insights into the performance of the estimation strategy under varying conditions. In practical applications where achieving precise control over $\theta$ is challenging or not feasible, the significance of this total area is amplified as it mirrors the collective efficacy of the estimation approach across the broad expanse of $\theta$. Consequently, a larger total area indicates a more robust and adaptable estimation process, capable of providing reliable parameter estimates even in situations where precise calibration of $\theta$ is hindered. In essence, the total area under the CFI curve serves as a comprehensive measure of the estimation strategy's overall performance and adaptability, highlighting its ability to deliver accurate parameter estimates in diverse and challenging scenarios.

{\color{black}

\subparagraph{Comparison of the CFI of symmetric ON states and SCS against squeezed state with binary detection}

To clarify the role of asymmetry for surpassing the fundamental Gaussian QFI bound, it is instructive to compare the performance of symmetric non-Gaussian probes against a more practical benchmark: a squeezed vacuum state measured with a simple binary projection onto itself. 
For a fixed average number of quanta $N_\text{av}$, we compare the CFI of symmetric non-Gaussian probes under a binary projective measurement scheme against CFI of Gaussian states.
For a symmetric ON state, defined as $|\psi\rangle_\text{ON,sym} \propto |0\rangle + |n\rangle$ with the Fock number $n=2N_\text{av}$, a binary projection onto the balanced state $(|0\rangle+|n\rangle)/\sqrt{2}$ yields a CFI from (\ref{eq:ONCFIloss}).
This CFI exhibits $2n=4N_\text{av}$ peaks as a function of the phase $\theta$, and its maximum value, calculated as $\frac{16 N_{\text{av}}^2 \eta ^{2 N_{\text{av}}}}{2 \eta ^{2 N_{\text{av}}}-\eta ^{4 N_{\text{av}}}+3}$, can surpass the CFI of Gaussian CFI in (\ref{eq:CFIgaussianloss}) in an intermediate loss regime of $0.8\lesssim\eta\lesssim 0.9$, demonstrating a practical non-Gaussian advantage.
We obtain similar tendency for the CFI a symmetric SCS which takes a complex form, surpassing the Gaussian strategy using binary detection.
This advantage, however, stems from the sub-optimality of the binary detection for the Gaussian strategy, disappearing if the homodyne detection can be utilized.

\begin{figure}
    \centering
    \includegraphics[width=0.8\linewidth]{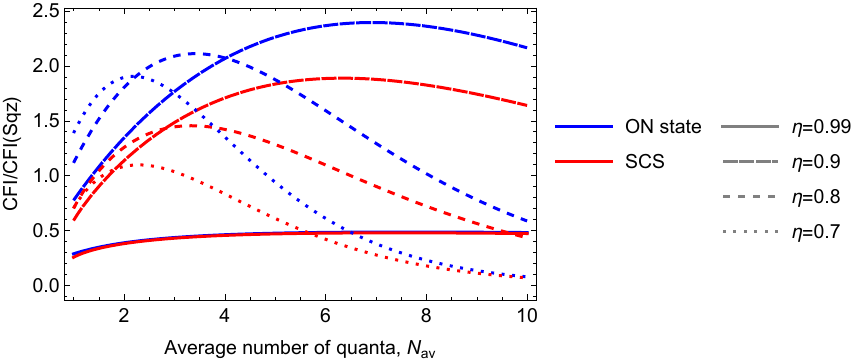}
    \caption{{\textbf{Practical non-Gaussian Advantage of Symmetric Non-Gaussian Probes.} 
Ratio of the CFI for symmetric ON states (blue) and symmetric SCS (red) to the CFI of a squeezed vacuum state measured via binary detection. 
The ratio is plotted against the average number of quanta $N_{\text{av}}$ for various loss levels $\eta$. 
The plot shows that for high transmission efficiencies ($0.6\lesssim\eta \lesssim 0.9$), symmetric non-Gaussian probes can outperform this practical Gaussian benchmark in an intermediate $N_{\text{av}}$ regime. 
This highlights that the choice of measurement for the benchmark state is critical, as a suboptimal measurement can also open a window for  non-optimized non-Gaussian states to provide an advantage.}}
\label{fig:appendix_cfi_ratio}
\end{figure}

To quantify this comparison, we evaluate the CFI for symmetric ON states (using $\ket{0} + \ket{n}$ with $n = 2N_{\text{av}}$) and symmetric SCS (using $\ket{0} + \ket{\alpha}$ with $\epsilon=1$) and plot their performance relative to this practical Gaussian benchmark in Fig.~\ref{fig:appendix_cfi_ratio}. 
The figure reveals a key practical insight: in the intermediate-loss regime ($0.6\lesssim\eta \lesssim 0.9$) and for an intermediate range of average quanta $N_{\text{av}}$, the CFI ratio exceeds one, indicating that both symmetric ON and SCS states can outperform this specific Gaussian strategy. 
The advantage is most pronounced for the symmetric ON state, which can offer more than double the Fisher information of the benchmark protocol around $N_{\text{av}} \approx 6$ for 1\% loss. 
This result does not contradict the conclusions of the main text; rather, it underscores the critical role of the measurement choice for the benchmark state. 
The sub-optimality of the binary detection for the squeezed state creates an operational window where even simpler, symmetric non-Gaussian states offer a tangible metrological advantage. 

\subparagraph{Optimized Projective Measurement for ON States and SCS}

Here, we analyze the projective measurement for quantum phase estimation using probe states that are superpositions of two basis states, such as ON states $|\psi\rangle_{\text{ON}} \propto |0\rangle + \epsilon|n\rangle$ or SCS states $|\psi\rangle_{\text{SCS}} \propto |0\rangle + \epsilon|\alpha\rangle$. 
We consider a measurement onto a state of the same form, $|\phi\rangle \propto |0\rangle + \epsilon'|n\rangle$ (or $|\phi\rangle \propto |0\rangle + \epsilon'|\alpha\rangle$ for SCS), to determine the optimized measurement parameter $\epsilon'$ for a given probe asymmetry $\epsilon$.
The parametrization representation in $N_\mathrm{av}$ and $n$ is straightforwardly obtained, using $N_\mathrm{av}=\frac{\epsilon^2}{1+\epsilon^2}n$.
All analysis below is done under the constraint of $N_\mathrm{av}$. 

The detection probability $P(\theta)$ after a phase rotation is a harmonic function of the form $A+B\cos(\frac{1+\epsilon^2}{\epsilon^2}N_\mathrm{av}\theta)$ for ON state, where $A=\frac{\epsilon ^2 \epsilon '^2 \eta ^{\frac{N_{\text{av}}}{\epsilon ^2}+N_{\text{av}}}+1}{\left(\epsilon ^2+1\right) \left(\epsilon '^2+1\right)}$ and $B=\frac{2 \epsilon  \epsilon ' \eta ^{\frac{\left(\epsilon ^2+1\right) N_{\text{av}}}{2 \epsilon ^2}}}{\left(\epsilon ^2+1\right) \left(\epsilon '^2+1\right)}$. 
For SCS, it takes a non-harmonic form, but the general trends agree well with those of ON states.
The effectiveness of the measurement is characterized by key metrics, whose optimal conditions conflict:
\begin{itemize}
    \item \textbf{Visibility ($V$):} The normalized contrast, $V = B/A$. 
It is maximized to $V=1$ when the measurement project on the state with the inverse asymmetry of the probe, i.e., $\epsilon' = \eta^{-{\frac{1+\epsilon^2}{\epsilon^2}N_\mathrm{av}}/2}/\epsilon$ that takes an increasing value for decreasing $\eta$.
    \item \textbf{Contrast ($C$):} The absolute range of the probability, $C=2B$. 
It is maximized to $\frac{2 \epsilon  \eta ^{\frac{\left(\epsilon ^2+1\right) N_{\text{av}}}{2 \epsilon ^2}}}{\epsilon ^2+1}$ when both measurements project on the states that are symmetric, i.e., $\epsilon' = 1$. 
At $\eta=1$, the symmetric probe $\epsilon=1$ has the largest $C$.  For $\eta<1$, $\epsilon>1$ is favored.
The maximum contrast $C=1$ can be achieved only at $\eta=1$.
The optimized contrast over $\epsilon$ takes a complex form, approximated very well by a heuristically found formula $\exp[0.002 \log[\eta]N_{\text{av}}^2+0.555 \log[\eta]N_{\text{av}}]$ for all $\eta$ and $N_{\text{av}}$.

    \item \textbf{Local Sensitivity (Slope):} The slope is given $\partial P(\theta)/\partial\theta$, whose square is proportional to the CFI. For any given probe $\epsilon$, the maximum slope of $\frac{1+\epsilon^2}{\epsilon^2}N_\mathrm{av}B$ is maximized to $\frac{N_\mathrm{av} \eta ^{\frac{N_\mathrm{av} \left(\epsilon ^2+1\right)}{2 \epsilon ^2}}}{\epsilon }$ by a symmetric projective measurement  $\epsilon'=1$. 
The magnitude of this maximal slope is, in turn, largest for highly asymmetric probe states ($\epsilon \to 0$) under the the constraint of $N_\mathrm{av}$ at $\eta=1$. 
For $\eta<1$, due to the attenuation, the largest maximal slope is given as $\sqrt{-\frac{N_{\text{av}} \eta ^{N_{\text{av}}}}{e \log (\eta )}}$, achieved at small but finite $\epsilon$.
\end{itemize}

This analysis reveals a trade-off between the slope and contrast (or visibility). 
The main text demonstrates that the highest fundamental precision (CFI) is achieved with highly asymmetric probe states ($\epsilon \ll 1$) for a fixed average number of quanta. 
However, the results above show that these  asymmetric states have also lower contrast when a simple, fixed projective measurement is used, among which a symmetric projection ($\epsilon'=1$) is a robust and often optimal choice for the measurement basis. 
Conversely, symmetric states ($\epsilon=1$), which are suboptimal for CFI, maximize the contrast and visibility.
Therefore, leveraging the full metrological advantage of highly asymmetric probe states may technically require more sophisticated techniques, such as adaptive \cite{RodriguezGarcia2024adaptivephase, deNeevePRR2025Timeadaptive}, high-resolution measurements \cite{LiuPhysRevLett2023FullPeriod} or advanced error correction schemes \cite{Kessler2014PhysRevLettQECMetrology,ZhouNatComm2018Heisenberg}, to reach predicted CFI despite of lower signal contrast.
These behavior in contrast and visibility is independent of the average number of quanta in the probe $N_\mathrm{av}$, and their maxima is simply obtained when $N_\mathrm{av}=2n$, while the slope increase linearly with it at $\eta=1$.

}

\FloatBarrier

\section{CFI of displaced Fock states}
\label{append:displaced Fock}


In lossless scenarios,  the preparation and measurement setup is considered in an abstract way, the CFI of displaced Fock state $D[\alpha]\ket{n}_O$ is calculated as 
\begin{align}
    F_C=\frac{4 \alpha ^4 \sin ^2(\theta ) \left(L_n\left(-2 \alpha ^2 (\cos (\theta )-1)\right)+2 L_{n-1}^1\left(-2 \alpha ^2 (\cos (\theta
   )-1)\right)\right)^2}{e^{-2 \alpha ^2 (\cos (\theta )-1)}-L_n\left(-2 \alpha ^2 (\cos (\theta )-1)\right)^2},
\end{align}
where $L_n$ is the Laguerre polynomial. It reaches the QFI in the asymptotic limit of $\theta=0$.
This CFI peak is very fragile to noise and loss, and is reduced to zero at $\theta=0$ rapidly even though the loss and noise is extremely weak, similarly to ON states below but with worse performance.

\FloatBarrier

\section{Effects of thermal noise}
\label{appendix:thermal}

\begin{figure}
    \centering
    \includegraphics[width=1\linewidth]{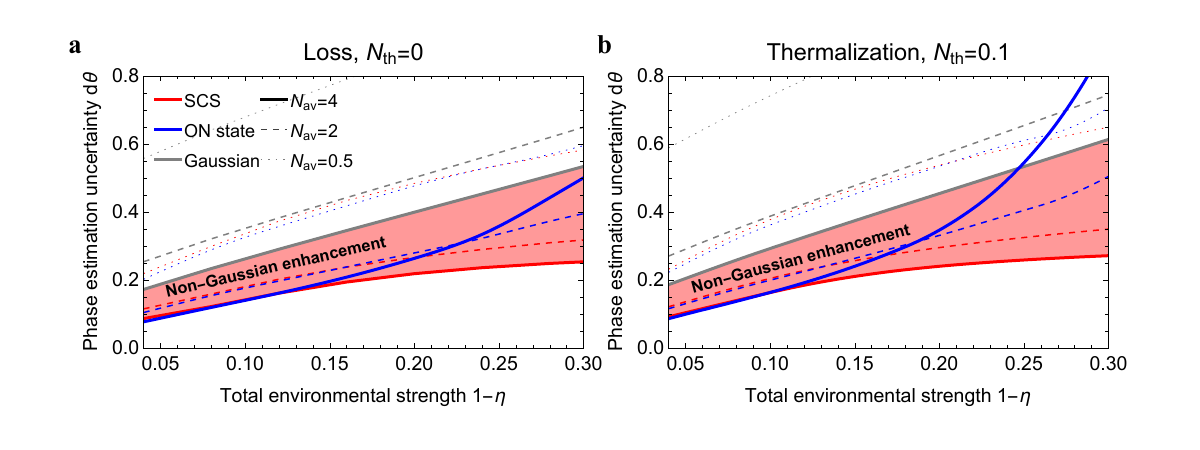}
    \caption{Phase estimation uncertainty $d\theta $ predicted  by QFI as a function of total environmental strength $\eta $ for various quantum probes under \textbf{a}) loss ($ N_\mathrm{th} = 0 $) and \textbf{b}) thermal noise ($ N_\mathrm{th} = 0.1 $). SCS demonstrate superior robustness with the lowest estimation uncertainty (red) compared to the ON states (blue) and Gaussian states (gray) across both loss and thermal noise conditions. This highlights the resilience of SCS {for sensing} in realistic noisy environments. The analysis shows that SCS maintain better precision in the strong environmental effects, particularly as environmental effects intensify due to the fragility of ON states.
 }
    \label{fig:figlossthermalvsT}
\end{figure}

We analyze the performance of SCS, comparing them with Gaussian states and ON states under thermal noise conditions. Keeping the average quanta number $N_\mathrm{av}$ constant, we find that SCS exhibit superior resilience to thermal noise. Figure \ref{fig:figlossthermalvsT} highlights the superior resilience of SCS to thermal noise.  Under the impact of thermal noise, modelled by an interaction with a thermal bath with a small average photon number  $N_\mathrm{th} = 0.1$, the maximum CFI for SCS decreases more slowly with increasing thermal noise compared to Gaussian states, while ON state quickly loses the non-Gaussian enhancement. The ability of SCS to maintain its quantum advantage under these conditions makes it a more promising candidate for robust quantum sensing compared to Gaussian and ON states which degrade more rapidly.

\begin{figure}
    \centering
    \includegraphics[width=0.5\linewidth]{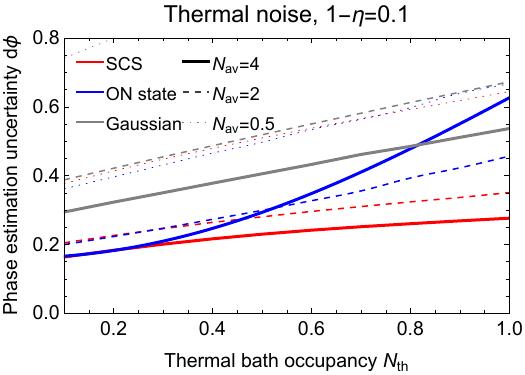}
    \caption{Phase estimation uncertainty $d\theta$ predicted by QFI as a function of thermal bath occupancy $N_\mathrm{th}$ under thermal noise at $1-\eta = 0.1$. The SCS (red) exhibits the lowest estimation uncertainty, outperforming the ON state (blue) and Gaussian states (gray) as the thermal bath occupancy increases. This result demonstrates the superior robustness of the SCS compared to Gaussian-bound probes and ON state which loses the non-Gaussian enhancement quickly for phase estimation in thermally noisy environments.}
    \label{fig:dphivsNth}
\end{figure}

Figure~\ref{fig:dphivsNth} illustrates the phase estimation uncertainty $d\theta$ predicted by QFI as a function of thermal bath occupancy $N_{\text{th}}$ under a fixed thermal noise strength $1-\eta = 0.1$. The SCS consistently exhibits the lowest estimation uncertainty compared to the ON state and Gaussian states  as $N_{\text{th}}$ increases. This superior performance of SCS highlights their enhanced robustness against thermal noise, making them more effective for phase estimation in thermally noisy environments. This demonstrates that the advantage of SCS over Gaussian probes persists even under strong thermal effects. The robustness of SCS makes them particularly suitable for systems where thermal noise is significant, such as in trapped ions, ensuring enhanced phase estimation precision despite environmental challenges.

\begin{figure}
    \centering
    \includegraphics[width=1\linewidth]{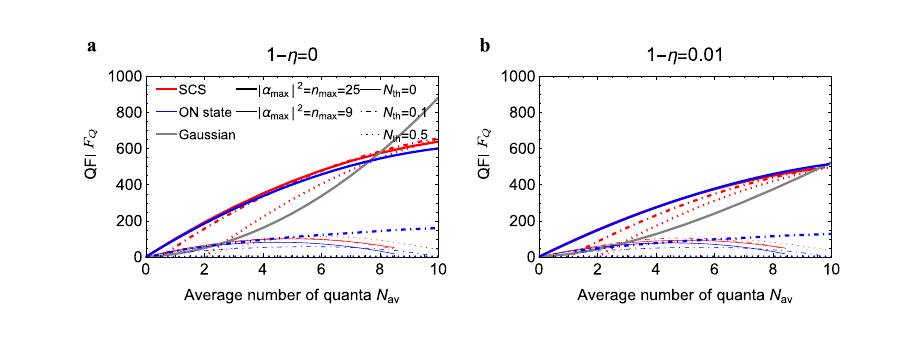}
    \caption{Impact of initial thermal occupation and loss on the QFI of different quantum states. \textbf{a}) QFI as a function of the average number of quanta ($N_\mathrm{av}$) in the absence of loss ($1-\eta=0$). 
    Initial thermal occupation reduces the QFI, particularly for ON states. \textbf{b}) QFI versus $N_\mathrm{av}$ with a weak loss of $1-\eta=0.01$. The presence of even a small loss significantly degrades the QFI of ON states, especially at higher $N_\mathrm{th}$ values, while SCS maintain a higher QFI, demonstrating superior robustness against combined thermal noise and loss. }
    \label{fig:initialthermal}
\end{figure}

Figure~\ref{fig:initialthermal}\textbf{a} illustrates the dependence of the QFI on the average number of quanta $N_\mathrm{av}$, for varying thermal peak displacements $\alpha_\mathrm{max}$ and initial thermal occupancies $N_\mathrm{th}$. Higher values of $\alpha_\mathrm{max}$ generally lead to increased QFI even for initial thermal noise, while larger initial thermal occupancies $N_\mathrm{th}$ tend to decrease it, although not dramatically. Figure~\ref{fig:initialthermal}\textbf{b} shows when a weak loss has occurred. The SCS responds to it moderately, while the ON state rapidly loses the non-Gaussian enhancement.

\FloatBarrier

{\color{Black}

\section{Estimation Bias due to Rabi Strength Errors}
\label{sec:appendix_bias}

In practical implementations, imperfections in control parameters can lead to systematic errors, or bias, in the estimated parameter. Bias, defined as $\langle \theta' \rangle - \theta$ (where $\theta'$ is the estimated phase and $\theta$ is the true phase), quantifies the average deviation of the estimate from the true value. Here, we analyze the bias introduced specifically by a simulated systematic error of 1\% in the Rabi strengths used in the preparation and measurement gates of our protocol employing asymmetric SCS probes ($\ket{0} + \epsilon \ket{\alpha}$), even if we have an infinite number of probes.

 \begin{figure}[htbp]
     \centering
     \includegraphics[width=0.6\linewidth]{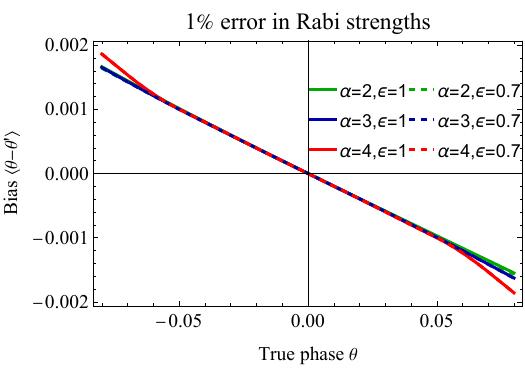} 
     \caption{Simulated estimation bias $\langle \theta - \theta' \rangle$ versus the true phase $\theta$ for asymmetric SCS probes under a 1\% systematic error in Rabi strengths. Curves for different coherent amplitudes ($\alpha=2, 3, 4$) and asymmetry parameters ($\epsilon=1, 0.7$) are shown, indicating weak dependence on these state parameters for this specific error model, while different parameters do not affect the linear dependence critically.}
     \label{fig:bias_appendix}
 \end{figure}

Figure~\ref{fig:bias_appendix} (corresponding to the user-provided plot) illustrates the calculated bias, as a function of the true phase $\theta$ near $\theta=0$. The analysis considers different SCS probe states characterized by coherent amplitudes $\alpha \in \{2, 3, 4\}$ and asymmetry parameters $\epsilon \in \{1, 0.7\}$.
 The bias is small, on the order of $10^{-3}$ for true phases up to $|\theta| = 0.05$, approximately about 1\% of the true value. The bias is approximately zero at $\theta=0$ and exhibits a nearly linear dependence on $\theta$ in this range, with a negative slope (i.e., $\langle \theta - \theta' \rangle$ is positive for $\theta < 0$ and negative for $\theta > 0$). This indicates that the estimator tends to slightly overestimate the magnitude of the phase shift under this error model. A systematic error in Rabi strength likely causes a small, phase-dependent rotation in the effective measurement basis or state preparation, leading to this bias.
 Crucially, the curves for different values of $\alpha$ and $\epsilon$ nearly overlap. This suggests that, for this specific error mechanism (1\% Rabi strength error) and within this range of parameters ($\alpha \ge 2$, $\epsilon \in \{0.7, 1\}$), the introduced bias is only weakly dependent on the specific coherent amplitude or asymmetry of the SCS probe state. The small overlap between the vacuum and coherent state components for $\alpha \ge 2$ might render the bias mechanism relatively insensitive to further increases in $\alpha$ or moderate changes in $\epsilon$.
The small magnitude of the bias and its weak dependence on the state parameters $\alpha$ and $\epsilon$ are encouraging results for the practical feasibility of the protocol. It suggests a degree of robustness against small systematic errors in Rabi strength, at least concerning the resulting estimation bias near $\theta=0$.

However, this analysis is specific to systematic errors in Rabi strengths. Other potential error sources, such as photon loss, dephasing during phase accumulation, qubit state preparation and measurement errors, or errors in other control parameters, could introduce different biases with potentially stronger dependencies on the SCS state parameters ($\alpha, \epsilon$). Furthermore, the analysis is limited to a small range around $\theta=0$. A comprehensive assessment of the protocol's robustness requires simulating various error types and magnitudes across a wider operational range.

In conclusion, while small systematic errors in Rabi strengths introduce a bias in the phase estimation, this bias is found to be small and relatively insensitive to the specific choice of $\alpha$ (for $\alpha \ge 2$) and $\epsilon$ (around 1) for the asymmetric SCS probes near zero phase shift.
Analysis of the bias by the ON states also exhibit a comparable level of bias.
}

\FloatBarrier

\section{Simulation of SCS Preparation with Optimized Pulses}
\label{app:pulse_simulations}

To validate the practical feasibility of preparing the advantageous asymmetric SCS discussed in the main text using the proposed two-gate sequence, we performed numerical simulations incorporating optimized control pulses in bosonic circuit quantum electrodynamics (cQED) platform.

\subsection{Methodology}
In a typical bosonic cQED system~\cite{blais2004cavity}, the Hamiltonian of the qubit-oscillator system can be written as 
\begin{equation}
    \hat H = \hat H_0 + \hat H_d,
\end{equation}
where the bare term reads
\begin{eqnarray}
    \hat H_0 &=&\Delta \hat a^{\dagger} \hat a- \chi|e\rangle \langle e| \hat a^{\dagger} \hat a -\frac{K}{2} \hat a^{\dagger}\hat a^{\dagger}\hat a\hat a-\frac{\chi^{\prime}}{2}|e\rangle \langle e|\hat a^{\dagger}\hat a^{\dagger}\hat a\hat a, \label{EQ_H_bare}
\end{eqnarray}
and the driving term reads
\begin{eqnarray}
    \hat H_d &=&(\varepsilon(t)\hat \sigma_-+\text{h.c.})+(\Omega(t)\hat a+\text{h.c.}). 
\end{eqnarray}
The Hamiltonian above is written in a rotating frame with the frequencies of the qubit and oscillator drives. We have used $\Delta$ to denote the detuning between the oscillator and its drive, $\chi$ ($\chi^{\prime}$) the first (second) order dispersive qubit-oscillator coupling, $K$ the self-Kerr of the oscillator, $\hat a$ $(\hat a^{\dagger})$ the annihilation (creation) operator of the oscillator, $\hat \sigma_-=|g\rangle \langle e|$, $|g\rangle$ ($|e\rangle$) the ground (excited) state of the qubit, $\varepsilon(t)$ and $\Omega(t)$ are respectively the time-dependent drive amplitudes of the qubit and oscillator.
We have chosen a resonant drive for the qubit, whereas for the cavity the detuning is chosen as $\Delta = \chi/2$ such that the first two terms in Eq.~(\ref{EQ_H_bare}) equals to $(\chi/2)\hat \sigma_z \hat a^{\dagger}\hat a$, which is responsible for encoding the phase rotation after the state preparation~\cite{Pan2024}. The Hamiltonian parameters, together with the lifetimes of the system, are taken from an experimental device in Ref.~\cite{krisnanda2025demonstrating}, which are summarized in table~\ref{table_H_para}. 

\begin{table*}[h]
    \centering
    \begin{tabular}{|l|c|c|}
        \hline
        \hline
        Oscillator self-Kerr  & $K/2\pi$ & $6$ \rm{kHz}\\
        1st order qubit-oscillator coupling & $\chi/2\pi$ & 1.423 MHz\\
        2nd order qubit-oscillator coupling  & $\chi^{\prime}/2\pi$ & 16 \rm{kHz}\\
        Oscillator single-photon lifetime  &$T_{c,1}$ & 992 $\mu s$\\
        Qubit energy lifetime  & $T_{q,1}$ & 113 $\mu s$\\
        Qubit coherence lifetime & $T_{q,2}$ & 48 $\mu s$\\
        \hline
        \hline
    \end{tabular}
    \caption{System parameters and lifetimes.}
    \label{table_H_para}
\end{table*}

Here, we follow the protocol for phase estimation in cQED that starts with the initial state $|g\rangle |0\rangle$ followed by a unitary $\hat U_1$, phase rotation $\exp(\ii\hat a^{\dagger} \hat a \theta)$ by the qubit-cavity coupling with $\theta=\chi t/2$, a second unitary $\hat U_2$, and measurements of the projector $|g\rangle\langle g|\otimes|0\rangle\langle0 |$~\cite{Pan2024}. Any two unitary operations that implement $\hat U_1|g\rangle | 0\rangle=|g\rangle| \psi(\alpha)\rangle_{\text{SCS}}$ and $\hat U_2|g\rangle| \psi(\alpha)\rangle_{\text{SCSb}}=|g\rangle | 0\rangle$, where $| \psi(\alpha)\rangle_{\text{SCS}}=N_{\alpha,\epsilon}(|0\rangle+\epsilon |\alpha \rangle)$ and $| \psi(\alpha)\rangle_{\text{SCSb}}=N_{\alpha}(|0\rangle+\ii |\alpha \rangle)$, are sufficient. The expectation value of the projector can then be written as
\begin{eqnarray}
    p(\theta) &=& |\langle g|\langle 0| \hat U_2e^{i\hat a^{\dagger} \hat a \theta} \hat U_1|g\rangle |0\rangle|^2\nonumber \\
    &=& |\langle\psi(\alpha)|_{\text{SCSb}} e^{i\hat a^{\dagger} \hat a \theta} |\psi(\alpha)\rangle_{\text{SCS}}|^2.
\end{eqnarray}

To implement the two unitary operations in the protocol, we employ GRadient-Ascent Pulse Engineering (GRAPE)~\cite{heeres2017implementing} to optimize the time-dependent drive amplitudes $\varepsilon(t)$ and $\Omega(t)$ for a duration of 3 $\mu$s. 
Given an initial state and Hamiltonian of the system, the algorithm optimizes the fidelity between the final state resulting from the dynamics and the target state. 
Note that the optimization is performed under ideal condition, i.e., no decoherence. 
{The simulation of the gate dynamics also includes realistic hardware decoherence, with rates taken from the experimental parameters in Table~\ref{table_H_para}. 
These channels include not only oscillator single-photon loss (\(T_{c,1}\)), but also qubit energy relaxation (\(T_{q,1}\)) and dephasing (\(T_{q,2}\)). 
The combined effect of these processes during the gate execution introduces a degree of infidelity. }


\subsection{Results}
\begin{figure}[htbp]
\centering
\includegraphics[width=0.5\linewidth]{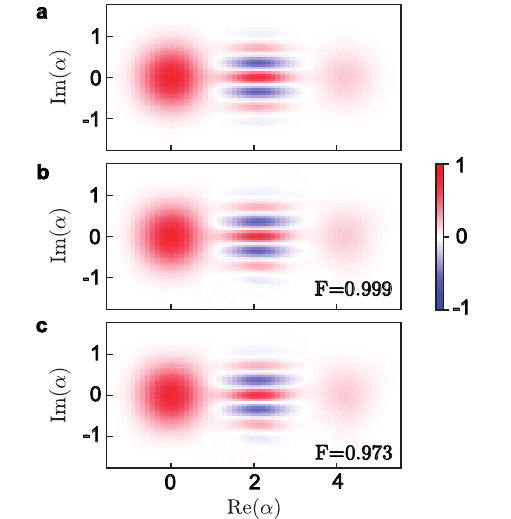} 
\caption{Wigner plots of an exemplary state $|\psi(\alpha)\rangle_{\text{SCS}}$ with $\alpha=4.20$ and $\epsilon=0.45$. \textbf{a}) shows the target state, while \textbf{b}) and \textbf{c}) show the simulated final state after applying drive amplitudes obtained via GRAPE without and with decoherence taken into account, respectively.}
\label{fig_wignerSCS}
\end{figure}

Figure~\ref{fig_wignerSCS} shows the Wigner function of an exemplary generated state $|\psi(\alpha)\rangle_{\text{SCS}}$ with $\alpha=4.20$ and $\epsilon=0.45$ from $\hat U_1$ implemented by GRAPE. Panel (a) shows the target state, panel (b) the final state after the dynamics without decoherence, and panel (c) the final state with decoherence (see Table~\ref{table_H_para}) in the dynamics. We see that the Wigner plot in panel (c) closely resembles the ideal case in panel (a), with reduced contrast in the fringes as a result of the decoherence. 

\begin{figure}[htbp]
\centering
\includegraphics[width=0.5\linewidth]{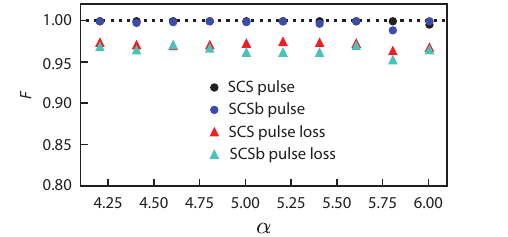} 
\caption{The fidelities of the generated final states $| \psi(\alpha)\rangle_{\text{SCS}}$ after applying $\hat U_1$ and $| \psi(\alpha)\rangle_{\text{SCSb}}$ after applying $\hat U_2$ implemented via GRAPE for different amplitudes of SCS $\alpha$. We plotted simulation results both without and with decoherence in the system taken into account.}
\label{fig_fidelitiesSCS}
\end{figure}

Furthermore, the fidelity of all the SCS used in the main text is plotted in Fig.~\ref{fig_fidelitiesSCS} against the SCS amplitude $\alpha$. Here, SCS pulse (loss) denotes the case where $\hat U_1$ is implemented to target $|g\rangle| \psi(\alpha)\rangle_{\text{SCS}}$ given the initial state $|g\rangle |0\rangle$ without (with) decoherence in the dynamics and SCSb pulse (loss) the case where where $\hat U_2$ is implemented to target $|g\rangle| 0\rangle$ given the initial state $|g\rangle | \psi(\alpha)\rangle_{\text{SCSb}}$ without (with) decoherence in the dynamics. In all cases, we have fidelities near unity for the pulse simulation without decoherence. On the other hand, the fidelities reduce to $\sim 0.97$ when decoherence is included. 


\subsection{Conclusion}
These simulation results, utilizing optimized control pulses within realistic parameter ranges for \(\alpha\) up to approximately 6, strongly support the practical viability of the asymmetric SCS probes analyzed in this work within this regime. 
The demonstrated high fidelities{, achieved despite the inherent realistic hardware decoherence (including both oscillator and qubit loss channels) during the gate dynamics,} confirm that the theoretical potential for enhanced phase estimation surpassing the Gaussian bound is accessible via experimentally relevant control techniques.
Further simulations and experimental investigations are necessary to thoroughly characterize the fidelity and resource requirements for preparing SCS with significantly larger coherent amplitudes (\(\alpha \gg 6\)) and to fully map out the practical boundaries of this approach for achieving maximal quantum-enhanced precision.

\bibliography{test2.bib}



\end{document}